\documentclass[12pt]{iopart}
\usepackage{graphicx}
\usepackage{bm}
\usepackage{natbib}
\newcommand{\dec}[2]{#1\,10^{#2}}
\newcommand{\EQ}{\begin{equation}}
\newcommand{\EN}{\end{equation}}
\newcommand{\EQA}{\begin{eqnarray}}
\newcommand{\ENA}{\end{eqnarray}}
\newcommand{\eq}[1]{(\ref{#1})}
\newcommand{\EEq}[1]{Equation~(\ref{#1})}
\newcommand{\Eq}[1]{equation~(\ref{#1})}
\newcommand{\Eqs}[2]{equations~(\ref{#1}) and~(\ref{#2})}

\newcommand{\Eqss}[2]{equations~(\ref{#1})--(\ref{#2})}

\newcommand{\Sec}[1]{\S\,\ref{#1}}

\newcommand{\Fig}[1]{Figure~\ref{#1}}
\newcommand{\FFig}[1]{Figure~\ref{#1}}
\newcommand{\Tab}[1]{Table~\ref{#1}}
\newcommand{\Figs}[2]{Figures~\ref{#1} and \ref{#2}}

\newcommand{\bra}[1]{\langle #1\rangle}

\newcommand{\meanemf}{\overline{\mbox{\boldmath ${\mathcal E}$}} {}}

\newcommand{\meanFFF}{\overline{\cal F}}
{}

\newcommand{\meanB}{\overline{B}}
\newcommand{\meanC}{\overline{C}}
\newcommand{\meanJ}{\overline{J}}

\newcommand{\meanU}{\overline{U}}

\newcommand{\meanO}{\overline{\Omega}}

\newcommand{\meanBB}{\overline{\mbox{\boldmath $B$}}}

\newcommand{\meanJJ}{\overline{\mbox{\boldmath $J$}}}

\newcommand{\meanUU}{\overline{\mbox{\boldmath $U$}}}

\newcommand{\meanuu}{\overline{\mbox{\boldmath $u$}}}

\newcommand{\meanAAAA}{\overline{\mbox{\boldmath ${\mathsf A}$}} {}}
\newcommand{\meanSSSS}{\overline{\mbox{\boldmath ${\mathsf S}$}} {}}
\newcommand{\meanAAA}{\overline{\mathsf{A}}}
\newcommand{\meanSSS}{\overline{\mathsf{S}}}
\newcommand{\hatBB}{\hat{\bm{B}}}

\newcommand{\hatB}{\hat{B}}
\newcommand{\pp}{\hat{\mbox{\boldmath $\phi$}} {}}

\newcommand{\uu}{\mbox{\boldmath $u$} {}}

\newcommand{\UU}{\mbox{\boldmath $U$} {}}
\newcommand{\bb}{\mbox{\boldmath $b$} {}}

\newcommand{\BB}{\mbox{\boldmath $B$} {}}
\newcommand{\NN}{\mbox{\boldmath $N$} {}}

\newcommand{\AAA}{\mbox{\boldmath $A$} {}}

\newcommand{\jj}{\mbox{\boldmath $j$} {}}
\newcommand{\RR}{\mbox{\boldmath $R$} {}}

\newcommand{\JJ}{\mbox{\boldmath $J$} {}}

\newcommand{\ff}{\mbox{\boldmath $f$} {}}
\newcommand{\EE}{\mbox{\boldmath $E$} {}}
\newcommand{\FF}{\mbox{\boldmath $F$} {}}
\newcommand{\TT}{\mbox{\boldmath $T$} {}}

\newcommand{\kk}{\bm{k}}
\newcommand{\xx}{\bm{x}}

\newcommand{\zz}{\bm{z}}
\newcommand{\rr}{\bm{r}}
\newcommand{\rrr}{\bm{\hat{r}}}
\newcommand{\zzz}{\bm{\hat{z}}}
\newcommand{\SSS}{\mbox{\boldmath $S$} {}}
\newcommand{\grav}{\mbox{\boldmath $g$} {}}
\newcommand{\nab}{\mbox{\boldmath $\nabla$} {}}
\newcommand{\OO}{\mbox{\boldmath $\Omega$} {}}
\newcommand{\oo}{\mbox{\boldmath $\omega$} {}}
\newcommand{\ttau}{\mbox{\boldmath $\tau$} {}}

\newcommand{\SSSS}{\mbox{\boldmath ${\sf S}$} {}}

\newcommand{\ii}{{\rm i}}

\newcommand{\DD}{{\rm D} {}}
\newcommand{\dd}{{\rm d} {}}
\newcommand{\const}{{\rm const}  {}}

\def\la{\mathrel{\mathchoice {\vcenter{\offinterlineskip\halign{\hfil
$\displaystyle##$\hfil\cr<\cr\sim\cr}}}
{\vcenter{\offinterlineskip\halign{\hfil$\textstyle##$\hfil\cr<\cr\sim\cr}}}
{\vcenter{\offinterlineskip\halign{\hfil$\scriptstyle##$\hfil\cr<\cr\sim\cr}}}
{\vcenter{\offinterlineskip\halign{\hfil$\scriptscriptstyle##$\hfil\cr<\cr\sim\cr}}}}}

\def\Tat{\mbox{\rm Ta}_{\rm turb}}

\def\Ma{\mbox{\rm Ma}}

\def\Pm{\mbox{\rm Pr}_M}
\def\Rm{\mbox{\rm Re}_M}
\def\Rmcrit{\mbox{\rm Re}_{M,{\rm crit}}}
\def\Rey{\mbox{\rm Re}}
\def\Pe{\mbox{\rm Pe}}
\def\vA{v_{\rm A}}
\def\cs{c_{\rm s}}
\def\kf{k_{\rm f}}
\def\urms{u_{\rm rms}}

\def\Brms{B_{\rm rms}}
\def\orms{\omega_{\rm rms}}

\def\etat{\eta_{\rm t}}

\def\half{{\textstyle{1\over2}}}
\def\threehalf{{\textstyle{3\over2}}}
\def\onethird{{\textstyle{1\over3}}}

\newcommand{\mG}{\,{\rm mG}}
\newcommand{\uG}{\,\mu{\rm G}}

\newcommand{\nHz}{\,{\rm nHz}}

\newcommand{\K}{\,{\rm K}}

\newcommand{\g}{\,{\rm g}}
\newcommand{\s}{\,{\rm s}}
\newcommand{\cm}{\,{\rm cm}}

\newcommand{\km}{\,{\rm km}}
\newcommand{\mpers}{\,{\rm m/s}}
\newcommand{\kms}{\,{\rm km/s}}

\newcommand{\Mm}{\,{\rm Mm}}
\newcommand{\pc}{\,{\rm pc}}
\newcommand{\kpc}{\,{\rm kpc}}

\newcommand{\yr}{\,{\rm yr}}
\newcommand{\Myr}{\,{\rm Myr}}
\newcommand{\Gyr}{\,{\rm Gyr}}
\newcommand{\erg}{\,{\rm erg}}

\newcommand{\dyn}{\,{\rm dyn}}

\begin{document}

\title{Astrophysical turbulence modeling}
\author{Axel Brandenburg$^{1,2}$ \& {\AA}ke Nordlund$^3$}
\address{1. NORDITA, Roslagstullsbacken 23, SE-10691 Stockholm, Sweden}
\address{2. Department of Astronomy, Stockholm University, SE-10691 Stockholm, Sweden}
\address{3. Niels Bohr Institute, Juliane Maries Vej 30, DK-2100 Copenhagen \O, Denmark}

\begin{abstract}
The role of turbulence in various astrophysical settings is reviewed.
Among the differences to laboratory and atmospheric turbulence we
highlight the ubiquitous presence of magnetic fields that are generally
produced and maintained by dynamo action.
The extreme temperature and density contrasts and stratifications
are emphasized in connection with turbulence in the interstellar medium
and in stars with outer convection zones, respectively.
In many cases turbulence plays an essential role in facilitating enhanced
transport of mass, momentum, energy, and magnetic fields in terms of
the corresponding coarse-grained mean fields.
Those transport properties are usually strongly modified by anisotropies
and often completely new effects emerge in such a description that have no
correspondence in terms of the original (non coarse-grained) fields.
({\it \today, $ $Revision: 1.125 $ \!\!$}).
\end{abstract}


\section{Introduction}

Astrophysical flows tend to be turbulent in the sense
of being highly irregular.
The study of astrophysical turbulence is important for several reasons.
Firstly, turbulence needs to be taken into account when modeling
most astrophysical systems.
It can provide enhanced turbulent viscosity, turbulent heating,
turbulent pressure, and leads to other effects, some of which can be
non-diffusive in nature.
Secondly, turbulence needs to be taken into account when interpreting
observations of such systems.
This is particularly evident in modeling line broadening and
line asymmetries.
Thirdly, astrophysical turbulence often spans an enormous range of length
scales, allowing unique insights into the scaling properties of
turbulence.

In many text books various definitions of turbulence are suggested.
However, none of them is quite without problems.
Throughout this review, turbulence will remain a loosely defined
property of flows that are highly irregular in space and time.

Astrophysical turbulence as such is in principle not any different
from ordinary turbulence.
What is characteristic about it is the extremes in some
parameters, e.g.\ huge Reynolds numbers, Prandtl numbers very different
from unity, and, in some cases, strong density stratification and/or very
high Mach numbers.
Also, of course, the gas is often ionized and hence electrically
conducting, so the interaction with magnetic fields cannot be neglected.
As a rule, astrophysical flows tend to be magnetized spontaneously
by self-excited dynamo action.

Contrary to laboratory and technical realizations of turbulence, where
the driving often comes from the interaction with boundaries, astrophysical
turbulence tends to be largely independent of explicit boundaries and
is facilitated by intrinsic instabilities.
Another difference between astrophysical and laboratory turbulence
is the fact that, with very few exceptions, in situ observations are
impossible and one has to rely on the radiative properties of the
gas to infer velocity, temperature, and magnetic fields, for example.
Yet another difference is that in some astrophysical flows the gas is
extremely tenuous and close to collisionless, so the fluid approximation
may actually break down.
In some cases, multi-fluid descriptions are possible, for example when
charged and neutral species move at different speeds,
have different temperatures, or when positive
and negative charge carriers, as well as dust particles need to be considered.
However, quite often the multi-fluid description is then also not sufficient
and it is better to employ more accurate techniques using, for example,
particle-in-cell (PIC) methods or to solve the underlying Vlasov equations.
This can be made more feasible by making use of the guide-field or gyro-kinetic
approximations, where one averages out the azimuthal particle position
around magnetic field lines.

Astrophysical turbulence has been discussed
in many excellent text books and reviews \cite{%
2003matu.book.....B,
1971ARA&A...9..323S,
1972ARA&A..10..261S,
1978RPPh...41.1929B,
1990ARA&A..28..263S,
1995tlan.book.....F,
1995ARA&A..33..283G,
2004tise.book.....D,
2004ARA&A..42..211E,
2004ARA&A..42..275S
}.
In recent years, however, high resolution numerical simulations have become
feasible and have added significantly to our understanding.
Furthermore, the availability of three-dimensional codes has helped
to make astrophysical turbulence a natural ingredient to many
astrophysical investigations.
The purpose of this review is to highlight recent progress in the field.
We will focus on hydrodynamic and magnetohydrodynamic (MHD) aspects,
but will try to keep the level of repetition with earlier reviews at
a minimum.
In particular, we shall not go in depth into aspects of dynamo theory,
but refer instead to the recent review of Brandenburg \& Subramanian \cite{%
2005PhR...417....1B
} on recent progress and in particular on the nonlinear saturation of dynamos.

\section{Commonly used tools and conventions}

Throughout this review we assume some basic level of familiarity with
commonly used tools and techniques in turbulence research.
Here we only review the essentials in simplistic terms.

\subsection{Spectra}
\label{Spectra}

Common tools include energy and helicity spectra, as well as structure
functions and structure function exponents.
These concepts become particularly useful if spatial homogeneity
can be assumed.
In simulations this usually means that one deals with triply periodic
boundary conditions.
Alternatively, one can apply these tools to just one or two periodic
directions (for example convection in a domain with periodic boundary
conditions in the horizontal directions).

In incompressible (or nearly incompressible) isotropic turbulence one usually
defines the spectral energy per unit mass,
\EQ
E(k,t)=\!\!\sum_{k_-<|\kk|\leq k_+}\!\!|\hat{\uu}(\kk,t)|^2,
\EN
where $k_\pm=k\pm\delta k/2$ mark a constant linear interval around
wavenumber $k$, and the hat on $\uu$ denotes the three-dimensional
Fourier transformation in space.
The spectral kinetic energy is normalized such that
\EQ
\int_0^\infty E(k)\,\dd k=\half\bra{\uu^2},
\EN
where angular brackets denote volume averaging.
This equation shows that the dimension of $E(k,t)$ is
$\cm^3\s^{-2}$, and $E(k)$ can be interpreted as the kinetic
energy per unit mass and wavenumber.

In turbulent flows spectra remain in general time-dependent, so one
is interested in spectra that are averaged over a sufficiently long
time span.
Such spectra can then also be compared with analytic predictions
where statistical averaging is adopted instead.

In strongly compressible flows one can also define the spectrum of
kinetic energy per unit volume as
\EQ
E_{2}(k,t)=\!\!\sum_{k_-<|\kk|\leq k_+}\!\!|\widehat{\rho^{1/2}\uu}|^2,
\EN
and the spectrum
\EQ
E_{3}(k,t)=\!\!\sum_{k_-<|\kk|\leq k_+}\!\!|\widehat{\rho^{1/3}\uu}|^2,
\EN
which does not have a simple physical interpretation, except that
$E_{3}(k,t)^{3/2}$, integrated over $k$,
has the dimension of an energy flux \cite{%
1955IAUS....2..121L
}.
In strongly compressible (e.g.\ supersonic) flows these various spectra
can become quite distinct.
The closest agreement between spectra for subsonic and supersonic
turbulence is achieved when using the quantity $E_{3}(k,t)$ \cite{%
2007ApJ...665..416K
}.

In anisotropic turbulence it is useful to consider the spectral energy
dependence along and perpendicular to the preferred direction of the
turbulence, i.e.\ $E(k_\perp,k_\parallel,t)$.
Examples where this is important include rotating turbulence and turbulence in
the presence of a strong magnetic field, but also inhomogeneous turbulence
such as stratified turbulence and convection where one usually considers
the spectral dependence on the horizontal wavenumber only.

The kinetic helicity spectrum is defined as
\EQ
F(k)=\!\!\sum_{k_-<|\kk|\leq k_+}\!\!
(\hat{\oo}^*\cdot\hat{\uu}+\hat{\oo}\cdot\hat{\uu}^*),
\EN
where $\oo=\nab\times\uu$ is the vorticity, and asterisks
denote complex conjugation.
The kinetic helicity spectrum is normalized such that
\EQ
\int_0^\infty F(k)\,\dd k=\bra{\oo\cdot\uu}.
\EN
This kinetic helicity spectrum obeys the realizability condition,
\EQ
|F(k)|\leq 2kE(k),
\label{Flt2kE}
\EN
which is easily demonstrated by decomposing velocity and vorticity
into positively and negatively polarized waves \cite{%
2005PhR...417....1B,
2002AN....323...99B
}.
Sometimes the helicity is defined with a 1/2 factor, just like the energy is.
In that case the factor $2$ in \Eq{Flt2kE} would disappear.

Equivalent concepts and definitions also apply to the magnetic
field $\BB$, where one defines spectra of magnetic energy $M(k)$,
magnetic helicity $H(k)$, and current helicity $C(k)$, which are
normalized such that $\int M(k)\,\dd k=\bra{\BB^2}/2\mu_0$, where $\mu_0$
is the vacuum permeability, $\int H(k)\,\dd k=\bra{\AAA\cdot\BB}$, and
$\int C(k)\,\dd k=\bra{\JJ\cdot\BB}$.
Here, $\AAA$ is the magnetic vector
potential with $\BB=\nab\times\AAA$ and $\JJ=\nab\times\BB/\mu_0$
is the current density.
The magnetic helicity and its spectrum are gauge-invariant because of
the assumed periodicity of the underlying domain.
In that case the addition of a gradient term, $\nab\Lambda$, in $\AAA$ has
no effect, because $\bra{\nab\Lambda\cdot\BB}=\bra{\Lambda\nab\cdot\BB}=0$,
where we have used the condition that $\BB$ is solenoidal.
Additional mathematical properties can be found in Ref.~\cite{%
1978mfge.book.....M
}.
Magnetic helicity is an important quantity, because it is conserved in
the limit of vanishing magnetic resistivity and in the absence of boundary losses.
Another similarly conserved quantity is the cross-helicity, $\bra{\uu\cdot\BB}$.
Its sign indicates whether Alfv\'en waves travel preferentially parallel
or antiparallel to the local magnetic field.

\subsection{Turbulent cascade}

The energy-carrying scale is often defined as the scale
$\ell_{\rm f}=2\pi/k_{\rm f}$, where $k_{\rm f}$ is the wavenumber
where the energy spectrum peaks.
It is close to the integral scale $\ell_{\rm I}=2\pi/k_{\rm I}$,
where $k_{\rm I}^{-1}=\int k^{-1} E(k)\,\dd k/\int E(k)\,\dd k$.

Turbulence is driven either by some explicit stirring
or by some type of instability.
Explicit stirring is frequently used in direct numerical simulations (DNS)
and large eddy simulations (LES).
Here, DNS means that one considers the original equations with the proper
diffusion term, as opposed to other schemes such as LES that are motivated
by numerical considerations and limited resolution.
An astrophysical example is the driving accomplished by supernova
explosions in the interstellar medium within each galaxy.
Examples of instability-driven turbulence include Rayleigh-B\'enard
convection, the magneto-rotational instability (MRI), and shear flow
instabilities with inflection points resulting from rigid surfaces such as the
accretion disc near the surface of a neutron star.
In the latter case the domain is obviously no longer periodic.

The driving usually occurs over a certain range of length scales
around the wavenumber $k_{\rm f}$.
The nonlinearity of the hydrodynamic equations produces power on
progressively smaller scales (larger wavenumbers).
Qualitatively, this leads to a cascade of energy from large to small
scales until energy is dissipated at scales corresponding to
the wavenumber $k_{\rm d}$.
The range of wavenumbers between $k_{\rm f}$ and $k_{\rm d}$ is called
the inertial range.
An important quantitative property of turbulence is the approximate
constancy of spectral energy flux $\epsilon$ throughout the inertial
range, where $\epsilon$ has dimensions $\cm^2\s^{-3}$.
Making the ansatz
\EQ
E(k)=C_{\rm K}\epsilon^a k^b,
\label{EkCK}
\EN
where $C_{\rm K}$ is the Kolmogorov constant, the values of the
exponents $a$ and $b$ are determined by matching the dimensions
for length (cm) and time (s) as follows: $3=2a-b$ and $2=3a$,
respectively.
This yields $a=2/3$ and $b=-5/3$,
so $E(k)=C_{\rm K}\epsilon^{2/3} k^{-5/3}$.

The length of the inertial range can be calculated by assuming that
$E(k)$ is finite only in the range $k_{\rm f}\leq k\leq k_{\rm d}$.
Thus, $u_{\rm rms}$ and $\epsilon$ are given by the two integrals
\EQ
\half u_{\rm rms}^2=\int_{k_{\rm f}}^{k_{\rm d}} E(k)\,\dd k
\approx{\textstyle{3\over2}}
C_{\rm K}\epsilon^{2/3} k_{\rm f}^{-2/3},
\EN
\EQ
\epsilon=\int_{k_{\rm f}}^{k_{\rm d}} 2\nu k^2 E(k)\,\dd k
\approx{\textstyle{3\over2}}\nu
C_{\rm K}\epsilon^{2/3} k_{\rm d}^{4/3},
\EN
which are just the normalization condition of $E(k)$ and the
definition of the energy dissipation, respectively.
Here, $\nu$ is the kinematic viscosity.
Eliminating $\epsilon$, and writing the result in terms
of the Reynolds number yields
\EQ
\mbox{Re}={u_{\rm rms}\over\nu k_{\rm f}}
\approx{\textstyle{3\over2}}\sqrt{3}\,
C_{\rm K}^{3/2} \left({k_{\rm d}\over k_{\rm f}}\right)^{4/3}.
\label{Rekfkd}
\EN
Thus, the length of the inertial range scales with the Reynolds number like
$k_{\rm d}/k_{\rm f}\approx\mbox{Re}^{3/4}$; see, e.g., Ref.~\cite{%
1951JAP....22..469C
}.

In astrophysics one often deals with extremely large Reynolds numbers,
and hence an extremely broad inertial range.
This is in practice not possible to simulate.
However, many aspects of interest are independent of the length
of the inertial range.
Those that are not exhibit a well-understood scaling with Reynolds number.
This is the main reason why it is at all possible to attempt simulating
astrophysical systems on the computer!

\subsection{Taylor hypothesis and one-dimensional spectra}
\label{TaylorHypothesis}

In laboratory and atmospheric turbulence, for example, one usually
measures time series which allow only one-dimensional spectra to be
determined.
This involves making the Taylor hypothesis, i.e.\ the assumption that
the temporal power spectrum, $\tilde{u}(\omega)$, can be associated with
a spatial one, $\hat{u}(k)$, via $\omega=U_0 k$.
Here, $U_0$ is the mean flow at the location of the detector.

It is important to realize that one-dimensional spectra can differ
from the fully three-dimensional spectra that are normally considered
in numerical simulations of turbulent flows.
The two agree only in regions of the spectrum where one has power law scaling,
i.e.\ where $E(k)\sim k^n$ with some exponent $n$,
This is evidently not the case near the dissipation subrange and near
the sub-inertial range at small wavenumbers.
This is probably the main reason why spectra from high resolution DNS show
a significantly shallower spectrum just before the dissipative subrange
than the one-dimensional spectra obtained using the Taylor hypothesis,
where a shallower part in the spectrum is essentially absent.
In the following we briefly explain this difference \cite{%
2003PhRvE..68b6304D
}.

Consider the case of a one-dimensional spectrum obtained by Fourier
transformation over the $z$ direction.
To relate this to the three-dimensional spectrum, we average over
the remaining $x$ and $y$ directions.
Thus, we compute for $k_z>0$
\EQ
E_{\rm 1D}(k_z)=\int\!\!\int|\hat{\uu}(x,y,k_z)|^2\, {\dd x\,\dd y\over L_xL_y}.
\EN
Next, using Parseval's relation for converting the averaging in real
space to an integration in spectral space, we can write
\EQ
E_{\rm 1D}(k_z)=\int\!\!\int|\hat{\uu}(k_x,k_y,k_z)|^2\, \dd k_x\,\dd k_y
=2\pi\int_0^\infty|\hat{\uu}(k_r,k_z)|^2\, k_r\,\dd k_r,
\EN
where we have assumed that $|\hat{\uu}|^2$ is statistically axisymmetric,
i.e.\ independent of the azimuthal angle about the $k_z$ axis.
Next, we use $k_r^2=k^2-k_z^2$ to replace the $k_r\,\dd k_r$ integration
by one over $k\,\dd k$ in the range from $k_z\leq k<\infty$, i.e.\
\EQ
E_{\rm 1D}(k_z)=2\pi\int_{k_z}^\infty |\hat{\uu}(k)|^2\, k\,\dd k
=\int_{k_z}^\infty{E(k)\over k}\,\dd k,
\EN
where we have used the fact that the three-dimensional spectrum can also be
written as $E(k)=2\pi k^2|\hat{\uu}(k)|^2$, where we have assumed averaging
over full shells in wavenumber space.
Thus, we see that one-dimensional spectra, $E_{\rm1D}(k)$, are related to
the fully three-dimensional spectra, $E(k)$, via integration, or via
differentiation for the reverse operation, i.e.\
\EQ
E_{\rm1D}(k)=\int_k^\infty {E(k')\over k'}\,\dd k'\quad\mbox{and}\quad
E(k)=-k{\dd E_{\rm1D}\over\dd k}.
\EN
We reiterate that, if one of the two spectra were a pure power law,
the other one would also be a pure power law.
However, this assumption breaks down near $k_{\rm f}$ and $k_{\rm d}$.
We mention this aspect here, because one of the unexpected results
obtained from a number of simulations over the last decade is a strong departure
from the Kolmogorov $k^{-5/3}$ slope near $k_{\rm d}$, where the spectrum
can be substantially shallower \cite{%
2003PhFl...15L..21K,
2003PhRvE..68b6304D,
2006PhFl...18g5106H
}.
This is now known as the bottleneck effect \cite{%
1994PhFl....6.1411F
} and was first noticed in atmospheric turbulence \cite{%
1968JPhE....1.1105W
}.
It is by far not as marked in one-dimensional spectra
as in three-dimensional spectra from recent high-resolution DNS
\cite{%
2003PhRvE..68b6304D
}.

\subsection{Intermittency}
\label{Intermittency}

The scaling of velocity differences over fixed distances is different
in different locations.
The flow is therefore said to be intermittent.
A related property is that the scaling of the structure functions,
\EQ
S_p(r,t)\equiv
\left\langle|\uu(\xx+\rr,t)-\uu(\xx,t)|^p\right\rangle,
\label{StructureFunctions}
\EN
with distance $r=|\rr|$ deviates from the scaling $r^{p/3}$
for all moments $p\neq3$, for both parallel ($\rr$ parallel to $\uu$)
and transverse ($\rr$ perpendicular to $\uu$) structure functions.
This property is quantified by the structure function exponents, $\zeta_p$,
which denote the slopes in graphs of $\ln S_p(r,t)$ with $\ln r$.
The averaging, denoted by angular brackets, is here taken to be
over the full volume.

In practice, approximate
scaling can only be identified in a rather limited range of $\ln r$.
Analytic theory predicts $\zeta_3=1$; see, e.g.\ Ref.~\cite{%
1995tlan.book.....F,
2004tise.book.....D
}.
This property is often used to improve the accuracy in the determination
of $\zeta_p$ for $p\neq3$ from numerical or experimental data by plotting
$\ln S_p(r,t)$ versus $\ln S_3(r,t)$.
This procedure is referred to as extended self-similarity or ESS \cite{%
1993PhRvE..48...29B
}.

Intermittency is linked to the property that the $\zeta_p$ deviate from a
linear dependence on $p$.
A completely non-intermittent behavior would mean $\zeta_p=p/3$.
A phenomenological relation that describes the behavior observed in
experiments and simulations is given by the She--Leveque relation \cite{%
1994PhRvL..72..336S
}
\EQ
\zeta_p=\frac{p}{9}+C\left[1-\left(1-\frac{2/3}{C}\right)^{p/3}\right],
\label{SheLeveque}
\EN
where $C$ is interpreted as the co-dimension of the dissipative
structures.
For weakly compressible or incompressible turbulence the dissipative
structures are one-dimensional tube-like structures, so the
co-dimension is $C=2$.
Under compressible conditions the dissipative structures tend to
become two-dimensional sheet-like structures, so $C=1$ \cite{%
2002ApJ...569..841B
},
which is also borne out by simulations of highly supersonic turbulence \cite{%
2002PhRvL..89c1102B,
2002ApJ...573..678B
}.
Sheet-like dissipative structures are also expected in hydromagnetic
turbulence, where these structures correspond to current sheets.
In that case one expects the same scaling as for supersonic turbulence \cite{%
1995PhRvE..52..636P
}.
However, in incompressible hydromagnetic turbulence
with constant density $\rho=\rho_0$, the relevant structure
functions are based on the so-called Elsasser variables
$\zz^\pm=\uu\pm\BB/\sqrt{\mu_0\rho_0}$.
In that case, analytic theory predicts that the mixed third-order
longitudinal structure functions of Politano \& Pouquet \cite{%
1998PhRvE..57...21P
},
\EQ
S_{3\parallel}^\pm(r)=\bra{\delta z_\parallel^\mp(r)
[\delta\zz^{\pm}(r)]^2},
\label{ThirdMag}
\EN
scale linearly with $r=|\rr|$.
Here, $\delta\zz^\pm(r)=\zz^\pm(\xx+\rr)-\zz^\pm(\xx)$,
$\delta z_\parallel^\pm(r)=\delta\zz^\pm\cdot\rrr$,
and $\rrr=\rr/r$ is the unit vector of $\rr$.

Simulations tend to give slightly different scalings for the
longitudinal and transverse structure functions.
This may be a consequence of different cascade speeds for
longitudinal and transverse velocity increments
\cite{%
2004PhRvE..70a5302S
}, but it may also just be an artifact of insufficient resolution and may
go away at larger resolution, as indicated by recent simulations at high
numerical resolution \cite{%
2007ApJ...665..416K
}.

The assumption of the constancy of the spectral flux is well confirmed,
but the correlation between energy injection and energy dissipation
displays significant scatter.
This is mostly because the spectral flux fluctuates significantly in time
and there is some delay before the spectral energy has reached the
dissipation scale.
By taking into account the appropriate delay the scatter can be
significantly reduced \cite{%
2004PhRvE..70e6301P
}.
The energy flux at large scales is characterized by
\EQ
\epsilon=C_\epsilon u_{\rm1D}^3/L,
\label{epsLarge}
\EN
where $C_\epsilon\approx0.5$.
It is customary to define the length scale as $L=3\pi/4\kf$, so
in terms of $k_{\rm f}$ and $u_{\rm rms}^2=3u_{\rm1D}^2$ we
can then write
\EQ
\epsilon\approx0.04\,k_{\rm f} u_{\rm rms}^3.
\label{Epsilon}
\EN
This formula will be useful later in connection with turbulence in
interstellar and intergalactic media.

\section{Sites of astrophysical turbulence}

The following discussion is concerned mainly with observations and simulations
covering a range of astrophysical settings where turbulence occurs.
In some cases strong theoretical evidence is used to
argue for the existence of turbulence, as for example in accretion discs
where turbulence has not yet been observed explicitly \cite{%
2006Natur.441..953M
}.

\subsection{Solar wind}

The gas above the visible surface of the Sun is not in hydrostatic equilibrium.
Instead, because of geometrical constraints and because of a gravitational potential
inversely proportional to the radial distance, there is the possibility
of a critical point, where the radial velocity equals the sounds speed.
The theory of such flows was first understood by Parker \cite{%
1958ApJ...128..664P
} in 1967 and is now explained in a number of text books on
compressible flows or on astrophysical fluid dynamics \cite{%
1992aita.book.....S,
1998pfp..book.....C
}.
Other transonic flows of this type include those through a Laval nozzle,
as well as Roche-lobe overflow between binaries, astrophysical jets
from accretion discs, and Bondi accretion.
In the case of the Sun the gas reaches speeds of around $400\km\s^{-1}$ in
the equatorial plane and $800\km\s^{-1}$ at higher latitudes \cite{%
1995AdSpR..16...85P
}.
The solar wind is turbulent and fluctuates between 300 and $800\km\s^{-1}$
on time scales ranging from seconds to hundreds of hours \cite{%
1995ARA&A..33..283G
}.

\begin{figure}[t!]\begin{center}
\includegraphics[width=.51\textwidth]{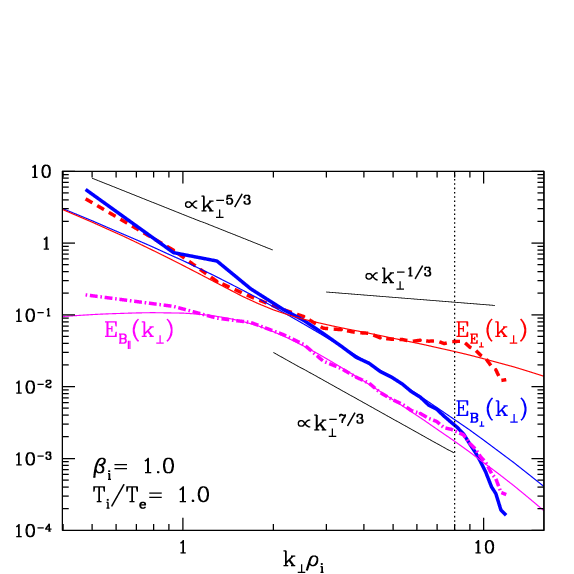}
\includegraphics[width=.47\textwidth]{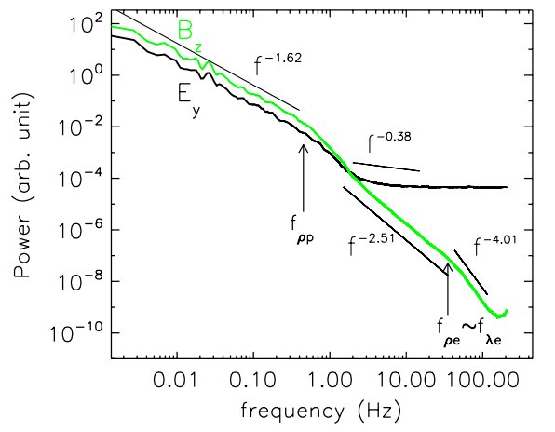}
\end{center}\caption[]{
Spectra of electric and magnetic fields from a gyrokinetic simulation \cite{%
2008PhRvL.100f5004H
} (left) compared with those obtained from the Cluster spacecraft \cite{%
2009PhRvL.102w1102S
} (right).
Note the approximate $k^{-5/3}$ spectrum below the Doppler-shifted
inverse proton Larmor radius
and an approximate $k^{-7/3}$ spectrum for the magnetic field
(solid/blue on the left and light shade/green on the right)
between the Doppler-shifted inverse proton and electron Larmor radii
(in the right hand plot referred to as $f_{\rho{\rm p}}$ and
$f_{\rho{\rm e}}$, respectively),
followed by a steeper dissipation subrange.
Above the inverse Doppler-shifted electron Larmor radius
the electric field spectrum develops
a shallower subrange consistent with $k^{-1/3}$
 (dashed/red on the left and black on the right).
Courtesy of Gregory Howes \cite{%
2008PhRvL.100f5004H
},
as well as Fouad Sahraoui and Melvyn Goldstein \cite{%
2009PhRvL.102w1102S
}, and Copyright (2010) by The American Physical Society.
}\label{Goldstein09}\end{figure}

In the case of the solar wind, spectral information can be obtained
under the Taylor hypothesis that was discussed in \Sec{TaylorHypothesis}.
Using this hypothesis the following properties have been inferred:
\begin{itemize}
\item An approximate $k^{-5/3}$ energy spectrum both for velocity and
magnetic field \cite{%
1995SSRv...73....1T
}.
\item Below the ion Larmor radius a steeper spectrum (between $k^{-2}$
and $k^{-4}$) is found for the magnetic field \cite{
2006ApJ...645L..85S
}.
In view of theoretical expectations the transition to a $k^{-7/3}$
spectrum for the magnetic field together with a $k^{-1/3}$ spectrum
for the electric field is particularly interesting \cite{%
2008PhRvL.100f5004H,
2009PhRvL.102w1102S
}; see \Fig{Goldstein09}.
\item Finite magnetic helicity (negative in the northern hemisphere and
positive in the southern hemisphere), possibly with a $k^{-7/3}$ spectrum
\cite{%
1982PhRvL..48.1256M
}.
\item Finite cross helicity of positive sign, indicating
outward travelling waves \cite{%
1982JGR....87.6011M
}.
\item Decay of turbulence with distance and evidence for additional heating
\cite{%
1995ARA&A..33..283G,
1995SSRv...73....1T,
1988GeoRL..15...88F,
2001JGR...106.8253S
}.
\end{itemize}
A possible connection between a $k^{-7/3}$ tail in the energy spectrum
at small scales (below the scale of the ion Larmor radius) and so-called
electron MHD as a model for collisionless plasmas such as the solar corona
and the Earth's magnetosphere has been discussed \cite{%
1996PhRvL..76.1264B
}.
A similar slope has now also been seen in simulations using the gyrokinetic
equations \cite{%
2008PhRvL.100f5004H
}.
These equations emerge from the Vlasov equations for a collisionless plasma by
averaging over the azimuthal angle of the gyrokinetic motions \cite{%
2009ApJS..182..310S
}.
Let us also mention here the possibility of obtaining spectra steeper than
$k^{-7/3}$ using electron MHD when equipartition between kinetic and magnetic
energies is not satisfied \cite{%
2007ApJ...656..560G
}, or when compressible effects are included \cite{%
2008ApJ...674.1153A
}.

In view of our discussion in \Sec{TaylorHypothesis}, it should be noted
that near the break point where the spectral index changes, the spectra
inferred by using the Taylor hypothesis are not exactly representative of
the three-dimensional spectra obtained from simulations.
However, in view of other general uncertainties, the changes in the
spectral slopes are probably sufficiently weak to be ignorable.

\subsection{Solar convection}
\label{SolarConvection}

The visible surface of the Sun is the photosphere, from where
photons can reach the Earth in a direct path.
Deeper inside the Sun the gas is opaque and photons are continuously
absorbed and re-emitted, following approximately a diffusion-like process.
At the surface, the Sun exhibits a granular pattern that can already be
seen with small amateur telescopes.
The pattern is irregular and changes on a time scale of around 5 minutes.
The horizontal pattern size is 1--$2\Mm$.
Here and elsewhere we use $1\Mm=1000\km$ as a convenient length scale.
The visible granulation is just a thin layer on top of a $200\Mm$ deep
convection zone.
The convection zone covers the outer 30\% of the Sun by radius.
The inner 70\% are convectively stable.
This region is referred to as the radiative interior.

\begin{figure}[t!]\begin{center}
\includegraphics[width=.98\textwidth]{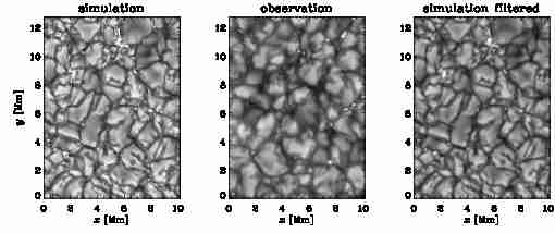}
\end{center}\caption[]{
Comparison between a granulation pattern from a simulation with 12 km
grid size (left), an observed granulation pattern from the
Swedish 1-meter Solar Telescope at disk center
(middle), and the simulated one after convolving with
the theoretical point spread function of a 1 meter telescope.
The simulation images are for wavelength integrated light intensity
while the observed image is for a wavelength band in the near UV.
The image was taken on 23 May 2010 at 12:42 GMT with
image restoration by use of the multi-frame blind de-convolution
technique with multiple objects and phase diversity
\cite{%
2005SoPh..228..191V
}.
Courtesy of V.~M.~J.~Henriques and G.~B.~Scharmer.
}\label{GranulationFig}\end{figure}

Its overall dynamics can be understood through simulations and turbulence
theory (i.e.\ mixing length theory) \cite{%
1990ARA&A..28..263S
}.
Excellent agreement between observed and simulated granulation patterns
has been obtained; see \Fig{GranulationFig}.
By calculating diagnostic spectra in the visible light and comparing
with observations one can determine the abundance of
chemical elements \cite{%
2005ARA&A..43..481A,
2008A&A...488.1031C,
2009ARA&A..47..481A
}.
The chemical element abundances are important for determining the opacity
of the gas which, in turn, determines the radial structure of the Sun.
This will be discussed in more detail in \Sec{Granulation}.

 From the viewpoint of turbulence theory, this type of convection is
special -- not so much because the Rayleigh number is extremely large
($\sim10^{30}$), but mainly because the density and temperature
stratifications are extreme, covering 6 orders of magnitude
of change in density and a factor of $300$ in temperature.
This huge stratification implies that the turbulence characteristics
become strongly depth-dependent.
It has long been anticipated that the energy-carrying scale varies with
depth in such a way that it is proportional to the local pressure scale
height, $H_p$.
The pressure scale height is proportional to the temperature and varies
from about 200\,km at the top of the convection zone to about 60\,Mm at
the bottom.
The typical correlation time of the turbulence is expected to be
proportional to the local turnover time, $H_p/\urms$, where $\urms$
is the rms velocity of the turbulence.
Estimating the convective energy flux as $F_{\rm conv}\sim\rho\urms^3$,
we expect $\urms$ to vary by a factor of 100 from about $4\kms$ at the top
of the convection zone to about $40\mpers$ at the bottom.
Thus, the turnover times vary by more than 4 orders of magnitude, from
minutes at the top of the convection zone to about a month at
the bottom.

\begin{figure}[t!]\begin{center}
\includegraphics[width=.9\textwidth]{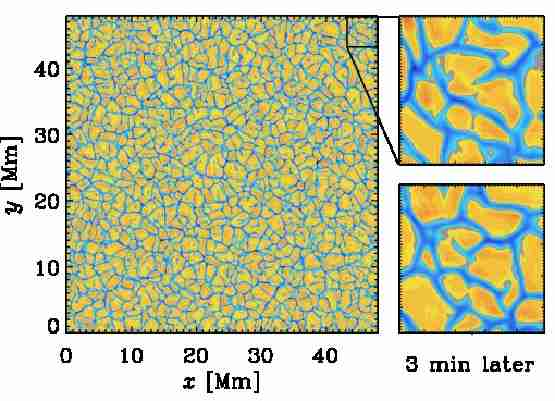}
\end{center}\caption[]{
Vertical velocity near the surface from a solar convection simulation.
The right hand panels show part of the domain at two different times.
}\label{paake_conv}\end{figure}

\begin{figure}[t!]\begin{center}
\includegraphics[width=.9\textwidth]{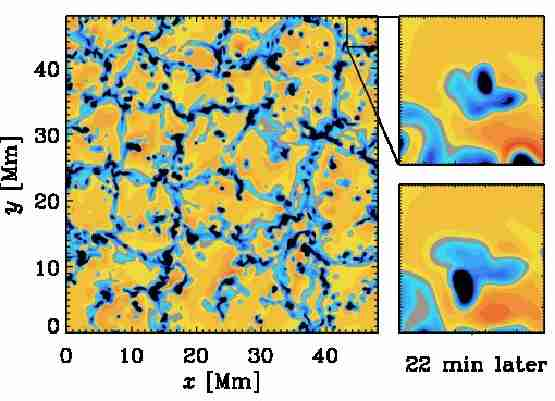}
\end{center}\caption[]{
Same as \Fig{paake_conv}, but for a deeper level, about 4 Mm away from the surface.
}\label{paake_deeper}\end{figure}

A general difficulty in carrying out simulations of the deep solar
convection zone is the long Kelvin-Helmholtz time in deeper layers.
The Kelvin-Helmholtz time can be defined as the ratio of thermal energy
density to the divergence of the energy flux or (operationally more
convenient) as the total thermal energy above a certain layer divided
by the solar luminosity.
This time scale determines the thermal adjustment time and can be
rather long.
However, by preparing initial conditions such that the mean stratification
as well as the fluctuations are close to those in the final state, the
difficulty with long time adjustment times can be alleviated.

\FFig{paake_conv} shows an example from radiation hydrodynamics
simulations of the horizontal pattern of the vertical velocity
near the surface, and \Fig{paake_deeper} the same at a depth of about $4\Mm$.
One sees clearly that the number of cells has decreased and that the
horizontal scale of the cells changes from about $2\Mm$ near the top
to about $10\Mm$ at a depth of about $3\Mm$.  This illustrates two important
properties:  (i) The horizontal cell size below the surface is typically
a few times the distance from the surface (which really reflects that it
is several times the local pressure and density scale heights), and (ii)
the structure size increases so rapidly with depth that even using the
concept of `` cells'' may be misleading.
One also sees that the typical cell life time changes rapidly with depth.
Near the surface the cell pattern shows some clear changes after only
$3\min$, while at $4\Mm$ depth the changes remain more limited even
after $20\min$.
The scales of the patterns and their rates of change are thus generally
consistent, at a semi-quantitative level, with assumptions that have
generally been made in simplified (mixing-length) models of convection:
\begin{itemize}
\item The energy-carrying scales of the turbulence are of the order of
the local vertical pressure scale height, $H_p=|\nab\ln p|^{-1}$.
\item The turbulence varies on time scales comparable to the turnover
time defined as $H_p/u_{\rm rms}$.
\end{itemize}
One should, however, not conclude that the numerical results `confirm' a scaling
with the pressure scale height.  Mass conservation really involves the density
scale height rather than the pressure scale height, and the main reason that
analytical theories of convection have generally tended to avoid using the density
scale height is that, because of a rapid change of the degree of hydrogen
ionization there is a narrow layer close to the surface of stars where the
density scale height may tend to infinity.

Many of the qualitative expectations from
mixing length theory are borne out by simulations.
This also includes the scaling of velocity and the temperature
fluctuations with convective flux and hence with depth.
Indeed, one finds that the convective energy flux (or enthalpy flux),
$F_{\rm conv}$, is proportional to the negative specific entropy gradient.
Velocity and temperature fluctuations scale like $F_{\rm conv}^{1/3}$
and $F_{\rm conv}^{2/3}$, respectively; see Fig.~11 in Ref.~\cite{%
2005AN....326..681B
}.

Early ideas about distinctively different modes of convection at different
scales are mostly due to differences in observational techniques rather than
real physical differences in the convection.
Supergranulation, for example, refers to a convection pattern with
a horizontal scale of about $30\Mm$, which is seen in Dopplergrams
measuring the line of sight velocity.  When plotting the horizontal
velocity amplitude as a function of horizontal size the supergranulation
scales appear to be just a part of a rather featureless power law extending
over many orders of magnitude in size \citep{2009ASPC..416..421S}.
Banana cells, on the other hand, refer to a theoretically expected pattern
of convection in deeper layers.
This expectation is based on the Taylor--Proudman theorem \cite{%
1970JFM....44..441B
},
rather than an observationally established fact, but it remains a pronounced
feature of convection in rotating shells between $\pm30^\circ$ latitude \cite{%
2000ApJ...532..593M,
2010AN....331...73K
}.

\subsection{Other effects of solar turbulence}
\begin{figure}[t!]\begin{center}
\includegraphics[width=.8\textwidth]{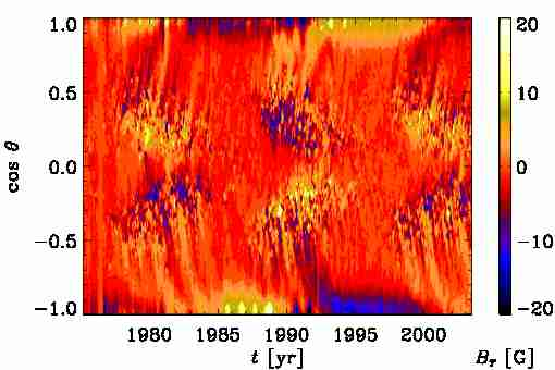}
\end{center}\caption[]{
Longitudinally averaged radial component of the observed solar
magnetic field as a function of cos(colatitude) and time.
Dark (blue) shades denote negative values and
light (yellow) shades denote positive values.
Note the sign changes both in time and across the equator.
Courtesy of R.~Knaack.
}\label{knaack}\end{figure}

There are a number of properties that occur on scales that
are larger than the energy-carrying scale.
These properties include:
\begin{itemize}
\item The angular velocity varies by about 30\% in latitude
(slow at the poles and fast at the equator) with approximate solid body
rotation below the convection zone and a general deceleration in the outer
5\% of the solar radius \cite{%
2003ARA&A..41..599T
}.
\item There is a large-scale magnetic field exhibiting a 22 year cycle
(11 years for the sunspot number) and a statistical antisymmetry
of the radial field with respect to the equator (\Fig{knaack}).
\item The solar surface exhibits a magnetic field that is strongest inside
sunspots, where it is seen through Zeeman splitting.
\item Magnetic and current helicity with strong fluctuations, but well-defined
averages: negative in the northern hemisphere and positive in the southern
hemisphere; see, e.g., Fig.~1 in Ref.~\cite{%
1995ApJ...440L.109P
}.
\end{itemize}

In addition to the convective motions of the Sun, there are
coherent wave patterns that correspond to discrete frequencies
in wavenumber and frequency space.
Using a technique called helioseismology \cite{%
1984ARA&A..22..593D,
1985SoPh..100...65G,
2002RvMP...74.1073C,
2008PhR...457..217B
}, the information contained in these modes can be used to infer the
depth dependence of sound speed and hence the radial dependence of the
temperature of the Sun.

Helioseismic constraints of the core temperature were important
in pinning down the origin of the low observed neutrino flux from the Sun
\cite{%
1988RvMP...60..297B,
2001PhRvD..64a3009S,
2005MNRAS.356..587C
} in terms of neutrino oscillations, i.e.\ the
Mikheyev-Smirnov-Wolfenstein effect \cite{%
1986NCimC...9...17M,
1978PhRvD..17.2369W
}.

Solar rotation lifts the degeneracy of modes with different azimuthal
order and allows a determination of the dependence of the internal
angular velocity on radius and latitude \cite{%
2003ARA&A..41..599T
}.
Rotation also causes the convection pattern to propagate in a prograde
direction \cite{%
2003Natur.421...43G
}.

At the equator, the Sun rotates with a period of about
$t_{\rm rot}=26\,\dd$, but at the poles it spins about 30\% slower.
This is referred to as differential rotation.
The angular velocity is $\Omega=2\pi/t_{\rm rot}$,
but in helioseismology one often talks about the rotation rate,
$\Omega/2\pi$, which is measured in $\nHz$.
The equatorial value at the surface is $452\nHz$.
The radiative interior is found to rotate rigidly \cite{%
2003ARA&A..41..599T
}.
The interface between the differentially rotating convection zone and
the rigidly rotating radiative interior is referred to as the tachocline
\cite{%
1992A&A...265..106S,
2007sota.conf.....H
}.

\subsection{Interstellar turbulence}
\label{InterstellarTurbulence}

The gas between the stars can be observed in absorption or emission
both at infrared and radio wavelengths.
The line of sight velocity component can be determined by Doppler
shifts of spectral lines; see, e.g., Ref.~\cite{2002A&A...390..307O}.
There is a general power law scaling of velocity amplitudes and
velocity differences with geometrical scale \cite{%
2002A&A...390..307O,
1979MNRAS.186..479L,1981MNRAS.194..809L
}. Velocity dispersions scale  with size to a power of about 0.4
from sub-parsec scales to scales of the order of about 1 kpc;
see Fig.~1 of Ref.~\cite{1979MNRAS.186..479L}.
The velocity scaling is practically the same in regions with
varying intensity of star formation, indicating that the velocity
scaling is inertial, and driven mostly by energy input at large scales,
rather than a result of direct, local driving by on-going star
formation \citep{%
2002ApJ...566..289B,2002ApJ...566..276B,2009A&A...504..883B
}.
Direct evidence of turbulence on small length scales ($\sim10^{12}\cm$)
in the ISM comes from radio scintillation measurements \cite{%
1990ARA&A..28..561R,
2003ApJ...584..791B
}.

Galaxies such as our own have typical radii of $R\approx15$
kiloparsecs (kpc).
Here, $1\kpc=3\times10^{21}\cm$ is used as a convenient length scale.
The density decreases rapidly away from the midplane with a typical
density scale height of $H\approx70\pc$.
Near the midplane of a typical galaxy the 3D rms turbulent
velocities are around $15\km\s^{-1}$.
This implies a typical turnover time, $H_p/u_{\rm rms}$, of around
$5\Myr$ (megayears).

An important aspect is the occurrence of supernovae, which mark the
death of massive stars and provide a significant energy release into
the interstellar medium through thermal energy and momentum injection.
Traces of supernovae are seen as supernova remnants, which give
a qualitative idea about the nature of interstellar turbulence.

\begin{figure}[t!]\begin{center}
\includegraphics[angle=-90,width=.35\columnwidth]{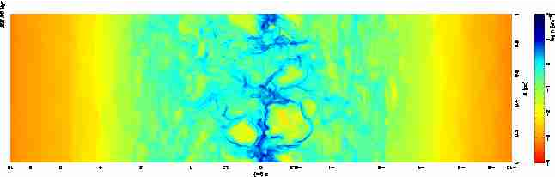}
\includegraphics[angle=-90,width=.35\columnwidth]{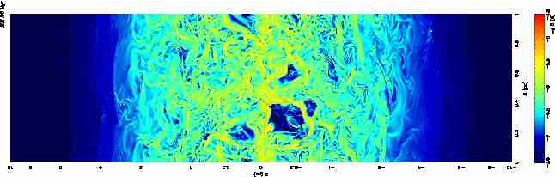}
\end{center}\caption[]{
Two-dimensional slices through a three-dimensional simulation domain
of supernova-driven turbulence in the interstellar medium showing
the vertical distribution of the density (left) and magnetic field (right).
Note the appearance of supernova remnants in density and magnetic fields
as well as an overall concentration around the midplane at $z=0$.
Courtesy of Miguel de Avillez \cite{%
2005A&A...436..585D
}.
}\label{Avillez}\end{figure}

Supernova explosions contribute about $E_{\rm SN}=10^{51}\erg$ per explosion.
With about 20 supernovae per million years per kpc$^2$ estimated
for the solar neighborhood this corresponds to an energy injection
per unit area of
\EQ
\int\epsilon_{\rm SN}\,\dd z\approx
\dec{20}{51}\erg / (\dec{3}{13}\s\times\dec{9}{43}\cm^2)\approx
\dec{7}{-5}\erg\cm^{-2}\s^{-1}.
\EN
This is almost two orders of magnitude more that what is required to
sustain the turbulent energy dissipation per unit area and time,
which, from \Eq{epsLarge}, may be estimated to be
\EQ
\int\epsilon\,\dd z\approx0.5\,\rho u_{\rm1D}^3 \approx
10^{-24}\g\cm^{-3} \, (10^{6}\cm\s^{-1})^3 \approx
10^{-6} \erg\cm^{-2}\s^{-1},
\EN
where the mean density of the interstellar medium is
$\rho\approx\dec{2}{-24}\g\cm^{-3}$ and the one-dimensional
rms velocity is $u_{\rm1D} \approx 10 \kms = 10^{6} \cm\s^{-1}$.
This is also in good agreement with simulations \cite{%
2005A&A...436..585D
}.
A visualization of density and magnetic field strength in such a simulation
is shown in \Fig{Avillez}.

The linear polarization properties of
synchrotron radiation can be used to infer the magnetic field
both along the line of sight via Faraday rotation and perpendicular
to it through the polarization plane projected onto the sky
\citep{2008ApJ...680..457T,2009ApJ...692..844C,2009RMxAC..36..107C}.
The field strength
is typically around $5\uG$ in the solar neighborhood of our Galaxy,
but it can be several $\mG$ in the galactic center \cite{%
1984Natur.310..557Y,
2004ApJS..155..421Y
}.
For many spiral galaxies large-scale magnetic fields have been found.
In many of them the magnetic field is approximately axisymmetric and
symmetric about the midplane \cite{%
1996ARA&A..34..155B
}.

\subsection{Accretion discs}

Accretion discs are disc-like structures through which gas gradually
spirals toward a central massive object while converting potential
energy into kinetic and magnetic energies that are dissipated and radiated away.
This conversion is believed to be of turbulent nature and may
be driven by the magneto-rotational instability \cite{%
2006Natur.441..953M,
1998RvMP...70....1B
}.
An alternative mechanism for disk dissipation is that the disk
functions as a self-regulating buffer.  As long as the disk accretion
towards  the central object is smaller than the rate of mass in-fall
onto the disk from the surrounding nebula, the mass density of the disk
increases.  When the surface density reaches a level sufficient for
gravitationally driven instabilities to develop, spiral waves starts
to grow, develop into spiral shocks, and dissipation in the shocks
then enhances the disk accretion enough to balance the rate of in-fall
onto the disk
\citep{2008ApJ...673.1138C}.

In order to allow material to spiral inward at a mass accretion rate
$\dot{M}$, half of the orbital potential energy must be converted viscously
and resistively into heat and radiation.
Therefore the total (bolometric) luminosity of an accretion disc is \cite{%
1992apa..book.....F
}
\EQ
L={GM\dot{M}\over2R_{\rm in}},
\EN
where $M$ is the mass of the central object and $R_{\rm in}$ is the
inner radius of the accretion disc.
Obviously, the further the disc stretches toward the central object,
i.e.\ the smaller the value of $R_{\rm in}$,
the more efficient the energy conversion will be.
Discs around black holes are most efficient in this respect, because
here the innermost stable orbit is 1--3 Schwarzschild radii,
i.e.\ (2--6)$\times GM/c^2$, where $c$ is the speed of light.
Thus, $L=0.1\times\dot{M}c^2$, which constitutes a much more
efficient conversion than nuclear fusion, where the efficiency is only
$0.007\times\dot{M}c^2$.
Here we have used for $\dot{M}$ the rate of hydrogen burning \cite{%
1992apa..book.....F
}.
Note that the factor 0.007 comes from the relative mass difference between
a helium atom (4.0026) and four hydrogen atoms (1.0078).

\subsection{Turbulence in galaxy clusters}

Galaxies themselves tend to cluster on Mpc scales.
There are typically around $10^4$ galaxies in a cluster, but some
clusters can be substantially smaller.
All clusters are generally strong X-ray emitters, but some are
also strong radio-emitters resulting from synchrotron emission
in the presence of magnetic fields.

Typical temperatures are around $10^8\K$ corresponding to a
sound speed of around $1000\km\s^{-1}$.
The implied velocity dispersion is also of that order,
as expected when the system is in approximate Virial equilibrium.
With typical length scales on the order of the density scale height,
$H_\rho=100\kpc$, the turnover time is $100\kpc/(1000\kms)=0.1\Gyr$.
This would also be the typical decay time of the turbulence in the
absence of mechanisms driving the turbulence.

Mechanisms for driving such turbulence include mutual encounters
of clusters \cite{%
1995ApJ...445...80L,
2008ApJ...680...17W
}.
Given that only a fraction of all galaxy clusters also have strong
radio halos \cite{%
1999PhR...310...97B
}, one may speculate that these clusters have undergone a recent encounter
or merger with another cluster within the last few gigayears.
Obviously, in this scenario one would just have decaying
turbulence between encounters.
In the context of galaxy clusters this subject has been studied by
various groups \cite{%
1999ApJ...518..603R,
1999ApJ...518..594R,
2006MNRAS.366.1437S
}.
Another mechanism that has been discussed in the literature is the driving
by individual galaxies moving through the cluster and producing a turbulent
wake behind them \cite{%
1980ApJ...241..925J,
1989MNRAS.241....1R,
1991ApJ...380..344G
}.

\subsection{Decaying turbulence in the early universe}
\label{DecayingTurbulence}

\begin{figure}[t!]\begin{center}
\includegraphics[width=.5\columnwidth]{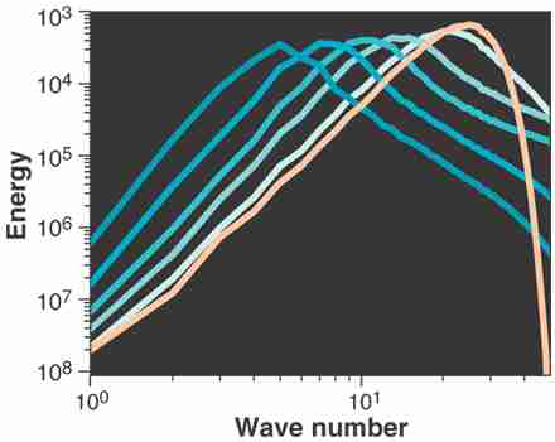}
\end{center}\caption[]{
Magnetic energy spectra at different times
(increasing roughly by a factor of 2).
The curve with the right-most location of the peak corresponds to
the initial time, while the other curves refer to later times (increasing
from right to left).
Note the temporal growth of spectral magnetic energy at wavenumbers
to the left of the peak and the associated
propagation of spectral energy to successively smaller wavenumbers,
i.e.\ to successively larger scales.
Adapted from Refs.~\cite{%
2001Sci...292.2440B,
2001PhRvE..64e6405C
}.
}\label{InvCasc}\end{figure}

Various mechanisms for the generation of ``primordial'' fields have been
proposed \cite{%
2001PhR...348..163G
}.
One problem is that the predicted magnetic field strengths are
extremely uncertain.
Another general problem is the small length scale of such fields.
For example, after the electroweak phase transition, about
$10^{-10}\s$ after the Big Bang, the horizon scale was around $3\cm$.
Magnetic fields generated during such a phase transition
may possess magnetic helicity, but this is also rather uncertain \cite{%
2001PhRvL..87y1302V
}.
However, during the subsequent decay of a helical field, energy is transformed
to larger scale by an inverse cascade of magnetic helicity \cite{%
1996PhRvD..54.1291B,
2001Sci...292.2440B
}.
\FFig{InvCasc} shows the evolution of the resulting magnetic power spectrum
at different times from a direct numerical simulation of the relevant
hydromagnetic equations \cite{%
2001PhRvE..64e6405C
}.
Simulations have demonstrated that turbulence decays in power law
fashion with the total energy being proportional to
$t^{-n}$, where $n=0.5$ for maximally helical fields
and $n=1$ for non-helical fields \cite{%
2001PhRvE..64e6405C
}. By comparison, nonhelical fluid turbulence leads to $n=1.2$ \cite{%
2003JFM...480..129K,
2004PhRvE..70b6405H
}.
As argued by Biskamp \& M\"uller \cite{%
1999PhRvL..83.2195B
}, helical fields may be more typical than non-helical ones.
Of course, the magnetic field generated in rotating bodies
(stars and galaxies, although neither is relevant to the early Universe)
tend to be helical, but of opposite sign in the two hemispheres,
so the net magnetic helicity would cancel to zero.
On the other hand, the helical contribution of a field generated
at an early phase transition will decay more slowly
than the non-helical contribution, and so the
{\it relative} importance of the helical fields will grow with time.

The question of the decay law of helical MHD turbulence is still not
fully settled.
It is generally believed that the magnetic energy, $E_{\rm M}$, follows
a power law decay, i.e.\ $E_{\rm M}\sim t^{-n}$, but proposals for the
value of $n$ range from $2/3$ to $1/2$, depending essentially on the
assumptions made about the evolution of the typical length scale $L$
of the energy-carrying motions.
If one assumes $L$ to be controlled by a resistive evolution of
magnetic helicity, $H_{\rm M}$, i.e.\
\EQ
-{\dd H_{\rm M}\over\dd t}={2\eta\over L^2}H_{\rm M},
\EN
then, for a power law evolution of $H_{\rm M}$ we have $L\sim t^{1/2}$,
and with $E_{\rm M}=H_{\rm M}/L$ and $H_{\rm M}\approx\const$ we find \cite{%
2005AN....326..393C
},
\EQ
E_{\rm M}\sim t^{-1/2}.
\EN
On the other hand, if one discards resistive effects, and assumes
instead that the decay is controlled by inertial range turbulence, i.e.\
\EQ
-{\dd E_{\rm M}\over\dd t}\equiv\epsilon\sim{U^3\over L}
\sim{E_{\rm M}^{3/2}\over L}\sim{E_{\rm M}^{5/2}\over H_{\rm M}},
\EN
then, after integration, we obtain \cite{%
1999PhRvL..83.2195B
}
\EQ
E_{\rm M}\sim t^{-2/3},
\EN
together with $L\sim t^{2/3}$.
Note than in either of the two proposals one has assumed that
$H_{\rm M}\sim LE_{\rm M}\approx\const$.
However, in the former approach $H_{\rm M}$ is not assumed to be
constant exactly, but to decay resistively like
$H_{\rm M}\sim t^{-2\eta t/L^2}$, which implies a corresponding speed-up
of the decay of $E_{\rm M}$ and hence an increase of $n$ from $1/2$ to
$1/2+2\eta t/L^2$.
This may explain why simulations at finite $\eta$ \cite{%
1999PhRvL..83.2195B,
2004PhRvD..70l3003B
} suggest exponents close to $n=2/3$.
This question needs to be followed up again in future at higher resolution,
but simulations at moderate resolution have confirmed the idea of a
correction factor proportional to $t^{-2\eta t/L^2}$ in the decay of
$E_{\rm M}$ \cite{%
2005AN....326..393C
}.

It is still unclear whether such primordial magnetic fields
would have a detectable effect on the polarization signal of
the cosmic background radiation and whether significant fields may have
been present when the first stars or galaxies were formed.
These questions are subject to current investigations \cite{%
2002A&A...387..383D,
2005JCAP...01..009D
}.
Another subject under active investigation concerns the production of
gravitational waves from the Maxwell stress associated with primordial magnetic
fields \cite{%
2005PhRvD..72l3509T,
2006PhRvD..74f3521C,
2008PhRvD..78l3006K,
2008PhRvL.100w1301K
}.

\section{Theoretical studies of turbulence}

\subsection{Incompressible turbulence}
\label{IncompressibleTurbulence}

Most turbulence research is restricted to incompressible turbulence, in
which case the Navier--Stokes equations take the form,
\EQ
{\DD\uu\over\DD t}=-\nab\tilde{p}+\ff+\nu\nabla^2\uu,
\quad\nab\cdot\uu=0.
\label{DuDtincompr}
\EN
Here, ${\rm D}/{\rm D}t=\partial/\partial t+\uu\cdot\nab$ denotes
the advective derivative.
It is this term that constitutes the important nonlinearity of the
Navier--Stokes equations.
In order to understand the nature of the nonlinearity it is useful to make
use of the vector identity $\uu\cdot\nab\uu=\oo\times\uu+\half\nab\uu^2$,
with $\oo=\nab\times\uu$.
Thus, we have
\EQ
{\partial\uu\over\partial t}=\uu\times\oo-\nab\tilde{p}+\ff+\nu\nabla^2\uu.
\label{DuDtincompr2}
\EN
Owing to incompressibility, we have $\rho=\const$, and only the reduced
pressure, $\tilde{p}=p/\rho+\half\uu^2$, enters in \Eq{DuDtincompr2}.
However, because of the solenoidality constraint, $\nab\cdot\uu=0$,
the pressure gradient also constitutes a quadratic nonlinearity
of the form
\EQ
\tilde{p}=\nabla^{-2}\nab\cdot(\uu\times\oo+\ff).
\EN
This relation follows directly from \Eq{DuDtincompr2}
after taking its divergence and noting that
$\nab\cdot\partial\uu/\partial t=\nab\cdot\nabla^2\uu=0$.

\subsection{Compressible fluid dynamics}

In the compressible case, the Navier--Stokes equation can be written
in the form
\EQ
\rho{\DD\uu\over\DD t}=-\nab p+\FF+\nab\cdot\ttau,
\label{DuDtcompr}
\EN
where $\ttau=2\rho\nu\SSSS$ is the stress tensor, here assumed to
be proportional to the kinematic viscosity $\nu$ and the
traceless rate of strain tensor, $\SSSS$, whose components are
\EQ
{\sf S}_{ij}=\half\left(u_{i,j}+u_{j,i}\right)
-\onethird\delta_{ij}\nab\cdot\uu,
\label{StrainDef}
\EN
where commas denote partial differentiation.
Note that the form of the stress tensor
above applies only to a monatomic gas.
In more general cases there may be additional contributions from the
bulk viscosity corresponding to terms proportional to $\delta_{ij}\nab\cdot\uu$.

To compare with the incompressible case, we evaluate
\EQ
{1\over\rho}\nab\cdot\ttau=\nu\left[\nabla^2\uu+\onethird\nab\nab\cdot\uu
+2\SSSS\cdot\nab\ln(\rho\nu)\right],
\label{Fvisc}
\EN
and note that, in addition to the $\nabla^2\uu$ term there is also
a term $\nab\nab\cdot\uu$, which vanishes in the incompressible case,
and a term ${\sf S}_{ij}\nabla_j\ln(\rho\nu)$, which vanishes when the
dynamical viscosity, $\mu=\rho\nu$, is constant.

\EEq{DuDtcompr} has to be solved together with the continuity equation
\EQ
{\DD\rho\over\DD t}=-\rho\nab\cdot\uu,
\label{DrhoDt}
\EN
and an energy or entropy equation,
\EQ
\rho T{\DD s\over\DD t}=2\nu\rho\SSSS^2.
\label{dsdt}
\EN
The heating term is generally given by $u_{i,j}\tau_{ij}$.
Splitting $u_{i,j}=s_{ij}+a_{ij}$ into symmetric and antisymmetric
parts, it is clear that only $s_{ij}$ contributes after multiplying
with another symmetric matrix, i.e.\ with $\tau_{ij}$.
Furthermore, since $\tau_{ij}$ is also trace-free, the result does
not change when adding or subtracting from $s_{ij}$ a term proportional
to $\delta_{ij}$, in particular $\onethird\delta_{ij}\nab\cdot\uu$.
Therefore, we have $u_{i,j}\tau_{ij}=2\rho\nu\SSSS^2$,
which is manifestly positive definite.

For a perfect gas the specific entropy $s$ is related to pressure
and density via
\EQ
s=c_v\ln p-c_p\ln\rho+s_0,
\EN
where $s_0$ is an additive constant.
(The specific entropy $s$ is not to be confused with $s_{ij}$ or ${\sf S}_{ij}$.)
It is important to realize that even in the inviscid limit, $\nu\to0$,
the term $2\nu\rho\SSSS^2$ cannot be neglected in \Eq{dsdt}.
For example across a shock there is always a well-defined increase in
specific entropy that is independent of the value of $\nu$.

In compressible fluid dynamics it is often advantageous to consider the
evolution equations in their conservative form.
This means that the rate of change of the density of a
conserved quantity, $X$, is given by the negative divergence of
its corresponding flux, i.e.\
\EQ
{\partial\over\partial t}\mbox{(density of $X$)}
=-\nab\cdot\mbox{(flux density of $X$)}
+\mbox{sources}-\mbox{sinks},
\EN
where the presence of sources and sinks indicates
additional processes whose detailed evolution is not captured
by the total energy equation within the same framework.
An example is radiation, which provides sources and sinks to
the energy equation as heating and cooling terms.
Alternatively, if the evolution of the radiation energy
is included in the total energy equation, any explicit
heating and cooling terms disappear, and only boundary
(flux divergence) terms remain \cite{%
2005ApJ...624..267T
}.
If there is no radiation, gravity, external forcing, etc, there are
no additional terms, so the conservative form of the equations is
\EQ
{\partial\rho\over\partial t}=-{\partial\over\partial x_j}
\left(\rho u_j\right),
\label{continuityeqn}
\EN
\EQ
{\partial\over\partial t}(\rho u_i)=-{\partial\over\partial x_j}
\left(\rho u_i u_j+\delta_{ij}p-\tau_{ij}\right),
\label{momentumeqn}
\EN
\EQ
{\partial\over\partial t}\left(\rho e+\half\rho\uu^2\right)=
-{\partial\over\partial x_j}
\left(\rho u_j h+\half\rho u_j\uu^2-u_i\tau_{ij}\right),
\label{energyeqn}
\EN
where $h=e+p/\rho$ is the specific enthalpy per unit mass.
For a perfect gas, $h$ and $e$ are proportional to temperature
with $h=c_p T$ and $e=c_v T$, where $c_p$ and $c_v$ are the specific
heats at constant pressure and constant volume, respectively.

The equations above show explicitly that the volume integrals of the terms
under the time derivative are conserved, i.e.\ constant in the absence
of fluxes in or out of the domain.
In one dimension, the terms in parentheses under the spatial derivatives
are constant and, in particular, uniform across a shock.
This allows shock jump conditions to be derived.
Note that, since viscosity acts only locally, these conditions are
independent of the width of the shock.
This is an important property that allows simulating highly supersonic
turbulence using a modified viscosity (Neumann--Richtmyer artificial
viscosity) for smearing out the shock \cite{%
1994dmiv.book.....R
}.
In the presence of source or sink terms in \Eqss{continuityeqn}{energyeqn}
this would no longer be possible.

\subsection{Anelastic approximation}
\label{AnelasticApproximation}

The advantage of making the assumption of incompressibility is not only
that one has one equation less to solve (the $\partial\rho/\partial t$
equation), but mainly that one eliminates sound waves, whose associated
wave speed is often much faster than the speed associated with other
processes.
This means that one can then focus more efficiently on the slower dynamics
of the system.

Incompressibility is normally associated with constant density.
In view of our earlier discussion regarding the strong density
stratification in stars, incompressibility would not be a useful
assumption, even though the sound speed can be much larger than
other speeds such as that associated with the convection itself.
It is then better to relax the condition $\nab\cdot\uu=0$ and use instead
$\nab\cdot\rho\uu=0$.
This is called the anelastic approximation \cite{%
1962JAtS...19..173O,
1969JAtS...26..448G
}.
It is important to realize that with this assumption one {\it replaces}
the original continuity equation \eq{continuityeqn}.
Consequently this equation can then no longer be used to argue that
$\partial\rho/\partial t=0$.
Indeed, $\rho$ is in general not constant in time and can evolve,
while $\nab\cdot\rho\uu=0$ is maintained at all times.
This technique is sometimes used in simulations of solar convection \cite{%
1981ApJS...45..335G,
1982A&A...107....1N,
1984JCoPh..55..461G,
2000ApJ...532..593M,
2004ApJ...614.1073B,
2005ApJ...620..432R
}.

Just like in the incompressible case, also here one has to solve a
Poisson-like equation that emerges when taking the divergence of
the evolution equation for the momentum density.
Taking the divergence of \Eq{momentumeqn} one obtains
\EQ
\nabla^2p=\nab\cdot\RR,
\label{Reqn}
\EN
where $\RR=-\rho\uu\cdot\nab\uu+\FF+\nab\cdot\ttau$ is the sum of
the advection term plus all the other terms on the right hand side of
\Eq{momentumeqn}, except for the pressure gradient term.
The $\FF$ term in \Eq{Reqn} refers to additional terms such as gravity and
Lorentz force terms in \Eq{DuDtcompr}.

The anelastic approximation is sometimes associated with
linearizing the equation of state \cite{%
2000ApJ...532..593M
}.
However, this is not necessary and one can just continue working
with the original, fully nonlinear equation of state \cite{%
1982A&A...107....1N,
1992A&A...265..328B
}.
The only difference is that in the fully compressible case one
would obtain the pressure from density and specific entropy, while in the
anelastic case one obtains the density from pressure and specific entropy,
if the latter is indeed the main thermodynamic variable.

\subsection{Large eddy and hyperviscous simulations}

The maximum achievable Reynolds number scales as
the number of mesh points in one direction, raised to the power
4/3; see \Eq{Rekfkd}.
With the largest attainable resolution being at present $4096^3$ \cite{%
2003PhFl...15L..21K
}, it is impossible to reach Reynolds numbers of $10^6$ and beyond.
In many engineering applications of turbulence one needs to calculate flows
at very large Reynolds numbers and one therefore uses large eddy simulations.
This involves some representation of the unresolved Reynolds
stress in terms of other flow variables.
This approach can be rather uncertain.
Unlike engineering applications, where
such models can be tested against measurements, this is usually
not possible in astrophysics, due to a large number of
additional complications (strong stratification, magnetic fields,
rotation, etc.) that are hard to realize in the laboratory.
The best one can therefore hope for is a rigorous comparison of large
eddy simulations with DNS.
Examples of this are discussed in \Sec{SimulationsOfTurbulence}.

One of the simplest subgrid scale models is the Smagorinsky model \cite{%
1963MWRv...91...99S
}.
This approach is strictly dissipative, i.e.\ the Reynolds stress of the
unresolved velocity fluctuations, denoted here by primes, is modeled by
a viscous stress of the form
\EQ
\overline{u'_iu'_j}=-2(C_{\rm S}\Delta x)^2\rho|\SSSS|{\sf S}_{ij},
\EN
where $C_{\rm S}$ is the Smagorinsky constant (between 0.1 and 0.2) \cite{%
1993PhFl....5.2306S,
2004PhyA..338..379A,
2007JFM...570..491L
},
and the rate-of-strain tensor $\SSSS$ was defined in \Eq{StrainDef},
and is here applied to the resolved motions $\meanuu$, i.e.\ excluding
the subgrid scale motions.
Another approach, which cannot be classified as large eddy simulation,
consists in using hyperviscosity.
In spectral space, the viscosity operator $-\nu k^2$ is simply replaced
by $-\nu_n k^{2n}$, where $n>1$ is the order of hyperviscosity.
Unlike the Smagorinsky model, the results from this approach are known
not to converge to the original Navier-Stokes equations, but the hope is
that in the inertial range the flow is unaffected by the unphysical
form of the diffusion operator.
This is indeed the case, as was demonstrated in Ref.~\cite{%
2004PhRvE..70b6405H
}.

\subsection{Turbulence simulations using Godunov/PPM-type schemes}

The Godunov scheme is a conservative numerical scheme for solving partial
differential equations.
In this method, the conservative variables are considered as piecewise
constant over the mesh cells at each time step and the time evolution
is determined by the exact solution of the Riemann shock tube problem
at the intercell boundaries.
This scheme consists of first defining a piecewise constant approximation
of the solution at the next time step.
The resulting scheme is usually first-order accurate in space.
This approximation corresponds to a finite volume method representation
whereby the discrete values represent averages of the state variables over
the cells.
Exact relations for the averaged cell values can be obtained from the
integral conservation laws.
Next, the solution for the local Riemann problem is obtained at the cell
interfaces.
This is the only physical step of the whole procedure.
The discontinuities at the interfaces are resolved as a superposition of
waves satisfying locally the conservation equations.
The original Godunov method is based upon the exact solution of
Riemann problems.
However, approximate solutions can be applied as an alternative.
Finally, the state variables are averaged after one time step.
The state variables obtained after the second step are averaged over each
cell defining a new piecewise constant approximation resulting from the
wave propagation during the time step.

Nowadays one uses often higher-order Godunov schemes for astrophysical
applications.
One such method is the piecewise parabolic method that is also referred to as PPM.
Examples of such codes include Athena \cite{%
2008ApJS..178..137S
}, Pluto \cite{%
2008A&A...488..429T
}, Nirvana \cite{%
2007CoPhC.176..652G
}, Ramses \cite{%
2002A&A...385..337T
}, Flash \cite{%
2000ApJS..131..273F
}, and Enzo \cite{%
2005ApJS..160....1O
}.
Such codes have been used for many astrophysical applications including supersonic,
isotropic homogeneous turbulence \cite{%
1992PhRvL..68.3156P
}.

\subsection{Analyzing and modeling turbulence with wavelets}

Wavelets are sometimes used both to analyze and to model turbulence.
In particular the wavelet technique has been used for extracting coherent
vortices out of turbulent flows.  The aim is to retain only the essential
degrees of freedom responsible for the transport.
It is intriguing that with this technique one can actually retain nearly
all velocity structure and dissipation information in turbulent flows by
using a relatively small selection of wavelets with non-zero amplitudes
\cite{1992AnRFM..24..395F}; 
see also Ref.~\cite{%
2007PhFl...19k5109O
}.
This method is related to the so-called proper orthogonal decomposition
of turbulent flows
\cite{%
1993AnRFM..25..539B
}.
This decomposition is statistically based and permits the extraction of
spatio-temporal structures that are judged essential according to
predetermined criteria.
It is not only useful in the analysis and synthesis of data from
simulations and experiments, but it also allows the construction of
low-order models from  the Navier-Stokes equations.
Finally, let us note that
the wavelet representation has been applied with success to simulations
of resistive drift-wave turbulence in magnetized plasma Hasegawa-Wakatani system \cite{%
2008PhPl...15g2305B
}.

\section{Extra ingredients to turbulence in astrophysical flows}

\subsection{Passive scalars: mixing and dust dynamics}

One of the simplest additional ingredients in fluid dynamics in general,
and in turbulence physics in particular, are passive scalars.
The passive scalar concentration per unit mass, $\theta$,
is governed by the equation
\EQ
{\partial\over\partial t}(\rho \theta)=-{\partial\over\partial x_j}
\left(\rho u_j\theta-\rho\kappa_\theta{\partial\theta\over\partial x_j}\right),
\label{drhoTheta}
\EN
where $\kappa_\theta$ is a diffusion coefficient for the passive scalar
concentration.
This equation describes the transport of chemicals in a gas.
Additional source and sink terms could be included to model production
and destruction of chemicals.
The non-conservative form of this equation can be written as
\EQ
{\DD \theta\over\DD t}=-{1\over\rho}\nab\cdot
\left(\rho\kappa_\theta\nab \theta\right),
\EN
where we have made use of the continuity equation \eq{DrhoDt}.
For $\kappa_\theta=0$, this equation gives $\DD \theta/\DD t=0$, which
shows that the concentration per unit mass is unchanged at each point
comoving with the flow.

Another class of scalars are inertial particles that are advected by their
own velocity $\uu_{\rm p}$ rather than the velocity of the gas $\uu$.
The evolution equation of $\uu_{\rm p}$ is similar to that of $\uu$, except
that it lacks the pressure gradient term and the Lorentz force.
However, such particles are strictly speaking active particles, because
of the mutual coupling between the two velocity fields.
Only in the limit of sufficiently light particles can the back-reaction
on $\uu$ be neglected.

In astrophysics one often finds the condensation of heavier elements into
solid dust.
Their evolution is described as a passive scalar or as passively
advected particles.
The inclusion of inertia can sometimes become important, because inertial
particles have a tendency to accumulate in anti-cyclonic vortices \cite{%
1995A&A...295L...1B,
1996Icar..121..158T,
1998A&A...330.1169H,
2004A&A...417..361J,
2006ApJ...649..415I
}.

\subsection{Active scalars: stratification and convection}
\label{Scalars}

\begin{figure}\centering\includegraphics[width=0.90\textwidth]{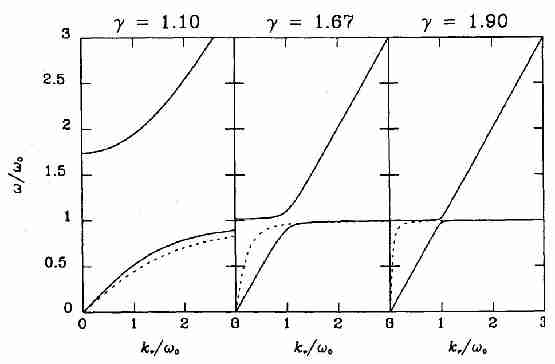}
\caption{
Dispersion relation $\omega=\omega(k_r)$, for $k_z=0$ with
$k_r=(k_x^2+k_y^2)^{1/2}$ being the horizontal wavenumber,
for different values of the ratio of specific heats, $\gamma$,
showing p-modes branch (upper branch) and g-models (lower branch)
compared with the case where the anelastic approximation has been
made (dashed line).
Length is given in units of $H_0=\gamma H_p/(1-\gamma/2)$ and time
in units of $T_0=H_0/\cs$, and $\omega_0=\sqrt{\gamma-1}/(1-\gamma/2)$
is the non-dimensional Brunt--V\"ais\"al\"a frequency.
In the plot, the break point at $k_r/\omega_0=1$ corresponds to
a critical horizontal wavelength of
$\ell_{\rm crit}=2\pi\gamma H_p/\sqrt{\gamma-1}$.
For $\gamma=5/3$ this means $\ell_{\rm crit}\approx12.8\,H_p$.
Adapted from Ref.~\cite{%
1988A&A...203..154B
}.
\label{B88}}
\end{figure}

In this context, the term ``active'' refers to the property that the scalar
quantity can affect the momentum equation, for example by exerting a
pressure gradient force.
An example is the advection--diffusion equation for the energy density
of low-energetic cosmic rays \cite{%
1990acr..book.....B,
2003A&A...412..331H
}.
Another example concerns temperature or specific entropy, which affect the
momentum equation by locally changing the relation between pressure and density.
In the presence of a gravity force, $\FF=\rho\grav$,
this can lead to an Archimedian buoyancy force.
Furthermore, with $\grav\neq0$ a new wave mode can exist known as
gravity waves (not to be confused with gravitational waves of the
space-time metric in general relativity; see comment at the end of
\Sec{DecayingTurbulence}).
The restoring force comes from the linearized buoyancy term,
$(\delta\rho/\rho_0)g\approx(\delta p/p_0+\delta s/c_p)g$.
Since the restoring force is related to gravity, these wave modes are
often referred to as g-modes, in contrast to p-modes or sound waves,
whose restoring force is related to the pressure gradient.
If pressure fluctuations may be neglected the essential terms are
\EQA
{\partial u_z\over\partial t}=...+\delta s\, g/c_p,\\
{\partial\delta s\over\partial t}=...-u_z {\partial\overline{s}\over\partial z},
\ENA
where $\overline{s}$
denotes the specific entropy of the background stratification.
The oscillation frequency $N_{\rm BV}$ (for Brunt--V\"ais\"al\"a frequency)
is given by
\EQ
N_{\rm BV}^2=-\grav\cdot\nab\overline{s}/c_p.
\label{NBVdef}
\EN
While this pair of equations represents the basic feedback loop
correctly, it ignores the fact that buoyancy is only possible when there
is lateral non-uniformity of density.
Indeed, solving the proper dispersion relation reveals that on large scales
the frequency increases linearly with wavenumber; see, e.g., Ref.~\cite{%
1974ARA&A..12..407S
} for a review.
In \Fig{B88} we show the dispersion relation as a function of the horizontal
wavenumber, $k_r=(k_x^2+k_y^2)^{1/2}$, for $k_z=0$ and different values of
the ratio of specific heats ranging from $\gamma=1.1$ to $1.9$.
The p-modes correspond to the upper branch while the g-modes to the
lower one.
Also shown are the g-modes obtained using the anelastic approximation
discussed in \Sec{AnelasticApproximation}.
Note that this approximation yields correct results for $\gamma$ close
to one and for large horizontal wavenumbers, i.e.\ on scales that are
small compared with the pressure scale height \cite{%
1988A&A...203..154B
}.

Given that gravity points downward, $N_{\rm BV}^2$ is positive
(i.e.\ the frequency is real) when the specific entropy increases
in the upward direction.
If it decreases with height, the system is unstable to the onset of
convection with an approximate growth rate given by $\mbox{Im}|N_{\rm BV}|$.
Here we have omitted viscous and diffusive effects that could slow down
the growth and even stabilize the system.
This is quantified by the value of the Rayleigh number that will be
defined and discussed in more detail in \Sec{Convection}.
However, in astrophysics viscosity and diffusivity are comparatively
small and one uses just the condition
$\grav\cdot\nab\overline{s}>0$ for instability.
This is known as the Schwarzschild criterion and corresponds to
saying that the Rayleigh number is positive (convection is discussed
in more detail in \Sec{Convection}).

In the presence of strong vertical density stratification, the convection flow
tends to develop an interconnected network of downdraft lanes, with isolated
tube-like stronger downdrafts at network vertices.  With depth, the downdrafts
merge and the network size increases \cite{%
1989ApJ...342L..95S
}; cf.\ also Figs.\ \ref{paake_conv} and \ref{paake_deeper}.
At large Reynolds numbers the flow is of course turbulent, but with
the intensity of turbulence strongly influenced by stratification
effects:  Because ascending flows are strongly divergent, turbulence is
suppressed there, while in downflows, which are converging, turbulent
intensity is enhanced.

In the Sun, the Prandtl number, $\mbox{Pr}=\nu/\chi$, is far below
unity (around $10^{-5}$).
This means that velocity or vorticity structures can be much thinner
than temperature structures.
As a consequence, thin vortex tubes can develop within downdrafts.
The dynamical pressure associated with vortex tubes allows locally
a lower gas pressure and hence a lower density, making vortex tubes buoyant.
As a result, the downdraft speed is slowed down (``vortex braking'')
\cite{%
1996JFM...306..325B,
1996ApJ...473..494B
}.
This is a particular property of low Prandtl number dynamics which,
at the same time, requires compressibility.

Compressibility leads to yet another interesting effect in convection.
The pressure gradient associated with driving the horizontal expansion
of upwelling motions works in all directions, and in particular also in
the downward direction.
This tends to brake the upwellings.
This phenomenon is known as buoyancy braking \cite{%
1984ApJ...282..557H
}.

Another important effect caused by compressibility is the production of
vorticity by the baroclinic term, i.e.\ the curl of $\rho^{-1}\nab p$.
The curl of this term is finite if the surfaces of constant $\rho$
and $p$ are inclined relative to each other.
Another way of writing this term is by using the thermodynamic relation
for the differential of enthalpy, $\dd H=T\dd S+V\dd p$.
With this we can write the pressure gradient term in terms of specific
enthalpy, specific entropy $s$, and specific volume $\rho^{-1}$ as
\EQ
-\rho^{-1}\nab p=-\nab h+T\nab s.
\label{enthalpy}
\EN
This formula shows that the baroclinic term is just given by
\EQ
\nab\times\left(-\rho^{-1}\nab p\right)=\nab T\times\nab s.
\label{baroclinic}
\EN
This relation will becomes useful later in connection with the
Taylor--Proudman theorem and ideas to understanding departures from it.
The baroclinic term vanishes under isothermal ($T=\const$),
isentropic ($\s=\const$), or barotropic [$p=p(\rho)$] conditions.
In all these cases, \Eq{enthalpy} can be written purely as a gradient term,
$-\nab\tilde{h}$, where $\tilde{h}$ is then called the pseudo-enthalpy
and it is proportional to $h$ which, in turn, is proportional to the
temperature.
In the irrotational case, $\oo=\bm{0}$, the only nonlinearity comes
from the $\half\uu^2$ term in the reduced pressure.

\subsection{Rotation and shear}

It is often convenient to solve the governing equations in a rotating
frame of reference.
In that case, Coriolis and centrifugal forces as well as possibly
the Poincar\'e force have to be included
on the right hand side of the Navier-Stokes equation, so the equation
takes the form
\EQ
{\DD\uu\over\DD t}=...-2\OO_0\times\uu-\OO_0\times(\OO_0\times\rr)
-\dot{\OO}_0\times\rr,
\label{rotation}
\EN
where $\rr$ is the position vector with respect to a point on the rotation
axis and $\OO_0=\const$ is the angular velocity vector of the reference frame.
The Poincar\'e force, $\dot{\OO}_0\times\rr$ can drive flows and even turbulence
in precessing bodies with boundaries.
This has been discussed in attempts to explain the flows that drive the
geodynamo \cite{%
1968Sci...160..259M,
1975PEPI...11...43L,
1991GApFD..59..209V,
2005PhFl...17c4104T
}.

An important effect of the Coriolis force is to suppress
variations of the azimuthal velocity in the axial direction.
This can be seen by taking the curl of the Coriolis term,
\EQ
\nab\times\left(-2\OO_0\times\uu\right)=2\Omega_0\left(
{\partial\uu_\perp\over\partial z}-\zzz\nab\cdot\uu_\perp\right),
\EN
where $\uu_\perp=\uu-(\uu\cdot\zzz)\zzz$ is the velocity in the
direction perpendicular to the rotation axis and $\zzz$ is the
unit vector along the direction of $\OO_0$.
Taking the curl of the evolution equation for $\partial\uu/\partial t$
in cylindrical polar coordinates, $(R,\phi,z)$,
and projecting on the $\phi$ component, yields
\EQ
\pp\cdot{\DD\oo\over\DD t}=2\Omega_0{\partial u_\phi\over\partial z}
-\pp\cdot\left(\nab T\times\nab s\right)+F_\phi^{\rm visc},
\EN
where $F_\phi^{\rm visc}$ is the $\phi$ component of the viscous force.
This shows that, when rotation is important, i.e.\ $\Omega_0$ is large,
$\partial u_\phi/\partial z$ must be small.

Of course, the physics is independent of the coordinate system one
is working in.
If one is working in the inertial (non-rotating) frame, there is
no Coriolis force, but one can then write $\uu_\phi=R\Omega$,
where $\Omega=\Omega(R,\phi,z)$ is now the local angular velocity,
which is not to be confused with $\OO_0$.
If the velocity has only an azimuthal component, $\uu=\pp R\Omega$,
one can write the curl of the $\uu\cdot\nab\uu$ term as
\EQ
\nab\times\left(-\uu\cdot\nab\uu\right)=R{\partial\Omega^2\over\partial z}.
\label{curl_ugradu}
\EN
We will return to the astrophysical consequences of this in \Sec{LambdaEffect}
when we discuss the angular velocity of the Sun.

\subsection{Active vectors: magnetic fields and dynamos}
\label{MagneticFields}

An important vector field to be included in the fluid equations is the
magnetic field, $\BB$.
It is an active vector because the Lorentz force, $\JJ\times\BB$,
backreacts through the momentum equation, so
\EQ
\rho{\DD\uu\over\DD t}=...+\JJ\times\BB,
\label{LorentzForce}
\EN
where $\JJ$ is the current density.
One makes here the assumption that there is no net charge in the fluid,
i.e.\ the density of positive and negative charge carriers balances
everywhere, and the currents are produced by the sum of the fluxes of
counterflowing positive and negative charge carriers.
The $\BB$ field is solenoidal and its evolution is governed by the
Faraday equation,
\EQ
{\partial\BB\over\partial t}=-\nab\times\EE,
\quad\mbox{with}\quad\nab\cdot\BB=0,
\label{Faraday}
\EN
where $\EE$ is given by Ohm's law,
\EQ
-\EE=\uu\times\BB-\JJ/\sigma,
\label{Ohm}
\EN
and $\sigma$ is the electric conductivity.
Its inverse is related to the magnetic diffusivity, $\eta=(\mu_0\sigma)^{-1}$,
and it has the same dimension as the kinematic viscosity $\nu$.
Their ratio is the magnetic Prandtl number $\Pm=\nu/\eta$.

Ampere's equation is used to express the current density in terms of the
magnetic field via
\EQ
\JJ=\nab\times\BB/\mu_0,
\label{Ampere}
\EN
where $\mu_0$ is the vacuum permeability.
Equation \eq{Ampere} is an approximation to the full Faraday equation which
includes also the displacement current.
Neglecting it corresponds to filtering out electromagnetic waves,
which is justified at finite electric conductivity and velocities
small compared with the speed of light.

Inserting \Eq{Ohm} into \Eq{Faraday} we obtain the induction equation
in the form
\EQ
{\partial\BB\over\partial t}=\nab\times\left(\uu\times\BB-\JJ/\sigma\right).
\label{Induction}
\EN
In its `uncurled' form this equation reads
\EQ
{\partial\AAA\over\partial t}=\uu\times\BB-\JJ/\sigma-\nab\phi,
\label{UncurledInduction}
\EN
where $\phi$ is the electrostatic potential.
By evaluating the time derivative of $\AAA\cdot\BB$ and integrating
over space we obtain the evolution equation for magnetic helicity,
\EQ
{\dd\over\dd t}\int\AAA\cdot\BB\,\dd V=-2\sigma^{-1}\int\JJ\cdot\BB\,\dd V
-\oint\FF_H\cdot\dd\SSS,
\label{MagneticHelicity}
\EN
where $\FF_H=\EE\times\AAA+\phi\BB$ is the flux of magnetic helicity.

Magnetic fields constitute an additional form of energy,
$E_{\rm M}=\int\BB^2/(2\mu_0)\,\dd V$, whose evolution is given by
\EQ
{\dd\over\dd t}\int{\BB^2\over2\mu_0}\,\dd V
=-\int\uu\cdot(\JJ\times\BB)\,\dd V-\sigma^{-1}\int\JJ^2\,\dd V
-\oint\FF_M\cdot\dd\SSS,
\label{MagneticEnergy}
\EN
where $\FF_M=\EE\times\BB/\mu_0$ is the Poynting flux.
\EEq{energyeqn} for the evolution of the total energy density
can be generalized correspondingly by adding $\BB^2/2\mu_0$ underneath the
time derivative and $\FF_M$ underneath the divergence term.

In this connection it might be useful to emphasize that in numerical
simulations one hardly uses the full energy equation in that form if
the magnetic energy becomes comparable to or in excess of the thermal energy.
Normally one would calculate the thermal pressure from the internal
energy, but in the magnetically dominated case this becomes a small
residual between total, kinetic, and magnetic energies, and so this calculation
becomes exceedingly inaccurate.

Another comment regarding simulations is here in order.
A commonly encountered difficulty is to preserve solenoidality of $\BB$.
One method is to use a staggered mesh and to evaluate the right-hand-side
of \Eq{Faraday} such that the numerical evaluation of the curl
produces zero divergence to machine accuracy.
Another method is to use $\AAA$ as dependent variable, which also
preserves $\nab\cdot\BB=0$, and it also allows for a straightforward
calculation of the magnetic helicity.
Yet another method is to write
\EQ
\BB=\nab\alpha\times\nab\beta,
\EN
where $\alpha$ and $\beta$ are the Euler potentials \cite{%
1970AmJPh..38..494S
}.
However, this method only works in the strictly ideal case, in which
case the evolution equations are just
\EQ
\DD\alpha/\DD t=\DD\beta/\DD t=0.
\EN
This approach is now quite popular in smoothed particle hydrodynamics
calculations, because then the values of $\alpha$ and $\beta$ are just
kept fixed at each Lagrangian particle \cite{%
2007MNRAS.379..915R,
2009MNRAS.397..733K
}.
Unfortunately, this method cannot even approximately capture non-ideal effects.
As a consequence, dynamo action (see below) is not possible in this
approach and energy spectra of MHD turbulence with imposed field become
too shallow \cite{%
2010MNRAS.401..347B
}.
Finally, there is the possibility of divergence cleaning, which requires
the solution of a Poisson-type equation for the correction term to the
numerically obtained $\BB$ field.
This approach is analogous to calculating the pressure under the constraint
that $\nab\cdot\uu=0$ or $\nab\cdot\rho\uu=0$; see \Sec{AnelasticApproximation}.
The disadvantage here is that this approach may introduce an unphysical
nonlocality as a consequence of invoking a Poisson-type equation.

The Lorentz force gives rise to various restoring forces that lead
to additional wave forms including Alfv\'en waves as well as fast
and slow magnetosonic waves.
The slow magnetosonic waves are particularly important in the presence
of shear and rotation, because those waves can be destabilized to give
rise to the magneto-rotational instability.
This will be discussed in more detail in \Sec{MagnetoRotational}.

One of the other new features allowed by the addition of magnetic fields is the
possibility of self-excited dynamo action, i.e.\ the spontaneous conversion of
kinetic energy into magnetic energy by work done against the Lorentz force.
This is an important process in astrophysics.
Magnetic fields observed in planets and stars with outer convection zones
are clear examples where dynamo action is required to sustain magnetic
fields against ohmic decay and to explain field reversals on time scales
short compared with the resistive time.
Galaxies and clusters of galaxies also harbor magnetic fields.
Many spiral galaxies show magnetic fields with a large-scale design
that is approximately axisymmetric.
One prominent exception is a galaxy with the name M81,
where the field is non-axisymmetric with
a strong $m=1$ component, i.e.\ the field is proportional to
$e^{\ii m\phi}$, where $\phi$ is the azimuthal angle.
Observations give direct indications about the turbulent nature of
galactic discs, so the magnetic field must be maintained against
turbulent decay in the vertical direction along the axis.
The relevant time scale is only about $10^7\yr$.
In the present review we discuss dynamos only insofar as they are
directly connected with understanding or clarifying astrophysical turbulence.

Details regarding dynamo theory as well as magnetic fields in
solar-like stars and galaxies have recently been reviewed
in Refs.~\cite{%
2005PhR...417....1B,
1996ARA&A..34..155B,
2006RPPh...69..563S
}.
One of the important recent developments concerns the realization that
the evolution of the large-scale magnetic field can be constrained
decisively by magnetic helicity evolution; see \Eq{MagneticHelicity}.
This has to do with the fact that large-scale magnetic fields tend to
be helical.
This point will be taken up briefly in \Sec{Alpha}, but for a more thorough
discussion we refer to Ref.~\cite{%
2005PhR...417....1B
} for a recent review.

In the incompressible case with constant density $\rho=\rho_0$,
it is convenient to write the MHD equations
using Elsasser variables $\zz_{\pm}=\uu\pm\BB/\sqrt{\mu_0\rho_0}$,
because then the evolution equations take a form similar to the usual
Navier-Stokes equations, i.e.\
\EQ
\frac{\partial \zz_{\pm}}{\partial t}+\zz_{\mp}\cdot\nab\zz_{\pm}
=-\nab \Pi + \nu\nabla^2\zz_{\pm},\quad\nab\cdot\zz_{\pm}=0.
\label{elsas}
\EN
Here, $\nu=\eta$ has be assumed for simplicity, and
$\Pi=(p+\BB^2/2\mu_0)/\rho$ is a pressure that ensures that
$\nab\cdot\zz_{\pm}=0$.

\subsection{Radiation: optically thick and thin}

Radiation transport describes the coupling to the photon field.
As far as the dynamics is concerned, the radiative flux gives rise to
a radiation force that can for example cause levitation
of the gas by radiation.
The radiative energy flux divergence enters the energy equation and
describes local heating and cooling.
Thus, the momentum and specific entropy equations are amended as follows,
\EQ
\rho{\DD\uu\over\DD t}=...+{\rho\kappa\over c}\FF_{\rm rad},
\label{radforce}
\EN
\EQ
\rho T{\DD s\over\DD t}=...-\nab\cdot\FF_{\rm rad}.
\label{radheating}
\EN
Here, $\kappa$ is the opacity, i.e.\ the photon cross-section per unit mass.
The cross-section per unit volume is $\rho\kappa$, which is also the
inverse mean free path of photons, $\ell=(\rho\kappa)^{-1}$.
If the mean free path is small compared with other relevant length scales,
a diffusion approximation may be used for $\FF_{\rm rad}$, which means
that it is proportional to the negative gradient of the radiation energy
density, $\FF_{\rm rad}=-\onethird\ell c\nab(aT^4)$,
and so it points in the direction of the negative temperature gradient.
The transition layer between optically thin and thick is an important
region in astrophysics, because it marks the effective surface of an
otherwise extended body.
In this transition region the diffusion approximation is no longer valid
and proper equations for the radiation intensity have to be solved
to obtain $\FF_{\rm rad}$; see Refs.~\cite{%
1982A&A...107....1N,
1978stat.book.....M
}.

\section{Simulations of turbulence}
\label{SimulationsOfTurbulence}

Astrophysical turbulence is frequently caused by instabilities.
However, many instabilities imply the presence of anisotropies.
For example in convection the vertical direction is a preferred one,
while in the case of the magneto-rotational instability the velocity
gradient matrix associated with the shear governs the anisotropy.
In the presence of magnetic fields, the otherwise isotropic turbulence
becomes at least locally anisotropic, because at every patch in the
turbulent flow the direction of the local mean field imprints
anisotropy on all smaller scales within this patch.
On the other hand,
much of turbulence theory is concerned with isotropic turbulence.
Computationally, isotropic turbulence can be modeled by adopting an
imposed forcing function.
Common applications of isotropically forced turbulence include
simulations of turbulent star formation, as well as turbulent mixing
and dynamo processes.
We begin by discussing some general aspects of isotropic turbulence
simulations.

\subsection{General aspects}

The concept of isotropic turbulence is a convenient and useful theoretical idealization.
Computationally, isotropy does not lead to any significant simplification, except
that periodic boundary conditions are possible and in many ways advantageous.
Isotropic turbulence needs to be forced by an isotropic body force,
unless an isotropic instability can be identified that would drive turbulence.
The thermal instability would be an example of an instability without preferred
direction, but simulations have not shown that it can lead to sustained
turbulence \cite{%
2002ApJ...569L.127K,
2007ApJ...654..945B
}.
The Jeans instability is another example, which is particularly relevant
to the problem of star formation through strong compressions by the
turbulence in the interstellar medium.
This problem is frequently tackled using smoothed particle
hydrodynamics \cite{%
1998MNRAS.298...93B,
1997MNRAS.288.1060B
}, while mesh-based techniques have explored mostly the case of
forced supersonic turbulence \cite{%
2002ApJ...576..870P,
2003ApJ...585L.131V
} and have only recently incorporated the effects of self-gravity,
augmented with so-called sink particles to account for the production
of high density cores that cannot be resolved with a fixed mesh \cite{%
2004ApJ...611..399K,
2007ApJ...656..959K,
2009ApJ...694.1161J,
2009arXiv0907.0248P
}.

In order to study more basic properties of turbulence one often resorts
to a random forcing function to simulate the effects of an instability
with a well-defined forcing strength and a well-defined length scale
of the driving.
Plane waves with randomly changing orientation is an obvious possibility
for driving turbulence.
To make the forcing divergence--free, one uses only transversal waves.

The idea of simulating turbulence on the computer developed during
the 1970ies.
Almost all simulations in those days utilized pseudo-spectral
methods, i.e.\ spatial derivatives are calculated in Fourier space by
multiplication with $\ii\kk$, but all nonlinear terms are calculated
in real space.
The main advantage of such methods is the small discretization error.
Furthermore, this technique also allows an efficient solution
of the Poisson-like equation for the pressure if one makes the
incompressible or the anelastic approximation, i.e.\ $\nab\cdot\uu=0$
or $\nab\cdot\rho\uu=0$, respectively.

Spectral methods have the disadvantage that one cannot easily deal
with arbitrary boundary conditions.
Also, the Fourier transformation is a nonlocal operation which is
not optimal when using many processors.
These are reasons why sometimes finite difference methods are used instead.
Finite difference methods are normally not as accurate as spectral
methods unless one uses a higher order scheme (e.g.\ fourth and sixth
order schemes are common choices).
On the other hand, many astrophysical flows develop shocks for which there
are a number of other dedicated methods (Riemann solvers, approximate
Riemann solvers, monotonicity schemes, Godunov schemes, and
Neumann--Richtmyer artificial viscosities \cite{%
1994dmiv.book.....R,
1992ApJS...80..753S
}).
These methods are frequently generalized to mesh refinement methods
that allow increased accuracy in specific locations of the flow.
Finally, there are also Lagrangian methods of which Smooth Particle
Hydrodynamics is an example \cite{%
1997MNRAS.288.1060B,
1992ARA&A..30..543M,
2002MNRAS.333..649S,
1996MNRAS.278.1005S
}.
A promising new Lagrangian method has been presented in Ref.~\cite{%
2010MNRAS.401..791S
}.

\subsection{Hydrodynamic turbulence}

\begin{figure}\centering
\includegraphics[height=6.8cm]{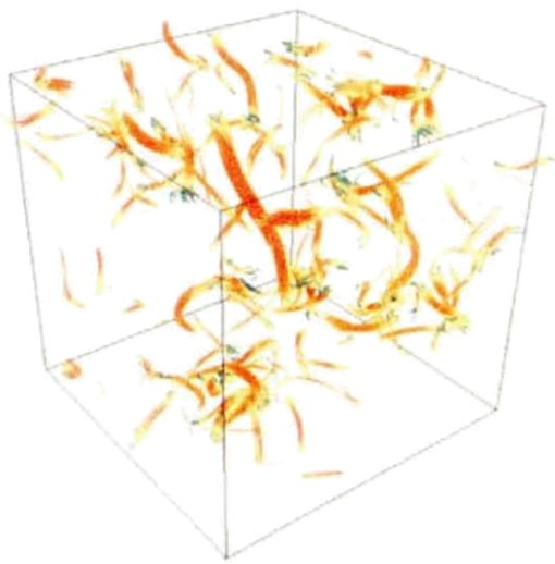}
\includegraphics[height=6.6cm]{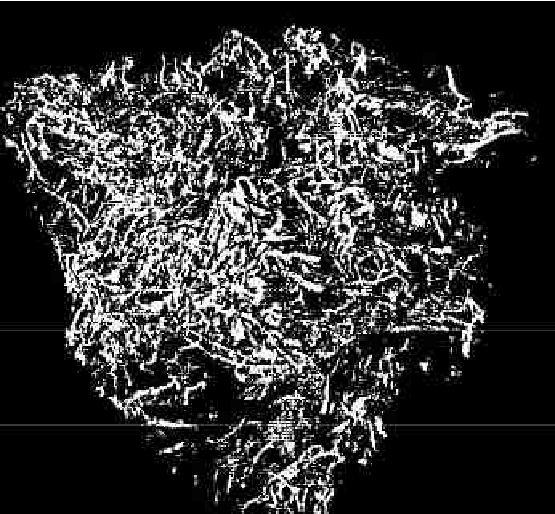}
\caption{Examples of vortex tubes in homogeneous turbulence.
Courtesy of Zhen-Su She (left figure) \cite{%
1990Natur.344..226S
} and Paul Woodward (right figure) \cite{%
1994PhFl....6.2133P
}.
\label{Vincent}}
\end{figure}

When simulations became able to resolve turbulence with around $128^3$
meshpoints, it became evident that much of the flow is governed by a
tangle of vortices; see, e.g., Refs.~\cite{%
1990Natur.344..226S,
1991JFM...225....1V,
1994PhFl....6.2133P
}.
The left-hand panel of \Fig{Vincent} shows examples of such vortices.
Their thickness is related to the viscous scale while their length
was often expected to be comparable with the integral scale.
However, in subsequent years simulations at increasingly higher Reynolds
numbers seem to reveal that the vortex turbulence become a less prominent
feature of otherwise nebulous looking structures of variable density
(see the right-hand panel of \Fig{Vincent}).

\begin{figure}\centering\includegraphics[width=0.60\textwidth]{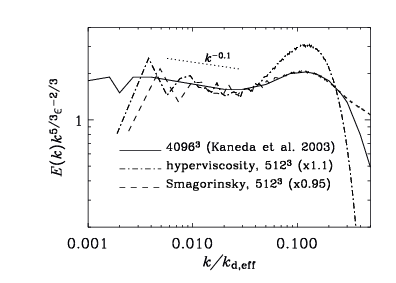}
\caption{Comparison of energy spectra of the $4096^3$ meshpoints run
\cite{
2003PhFl...15L..21K
} (solid line) and $512^3$ meshpoints runs with
hyperviscosity (dash-dotted line) and Smagorinsky viscosity (dashed line).
(In the hyperviscous simulation we use $\nu_3=5 \times 10^{-13}$.)
The Taylor microscale Reynolds number of the Kaneda simulation is 1201,
while the hyperviscous simulation of Ref.~\cite{%
2004PhRvE..70a6308H
} has an approximate
Taylor microscale Reynolds number of $340<\mbox{Re}_\lambda<730$.
For the Smagorinsky simulation the value of $\mbox{Re}_\lambda$ is
slightly smaller.
Courtesy of Nils E.\ Haugen \cite{%
2006PhFl...18g5106H
}.
\label{kan_hyp_smag}}
\end{figure}

Incompressible forced turbulence simulations have been carried out at
resolutions up to $4096^3$ meshpoints \cite{%
2003PhFl...15L..21K
}.
Surprising results from this work include a strong bottleneck effect
\citep{%
1994PhFl....6.1411F} 
near
the dissipative subrange, and possibly a strong inertial range correction of
about $k^{-0.1}$ to the usual $k^{-5/3}$ inertial range spectrum,
so that the spectrum is $k^{-1.77}$.
Note that the She--Leveque correction \eq{SheLeveque} is only $k^{-0.03}$,
so that the spectrum is $k^{-1.70}$.
Similarly strong inertial range corrections have also been seen in simulations
using a Smagorinsky subgrid scale model
\cite{%
2006PhFl...18g5106H
} ($512^3$ meshpoints, dashed line in \Fig{kan_hyp_smag}).
Here we also show the results of simulations with hyperviscosity,
i.e.\ the $\nu\nabla^2$ diffusion operator has been replaced by a
$\nu_3\nabla^6$ operator \cite{%
2004PhRvE..70a6308H
} ($512^3$ meshpoints, dash-dotted line).
Hyperviscosity greatly exaggerates the bottleneck effect,
but it does not seem to affect the inertial range significantly;
see \Fig{kan_hyp_smag}.

The nature of the $k^{-0.1}$ correction factor is currently not
understood.
It might be an artifact resulting from applying a forcing function
at a scale close to the scale of the box \cite{%
2004tise.book.....D
}.
Alternatively, the presence of a bottleneck might also lead to
the emergence of a dip just before the bottleneck.
In either case this would not be a true $k^{-0.1}$ correction in the
entire inertial range.

In virtually all astrophysical settings the relevant Reynolds numbers
are very large and the bottleneck is hardly important, because it is
located at very small length scales.
However, this is not the case in simulations which
show the bottleneck as a pronounced feature.
There are several important issues here.
Firstly, simulations at resolutions of $256^3$ meshpoints
give hardly any indication of a bottleneck effect, and only
at resolutions of $1024^3$ meshpoints and above does it really develop
its full strength.
For this reason the bottleneck effect has been studied more seriously
only in recent years.
Secondly, the bottleneck effect can affect certain
aspects of a simulation in a way that is not yet asymptotically
meaningful.
An example is the small-scale dynamo effect that is discussed below.

\subsection{Supersonic turbulence}

In the interstellar medium the gas can condense into more concentrated
regions called molecular clouds.
These clouds are so cold that molecules can form, which explains their name.
Because of low temperature in the range of 10--100\,K,
the flows in these clouds can become highly supersonic.
This in turn leads to even stronger mass concentrations that can become
gravitationally unstable and form stars.
This is why supersonic turbulence is commonly studied in connection with
star formation \cite{%
2003ApJ...585L.131V,
2004RvMP...76..125M
}.

With increasing Mach number, density fluctuations begin to become important.
In fact, in supersonic turbulence with an isothermal equation of state
it has been demonstrated that the standard deviation of the (linear) density,
$\sigma_{\rm linear}$, grows linearly with the Mach number
\cite{%
2002ApJ...576..870P,
1997MNRAS.288..145P,
1999ApJ...526..279P
}
\EQ
\sigma_{\rm linear}=\gamma\Ma,
\EN
where the Mach number is defined as $\Ma=u_{\rm rms}/c_{\rm s}$.
The density obeys a log-normal distribution, i.e.\ the probability
density function, $p(\ln\rho)$, with $\int p(\ln\rho)\,\dd\ln\rho=1$,
is given by
\EQ
p(\ln\rho)={1\over\sqrt{2\pi\sigma^2}}\exp\left[
-\half\left(\ln\rho-\overline{\ln\rho}\right)/\sigma^2\right],
\EN
where $\sigma$ is the standard deviation of the logarithmic density,
which is related to the Mach number like \cite{%
2002ApJ...576..870P
}
\EQ
\sigma=\ln\left(1+\gamma^2\Ma^2\right),
\EN
again with $\gamma\approx1/2$ to good accuracy.

\begin{figure}[t!]\centering
\includegraphics[width=0.49\textwidth]{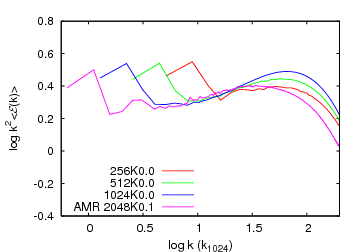}
\includegraphics[width=0.49\textwidth]{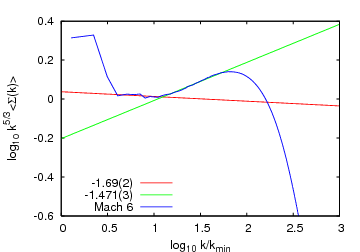}
\caption{
Left: time-averaged velocity spectra compensated by $k^2$ from uniform grid PPM simulations
at resolutions $256^3$, $512^3$, $1024^3$, and from an adaptive mesh refinement
simulation with effective resolution of $2048^3$ grid points.
The spectra demonstrate convergence for the inertial range of scales.
Right: time-averaged spectrum of $\rho^{1/3}\uu$ compensated by $k^{5/3}$.
The straight lines represent the least-squares fits to the data
within a suitable intermediate range of wavenumbers.
Courtesy of Alexei Kritsuk \cite{%
2007ApJ...665..416K
}.
}\label{kritsuk19}\end{figure}

As indicated in \Sec{Spectra} the spectra of $\uu$, $\rho^{1/2}\uu$, and
$\rho^{1/3}\uu$ begin to differ from each other at larger Mach number.
Observations of the line-of-sight velocity dispersion of molecular clouds
in the Perseus cluster also show that in the highly supersonic case the
velocity spectrum is not far from $k^{-1.8}$ \cite{%
2006ApJ...653L.125P
}, and thus deviates clearly from the
characteristic spectrum of shock turbulence \cite{%
1973SPhD...18..115K
}.
However, the density weighted spectra tend to become shallower.
In particular the spectra of $\rho^{1/3}\uu$ are very close to $k^{-5/3}$
\cite{%
2007ApJ...665..416K
}; see \Fig{kritsuk19}.
This appears to be connected with the fact that the kinetic energy flux,
i.e.\ the quantity that is constant throughout the inertial range at
scale $l$, is given by $\rho u_l^3/l$.
This idea goes back to an early paper by Lighthill \cite{%
1955IAUS....2..121L
}.

\subsection{Hydromagnetic turbulence}

The gas in many astrophysical settings is partially or fully ionized
and hence electrically conducting.
This means that the effects of magnetic fields cannot be neglected.
The full extent of associated behaviors is not yet well understood, nor is there
unambiguous evidence for universal and asymptotic scaling behavior
in the limit of large fluid and magnetic Reynolds numbers \cite{%
2009ApJ...702.1190B
}.
However, using decay simulations at moderate magnetic Reynolds numbers,
three types of behavior have been identified \cite{
2010PhRvE..81a6318L
}, depending essentially on the ratio of the initial magnetic to kinetic
energy densities.
The purpose of this section is to discuss the expected properties in these
three regimes.

It is convenient to introduce here the Alfv\'en speed
$\vA=\Brms/\sqrt{\mu_0\rho_0}$ associated with the random
magnetic field $\Brms$.
The case of sub-equipartition random fields with $\vA<\urms$
was studied by Iroshnikov \cite{
1964SvA.....7..566I
} and Kraichnan \cite{
1965PhFl....8.1385K
} who argued that the turbulence can still be treated as
isotropic and that the flux of energy $\epsilon$ down the turbulent
cascade will be modified by Alfv\'enic interactions and replaced by the geometric
mean of energy flux and Alfv\'en speed, i.e.\
\EQ
\epsilon\to(\epsilon\vA)^{1/2}.
\EN
The dimensional argument used in \Eq{EkCK} for the energy spectrum
of Kolmogorov turbulence gets correspondingly modified and is then of the form
\EQ
E(k)=C_{\rm IK}(\epsilon\vA)^{1/2} k^{-3/2}.
\label{EkCIK}
\EN
In the case of strong magnetic fields, $\vA\gg\urms$, the turbulence
becomes highly anisotropic, so the spectrum $E(k_\perp, k_\parallel)$ depends
on the wavenumbers parallel ($k_\parallel$) and perpendicular ($k_\perp$)
to the local direction of the field.
In this limit the turbulence can be treated as wave turbulence using
weak turbulence theory \cite{
2000JPlPh..63..447G
}, which leads to
\EQ
E(k_\perp, k_\parallel) \sim k_\perp^{-2}.
\EN
In the intermediate case, the kinetic energy of the turbulence is
comparable to that of the magnetic field.
This regime is referred to as {\it strong} turbulence -- not because the
field is strong, but because the $\uu\cdot\nab\uu$ nonlinearity cannot
be neglected.
The flow is still anisotropic, and energy is cascaded in $k_\perp$
at a rate $\epsilon$.
The resulting energy spectrum is \cite{
1995ApJ...438..763G,
1997ApJ...485..680G,
1994ApJ...432..612S,
2007ApJ...655..269L
}, i.e.\
\EQ
E(k_\perp, k_\parallel) =C_{\rm GS} \epsilon^{2/3} k_\perp^{-5/3}.
\EN
In the following we present a more detailed phenomenology that highlights
the essential physics behind the various regimes.

We consider as governing equations the MHD equations written for the
Elsasser variables $\zz_{\pm}$, see \Eq{elsas}, and denote by
$z_{k_\perp}$ the modulus of $\zz_{\pm}$ at wavenumber $k_\perp$.
In all cases the energy spectrum is given by
\EQ
E(k_\perp, k_\parallel)\sim z_{k_\perp}^2/k_\perp,
\label{E_z2k}
\EN
and the spectral energy flux is then given by an expression of the form
\EQ
\epsilon=z_{k_\perp}^2/\tau_{\rm casc},
\EN
where $\tau_{\rm casc}$ is the cascade time.
The main difference between the various regimes lies in the form of the
$\tau_{\rm casc}$; see also Refs.~\cite{
2005PhR...417....1B,
2003ApJ...582.1220L
}.

For strong magnetic fields, interactions are being accomplished by
wave packets traveling in opposite directions.
The duration of the interactions is given by
\EQ
\tau_{\rm A}=(\vA k_\parallel)^{-1},
\EN
where $k_\parallel^{-1}$ is the longitudinal extent of such a packet.
The fractional change in a wave packet is given by the ratio
\EQ
\chi=\tau_{\rm A}/\tau_{\rm NL}
\EN
of Alfv\'en time to the nonlinear interaction time
\EQ
\tau_{\rm NL}=(z_{k_\perp}k_\perp)^{-1}.
\EN
However, because the sign of each interaction is random, the net effect
grows only like the square root of the number of interactions.
Therefore, the effective fractional change associated with each interaction
is only $\chi^2$.
This means that the effective cascade time is
$\tau_{\rm casc}=\tau_{\rm A}/\chi^2$.
By contrast, in the strong turbulence regime the Alfv\'en and nonlinear
times are equal, i.e.\ $\chi=1$, and the cascade time is therefore just
$\tau_{\rm casc}=\tau_{\rm NL}$.
Since the $z_{k_\perp}$ in \Eq{E_z2k} enters also in the expression for
$\tau_{\rm casc}$, the resulting spectra are qualitatively different.
For weak turbulence we have
\EQ
\epsilon={z_{k_\perp}^2\over\tau_{\rm casc}}=z_{k_\perp}^2\,
{(z_{k_\perp}k_\perp)^2\over\vA k_\parallel},\quad\mbox{so}\quad
E(k_\perp, k_\parallel)\sim{z_{k_\perp}^2\over k_\perp}
\sim(\epsilon\vA k_\parallel)^{1/2}k_\perp^{-2},
\EN
while for strong turbulence we have
\EQ
\epsilon={z_{k_\perp}^2\over\tau_{\rm casc}}=z_{k_\perp}^2\,
(z_{k_\perp}k_\perp),\quad\mbox{so}\quad
E(k_\perp, k_\parallel)\sim{z_{k_\perp}^2\over k_\perp}
\sim\epsilon^{2/3}k_\perp^{-5/3}.
\EN
In the latter case, because of $\tau_{\rm NL}=\tau_{\rm A}$, we have
$k_\perp/k_\parallel=\vA/z_{k_\perp}=(\vA/\epsilon)k_\perp^{1/3}$,
so the degree of anisotropy increases toward smaller scales
until we have $k_\parallel\to\infty$.
For weak turbulence we have $k_\parallel\to0$, so the turbulence
is fully anisotropic at all scales.
Finally, for even weaker magnetic fields, the weak turbulence formalism
again applies, except that now the turbulence is isotropic, i.e.\ we
put $k_\perp=k_\parallel=k$ and thus recover \Eq{EkCIK}.
\Tab{Tspectra} summarizes the essential properties in the three regimes.

\begin{figure}[t!]\centering
\includegraphics[width=.7\textwidth]{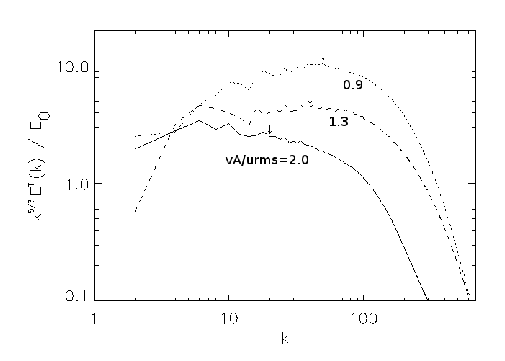}\caption{
Total energy spectra compensated by $k^{5/3}$ and averaged over
$\Delta t = 0.5$ (1.5 to 2 turnover times) about the maximum of
dissipation for three runs: solid line for super-equipartition initial fields
($\vA/\urms\approx2.0$, dashed for equipartition initial fields
($\vA/\urms\approx1.3$) and dots for sub-equipartition initial fields
($\vA/\urms\approx0.9$).
The three arrows indicate the magnetic Taylor scale.
Note that the three spectra follow noticeably
different spectral laws and possibly different scale-dependence for
their time scales as well.
In all cases the numerical resolution is $2048^3$.
Courtesy of E.\ Lee et al.\ \cite{%
2010PhRvE..81a6318L
}, and Copyright (2010) by The American Physical Society.
}\label{f_spectra_enr_times}\end{figure}

Using up to $2048^3$ simulations of decaying MHD turbulence with different
initial field strength, Lee et al.\ \cite{
2010PhRvE..81a6318L
} showed that all three scalings are indeed possible.
In \Fig{f_spectra_enr_times} we show compensated power spectra for three
runs with different initial field strengths with
$\vA/\urms\approx0.9$, $1.3$, and $2.0$,
that are consistent with the regimes of Iroshnikov--Kraichnan turbulence,
strong turbulence, and weak turbulence, respectively.

\begin{table}[htb]\caption{
Summary of the essential properties of the three regimes
of MHD turbulence.
}\vspace{12pt}\centerline{\begin{tabular}{c|ccc}
& Iroshnikov--Kraichnan & strong turbulence & weak turbulence \\
& (isotropic, sub-equip.) & (critically balanced) & (wave turbulence) \\
\hline
$\vA/\urms\sim\chi^{-1}$ & $<1$ & $\sim1$ & $>1$ \\
$\tau_{\rm casc}$ & $\chi^{-2}\tau_{\rm A}$ (with $k_\perp=k_\parallel$)
  & $\chi^{-1}\tau_{\rm A}\;(=\tau_{\rm NL})$ & $\chi^{-2}\tau_{\rm A}$ \\
$\epsilon$ & $z_k^4 k/\vA$  & $z_{k_\perp}^3 k_\perp$
  & $z_{k_\perp}^4 k_\perp^2/\vA k_\parallel$ \\
$k_\perp/k_\parallel$ & $1$ & $\propto k_\perp^{1/3}$ & $\to\infty$ \\
$E(k_\perp,k_\parallel)$ & $(\epsilon\vA)^{1/2}k^{-3/2}$
  & $\epsilon^{2/3}k_\perp^{-5/3}$ & $(\epsilon\vA k_\parallel)^{1/2}k_\perp^{-2}$ \\
\label{Tspectra}\end{tabular}}\end{table}

\subsection{Dynamo action}

\begin{figure}[t!]\centering
\includegraphics[width=0.5\textwidth]{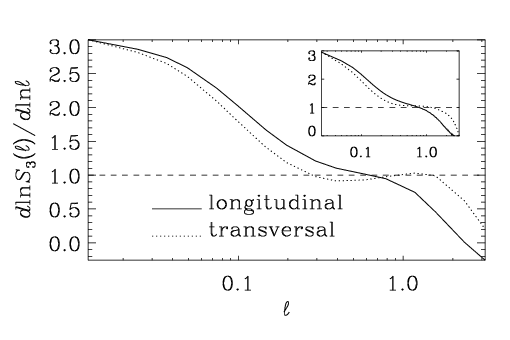}\caption{
Third order structure function, as defined in \Eq{StructureFunctions},
for hydromagnetic runs with $512^3$ meshpoints
\cite{%
2003ApJ...597L.141H
}.
The inset gives the result for $256^3$ meshpoints.
The scaling for transversal structure functions (dotted lines) tends
to be better than for the longitudinal ones (solid lines).
The statistics for the $256^3$ runs is somewhat better
than for the shorter $512^3$ runs.
Courtesy of Nils E.\ Haugen \cite{%
2003ApJ...597L.141H
}.
}\label{third_moment}\end{figure}

In the absence of an externally imposed magnetic field, it is possible that
the field-free state is unstable to the dynamo instability, which leads to a
conversion of kinetic to magnetic energy.
If dynamo action occurs, the magnetic field will grow exponentially
to become dynamically important.
The precise outcome regarding energy spectra and structure functions is
still uncertain, but there is mounting evidence that in the inertial
range they are similar to those in the purely hydrodynamic case \cite{%
2004PhRvE..70b6405H,
2003ApJ...597L.141H,
2007ApJ...665L..35D
}; see \Figs{third_moment}{structure-function}.
However, the largest resolutions obtained in MHD simulations so far
are still only between $1536^3$ \cite{%
2009arXiv0904.4860P
} and $2048^3$ mesh points \cite{%
2010PhRvE..81a6318L
}, and it is not necessarily surprising that
there is no evidence for a clear bottleneck effect, although
the spectra have always been seen to be slightly shallower than
$k^{-5/3}$ and closer to $k^{-3/2}$ \cite{%
2003ApJ...597L.141H,
2005ApJ...626L..37B,
2006PhRvL..96k5002B,
2006PhRvL..97y5002M,
2008ApJ...672L..61P
}.
However, it has been argued that, compared with fluid turbulence,
MHD turbulence is more nonlocal in spectral space \cite{%
2009ApJ...702.1190B
}.
The anticipated spectral bump would be more spread out, which might
explain the absence of a bottleneck at the resolutions
available so far, and that much larger resolution would be needed
to see it.
For supersonic MHD turbulence with dynamo action there is evidence that the mixed
longitudinal structure functions of Politano \& Pouquet \cite{%
1998PhRvE..57...21P
} in \Eq{ThirdMag} scale linearly with $r$, provided the Elsasser variables
are scaled with a $\rho^{1/3}$ factor \cite{%
2009JPhCS.180a2020K,
2009ASPC..406...15K
}.

\begin{figure}[t!]\centering
\includegraphics[angle=-90,width=.8\textwidth]{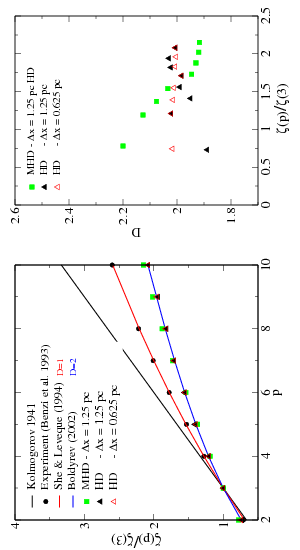}\caption{
Structure function exponents for interstellar turbulence simulations
(left) and fractal dimension of the dissipative structures (right)
from simulations. Courtesy of Miguel de Avillez \cite{%
2007ApJ...665L..35D
}.
}\label{structure-function}\end{figure}

In the absence of helicity and with full isotropy, a successful dynamo
(positive growth rate in the linear regime or finite amplitude in the
nonlinear regime) is referred to as small-scale dynamo.
This refers to the nature of the dynamo process rather
than just the typical scale of the magnetic field.
For example, a small-scale magnetic field that is just the result of
shredding of an imposed large-scale field is not the result of any dynamo
process.
On the other hand, in the presence of helicity, or with anisotropy
combined with a mean shear flow,
there is the possibility of large-scale dynamo action.

\subsection{Large-scale and small-scale dynamos}

\begin{figure}[t!]\begin{center}
\includegraphics[width=.46\textwidth]{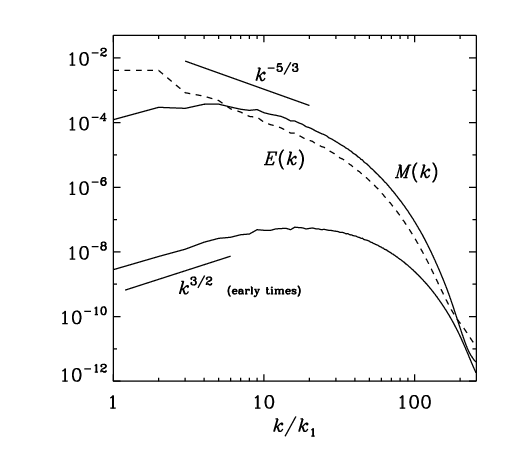}
\includegraphics[width=.46\textwidth]{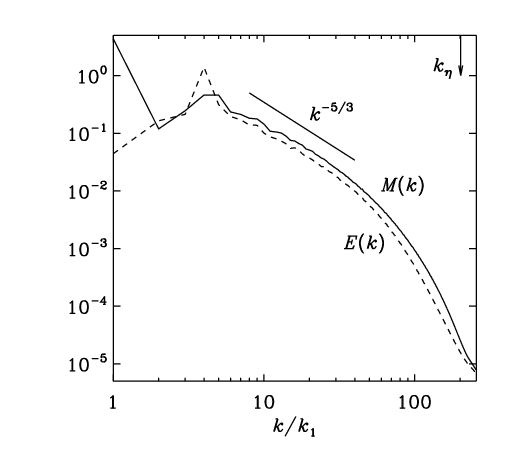}
\end{center}\caption[]{
Kinetic and magnetic energy spectra in a turbulence simulation without
net helicity (left) and with net helicity (right) for a magnetic Prandtl
number of unity and a mesh size is $512^3$ meshpoints.
Notice the pronounced peak of $M(k)$ at $k=k_1$ in the case with helicity.
Adapted from Refs.~\cite{%
2005PhR...417....1B
} and \cite{%
2009ApJ...697.1206B
}, respectively.
}\label{pspec_nohel512d2_pspec_PrM1}\end{figure}

\begin{figure}[t!]\begin{center}
\includegraphics[width=.45\textwidth]{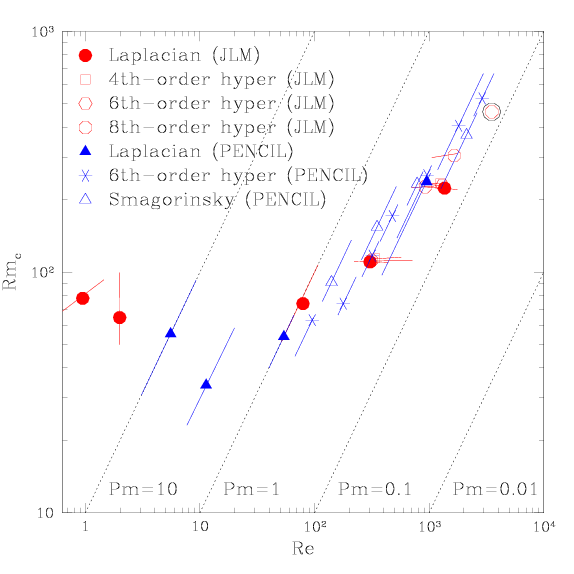}
\includegraphics[width=.45\textwidth]{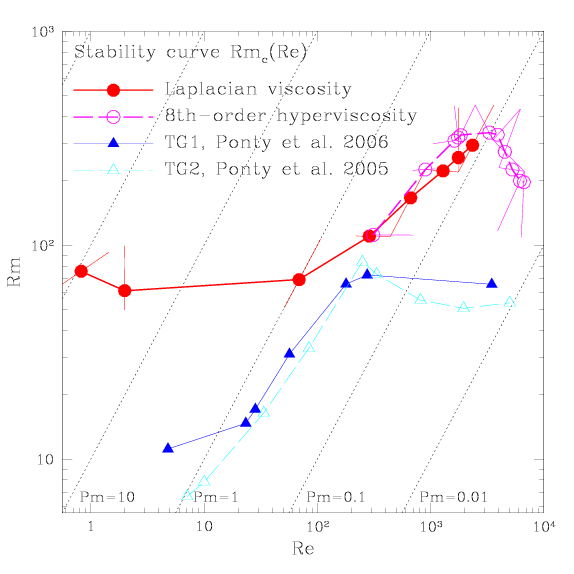}
\end{center}\caption[]{
Dependence of $R_{\rm crit}$ on $\mbox{Re}$.
The plot on the left hand side compares the results from spectral
simulations with those of meshpoint methods.
The plot on the right hand side shows the results of the most recent
simulations with magnetic Prandtl numbers down to 0.06 and compares
with the case of the Taylor-Green flow \cite{%
2005PhRvL..94p4502P,
2007NJPh....9..296P
}.
``JLM'' refers to simulations done with the incompressible spectral
code written by J.~L.~Maron: runs with Laplacian viscosity, 4th-, 6th-,
and 8th-order hyperviscosity (resolutions $64^3$ to $256^3$). In this
set of simulations, hyperviscous runs were done at the same values of
$\eta$ as the Laplacian runs, so the difference between the results
for these runs is nearly imperceptible.  ``PENCIL'' refers to weakly
compressible simulations done with the {\sc Pencil Code}: runs with
Laplacian viscosity, 6th-order hyperviscosity, and Smagorinsky large-eddy
viscosity (resolutions $64^3$ to $512^3$).
Courtesy A.\ A.\ Schekochihin \cite{%
2005ApJ...625L.115S
}.
}\label{prmcrit}\end{figure}

A typical large-scale dynamo produces magnetic energy on a scale larger
than the scale of the energy-carrying eddies.
A small-scale dynamo is one that generates magnetic energy on scales
smaller than the scale of the energy-carrying eddies.
The difference between large-scale and small-scale dynamos is demonstrated
in \Fig{pspec_nohel512d2_pspec_PrM1} where we compare kinetic and magnetic
energy spectra of turbulent dynamos with \cite{%
2009ApJ...697.1206B
} and without \cite{%
2005PhR...417....1B
} helicity.
Flows with a large-scale helical pattern of alternating sign, such as the
Taylor-Green flow \cite{%
2005PhRvL..94p4502P,
2007NJPh....9..296P
}, may be considered as an intermediate case between large-scale and
small-scale dynamos.

In the following we use the magnetic Reynolds number, defined analogously
to the fluid Reynolds number in \Eq{Rekfkd} by replacing $\nu$ by $\eta$,
\EQ
\Rm=u_{\rm rms}/\eta k_{\rm f}.
\EN
The ratio between kinematic viscosity and magnetic diffusivity is
referred to as the magnetic Prandtl number, $\Pm=\nu/\eta$.
The onset of a dynamo is characterized by $\Rm\ge\Rmcrit$,
where $\Rmcrit$ is the critical value.
An important difference between large-scale and small-scale dynamos is
the different dependence of $\Rmcrit$ on $\Pm$.
Establishing an asymptotic dependence of $\Rmcrit$
on $\Pm$ is important because, even though the computing power
will increase, it will still not be possible to simulate realistic
values of $\Pm$ in the foreseeable future.
Schekochihin et al.\ \cite{%
2005ApJ...625L.115S
} have compared the results from
two independent codes and show that there is a sharp increase of
$\Rmcrit$ with decreasing $\Pm$; see \Fig{prmcrit}
(where the two quantities are denoted as $\mbox{Rm}_{\rm c}$ and $\mbox{Pm}$).
Such a result was first derived analytically \cite{%
1997PhRvE..56..417R
}, well before it was seen also in simulations.

The reason for the increase of $\Rmcrit$ with increasing
$\mbox{Re}$ has been explained by Boldyrev and Cattaneo \cite{%
2004PhRvL..92n4501B
}
as being related to the fact that when $\Rm<\Rey$, the
resistive scale (i.e.\ where the magnetic power spectrum peaks
in the kinematic regime) shifts from the dissipative subrange into
the inertial range.
In the inertial range the velocity field is no longer smooth, but
it is rough in the sense that the exponent $\zeta_1$ (see \Sec{Intermittency})
in the scaling of velocity differences over distance $r$ is less than 1 \cite{%
2004PhRvL..92n4501B
}.
For $\zeta_1<1$, the velocity field becomes non-differentiable in the sense that
velocity gradients diverge like $r^{\zeta_1-1}$.
The smaller $\zeta_1$, the rougher the velocity field,
while $\zeta_1=1$ corresponds to a smooth velocity field.

More recent work \cite{%
2007PhRvL..98t8501I,
2007NJPh....9..300S
} suggests that the
threshold for small-scale dynamos is particularly high {\it only} in the range
$0.06\la\Pm\la0.2$, because then the resistive scale lies within
the range where the kinetic energy spectrum shows the bottleneck with
$\zeta_1\to0$, corresponding to an extremely rough velocity field
with very large critical magnetic Reynolds number.
However, when $\Pm\la0.06$, the resistive scale lies beyond the bump of the
bottleneck, i.e.\ well inside the inertial range,
and there the critical magnetic Reynolds number is again somewhat smaller.
Resolving this issue conclusively requires a numerical resolution well in excess
of $1024^3$ meshpoints, as well as long run times,
which is only now beginning to become feasible.
We may therefore expect further developments in this area in the near future.

\begin{figure}[t!]\centering
\includegraphics[width=\columnwidth]{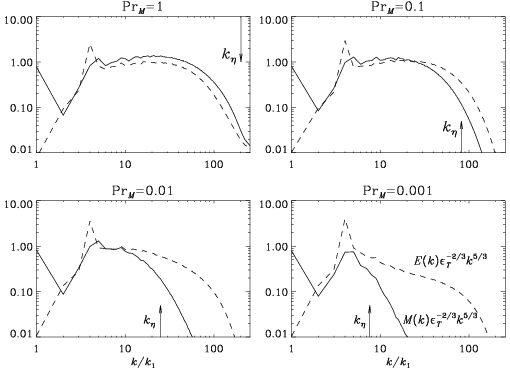}
\caption{
Compensated kinetic and magnetic energy spectra in the saturated regime for
$\Pm=1$ with $\Rey=450$, $\Pm=10^{-1}$ with $\Rey=1200$,
$\Pm=10^{-2}$ with $\Rey=2300$, and $\Pm=10^{-3}$ with $\Rey=4400$.
The spectra are compensated by $\epsilon_T^{-2/3}k^{5/3}$,
where $\epsilon_T$ is the sum of kinetic and magnetic energy dissipation rates.
The ohmic dissipation wavenumber, $k_\eta=(\epsilon_M/\eta^3)^{1/4}$,
is indicated by an arrow.
The viscous dissipation wavenumbers are 180, 290, 350, and 430 for
$\Pm=1$, $10^{-1}$, $10^{-2}$, and $10^{-3}$, respectively.
Adapted from Ref.~\cite{%
2009ApJ...697.1206B
}.
}\label{pspec_aver_comp_saturated2}\end{figure}

If there is large-scale dynamo action, the magnetic field grows
preferentially at scales large compared with the energy-carrying scale.
This process is non-local in spectral space \cite{%
2001ApJ...550..824B
}, although it has also been shown that an externally applied magnetic field
produces mainly local interactions \cite{
2007PhRvE..76e6313A
}.
On the other hand, large-scale dynamo action depends on velocity
and magnetic field correlations at the energy-carrying scale
(rather than the resistive scale).
The onset of this type of large-scale dynamo action is essentially
independent of $\Pm$ and occurs when $\Rm>\Rmcrit\approx1$.
The independence of the saturation strength of the large-scale dynamo
on the microscopic resistivity is demonstrated in
\Fig{pspec_aver_comp_saturated2},  where we show spectra of kinetic and
magnetic energies for different values of $\Pm$.

\subsection{Turbulent convection and stratification}
\label{Convection}

In certain layers of a star the opacity of the gas can become so large
that the energy flux can no longer be transported by radiative diffusion,
but by convection.
A phenomenological theory called mixing length theory allows one to make
reasonable estimates for the expected turbulent velocity.
As mentioned in \Sec{SolarConvection}, the convective energy flux is
approximately equal to the $\rho\urms^3$.
This gives a good estimate for the convective velocity in a star.

The basics of the convection instability was discussed in \Sec{Scalars}.
A necessary condition for convection is that the specific entropy
decreases with height, i.e.\ $N_{\rm BV}^2<0$; see \Eq{NBVdef}.
In addition, viscosity and thermal diffusion have to be small enough
compared with the height of the unstable layer, $d$, and the
Brunt--V\"ais\"al\"a frequency, $N_{\rm BV}$.
This is quantified by the Rayleigh number,
\EQ
\mbox{Ra}={d^4\over\nu\chi}\left(-N_{\rm BV}^2\right)_0,
\EN
which has to be above a certain critical value for the onset of convection.
Here, the subscript 0 refers to the requirement that the specific entropy
gradient has to be calculated for the associated hydrostatic equilibrium
solution, and not for the already convectively unstable solution.
Such solutions are not normally presented in the literature.
Also, the thickness of the outer layers of the Sun would be much smaller
in the hydrostatic reference state.
It is therefore not common to quote Rayleigh numbers in astrophysics,
except in idealized simulations whose hydrostatic reference solutions
tend to be polytropes where the initial density is related to the initial
temperature via $\rho\sim T^n$, where $n$ is the polytropic index.
Unlike the incompressible case, where the Rayleigh number is based on
the background gradient of temperature, in the compressible case
it is based on the gradient of specific entropy for the associated
hydrostatic solution \cite{%
2005AN....326..681B
}.

If the value of the Rayleigh number is increased sufficiently beyond
the critical value, the flow becomes turbulent.
Simulations of turbulent convection have been provided by many different
groups, both in the incompressible approximation \cite{%
1996JFM...310..139K,
1996JFM...322..243J,
2003PhRvL..91f4501H
}
as well as in the fully compressible case \cite{%
1989ApJ...342L..95S,
1991ApJ...370..282C,
1998ApJ...499..914S
}.
Typical Rayleigh numbers that are currently reached in simulations are
around $10^6$.
With rotation the onset of convection is delayed correspondingly, which
enables one to reach somewhat larger Rayleigh numbers in such cases.

The Nusselt number is another commonly used quantity in incompressible
and laboratory convection.
In that case it gives the ratio of the total heat flux to that transported
by heat conduction alone, using the same boundary conditions.
However, unlike laboratory convection, where the temperatures at top and
bottom are usually kept fixed, in many compressible simulations with a
polytropic background solution the energy flux at the bottom is actually
prescribed.
One compares therefore normally with the radiative solution
with a linear temperature profile that has the same top and bottom
temperatures as the convective solution.
One also subtracts out the flux that is transported by the adiabatic
stratification alone \cite{%
1984ApJ...282..557H
}.
Again, this value is nowadays not normally quoted for compressible simulation.
For many purposes, a more useful characterization of the turbulence is the
resulting value of the Reynolds number.

Another important difference to laboratory convection is the absence of
boundaries in astrophysical convection.
Convectively unstable layers are the result of a particular dependence
of opacity on temperature and density.
This has frequently been modelled by using prescribed spatial profiles of the
radiative conductivity.
In this way one can model convection in an unstable layer, sandwiched
between two stable layers \cite{%
1986ApJ...311..563H
}.
This makes the dynamics near the transition layer softer and allows
the flow to overshoot into the stably stratified layers.
This leads to the excitation of gravity waves in the stably stratified
layers \cite{%
1986ApJ...311..563H,
2004A&A...421..775D,
2005A&A...438..365D,
2005MNRAS.364.1135R,
2008MNRAS.387..616R,
2009A&A...494..191B
}.

\begin{figure}[t!]\begin{center}
\includegraphics[width=\textwidth]{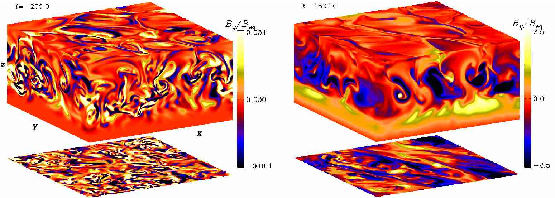}
\end{center}\caption[]{
Snapshots of $B_y$ in the early phase (left)
and saturated phase (right) of the dynamo.
The sides of the box show the periphery of the
domain whereas the top and bottom slices show $B_y$ at
top and bottom of the convectively instable layer, respectively.
Courtesy Petri J.\ K\"apyl\"a \cite{%
2008A&A...491..353K
}.
}\label{By_boxes_512b}\end{figure}

Convective flows can well support dynamo action.
As an example we mention here the result of a convection simulation
with horizontal shear which leads gradually to the development of a
large-scale magnetic field \cite{%
2008A&A...491..353K
}.
A result of such calculations is shown in \Fig{By_boxes_512b}, where
we visualize the toroidal field component at an early time when only
small-scale fields have been produced, and at a later time when also a
large-scale field is present.

The presence of large-scale fields is often characterized by energy spectra.
However, because of stratification it only makes sense to look at
horizontal spectra taken at a specific depth.
If the mean magnetic field depends mainly on depth, the horizontal
magnetic energy spectra will peak at wavenumber zero, which can only
be seen if one plots the spectral energy versus linear wavenumber;
see, for example, Fig.~12 of Ref.~\cite{%
2008A&A...491..353K
}.

\subsection{Global hydromagnetic dynamo simulations}

\begin{figure}[t!]\begin{center}
\includegraphics[width=\textwidth]{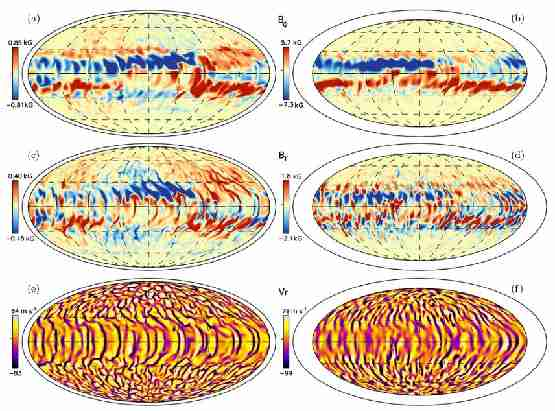}
\end{center}\caption[]{
Toroidal and radial magnetic field (first and second row) together with
radial velocity (bottom row) near the top of the convective shell (left
column at $r/R=0.95$) and in the middle (right column at $r/R=0.85$).
The mean magnetic field is approximately antisymmetric about the equator.
The radial velocity shows flows patterns elongated along the rotation axis
(so-called banana cells).
The resolution is $96\times256\times512$ mesh points or collocation points
in the radial, latitudinal, and longitudinal directions, respectively.
The magnetic Reynolds number based on the thickness of the convective shell
and without dividing by $2\pi$ is 86, and the Coriolis number, i.e.\ the
ratio of vorticity from the mean rotation to the rms vorticity of the
turbulence, is about 3.
Courtesy Benjamin P.\ Brown \cite{%
2010ApJ...711..424B
}.
}\label{Brown_etal10}\end{figure}

Simulations of global convection have demonstrated the generation
of differential rotation and magnetic fields \cite{%
1983ApJS...53..243G,
1985ApJ...291..300G
}.
However, with parameters relevant to the Sun such models have not yet
produced large-scale magnetic fields similar to those in the Sun \cite{%
2000ApJ...532..593M,
2004ApJ...614.1073B
}.
This is plausibly explained by the relevant dynamo numbers for coherent
large-scale field generation being still too small.
In that case, only small-scale magnetic fields are generated, while the
threshold for large-scale field generation has still not been reached.
This is different when the rotation rate of the sphere is increased to
several times the solar value \cite{%
2010ApJ...711..424B
}.
As an example we show here the results for a sphere that has
a stratification similar to that of the Sun, but it is rotating
about 3 times as fast; see \Fig{Brown_etal10}.

The rapid rotation is primarily responsible for producing the typical
convection patterns that are elongated in the direction of the rotation
axis.
This effect is especially obvious at low latitudes, outside the inner
tangent cylinder, i.e.\ the cylinder that is tangent with the bottom of
the convection zone.
The resulting convection pattern is often referred to as banana cells,
a concept that was widely discussed in the 1980s \cite{%
1986Sci...234...61H
}, but there has never
been observational evidence supporting this type of flow pattern for the Sun.
Banana cells occur as a consequence of rapid rotation, which is also
responsible for cylindrical angular velocity contours.
Although this does not apply to the Sun, it may well apply to some stars
that rotate much more rapidly than the Sun.

Simulations of rapidly rotating convection \cite{%
2010ApJ...711..424B,
2010AN....331...73K
} show that in the region with strong banana cell convection,
there is strong large-scale dynamo action with pronounced toroidal flux
belts on one or both hemispheres; see \Fig{Brown_etal10}.
This is partially reminiscent of the magnetic activity in the Sun,
although it would be premature to draw any conclusions from this
given that at present there is no explicit evidence of banana cell
convection in the Sun.

We mention here another line of research.
Instead of convection driving the flow one can apply an artificial
forcing function.
This has the advantage of producing a flow pattern whose typical
scale can be controlled.
In particular, it is possible to achieve turbulent scales that are
small compared with the radial extent of the domain, so as to
produce a well-defined scale separation \cite{%
2009ApJ...697..923M,
2010ApJ...719L...1M
}.
With such simulations it has been possible to to focus entirely
on the nature of the dynamo in spherical shell geometries and to
isolate its physics from many other effects that may still be important.
It turns out that even in the absence of global shear, oscillatory large-scale
fields can be generated \cite{%
2010ApJ...719L...1M,
2010AN....331..130M
}.
Such solutions show equatorward migration and are quite different in
nature from oscillatory solutions of $\alpha\Omega$ type.
It is well possible that these solutions have nothing to do with
those in the solar dynamo, but it serves as a reminder that the variety
of possibilities can be much larger than what is usually discussed.

\subsection{Interaction between convection and shear}
\label{ConvectionShear}

Simulations of rotating convection in spherical shells demonstrate
that there is equatorward acceleration of the mean flow.
This phenomenon is generally referred to as differential rotation and
will be discussed in more detail in \Sec{LambdaEffect}.
In addition, one sees that the convection pattern itself moves
differentially across the surface.
However, a more detailed inspection reveals that at the equator the
convection pattern can actually move still somewhat faster than the
mean flow.
This has been revealed both by linear theory \cite{
2006ApJ...641L..77G,
2007ApJ...665L..75G,
2007SoPh..245...27B
} and by nonlinear simulations \cite{
2007IAUS..239..457B
}, and may explain a phenomenon seen at the solar surface which shows that
magnetic tracers move at speeds faster than the speed of the plasma.
In fact, even very young sunspots tend to move not only faster than
the plasma at the surface, but they move also faster than the gas at any other place
in the Sun, as seen by global helioseismology; see Figure~4 of Ref.~\cite{
1999ApJ...517L.163B
}.

There is at present no universally accepted theory for the enhanced rotation
speed of magnetic tracers on the Sun.
It has however been pointed out that the enhanced pattern speed of magnetic
tracers might be understandable if the observed magnetic field (including that
responsible for producing sunspots) was generated in a layer not too far
below the surface \cite{
2005ApJ...625..539B
}.
This proposal would be in conflict with the generally adopted view
according to which the magnetic field responsible for the solar cycle is
generated near or even below the bottom of the convection zone of the Sun.

\subsection{Granulation, convection, and solar abundances}
\label{Granulation}

Simulations of solar granulation have reached a high level of realism
and have proved to be a viable and feasible alternative to earlier
one-dimensional models for calculating diagnostic spectra in visible light.
Strictly one-dimensional models always needed to incorporate ill-determined
parameterizations of what is known as micro and macro turbulence.
New realistic three-dimensional simulations of solar convection \cite{%
2005ARA&A..43..481A,
2008A&A...488.1031C,
2009ARA&A..47..481A,
2001ApJ...556L..63A,
2004A&A...417..751A,
2005ASPC..336...25A,
2009A&A...498..877C
} lead to diagnostic spectra that can be fitted to observed spectra without
invoking these ill-known parameterizations.
The use of 3-D models also results in abundances
derived from different spectral features (e.g.\ molecular and atomic lines) being
more consistent.

Initial efforts to derive updated solar abundances based
on 3-D models resulted in new abundance estimates for the heavier elements
in the Sun that were as low as only 60\% of previous estimates \cite{%
1989GeCoA..53..197A
}.
It should be noted, however, that even though these abundances are
often referred to as ``3-D abundances'', 3-D effects were {\em not} the
main cause of the systematic lowering of the abundance estimates, which
were instead a combined result of updated oscillator strengths, different
line fitting procedures, and choices made when estimating collision cross
sections important for non-LTE corrections for some spectral lines.  This
was elucidated by an independent analysis by a different group \citep{%
2008A&A...488.1031C,
2009A&A...498..877C
}, who confirmed that 3-D effects improve the consistency but do
not give rise to a significant systematic abundance effect for the
important heavy elements.

The abundances of the heavier elements determine the opacity of the
gas and thereby the detailed radial structure of the Sun.
On the other hand, the radial dependence of the sound speed and density
in the Sun can be determined independently through helioseismology \cite{%
1984ARA&A..22..593D,
1985SoPh..100...65G,
1988RvMP...60..297B,
2001ApJ...555..990B,
2006ApJS..165..400B
}, and helioseismology can thus provide important constraints on
the heavy element abundances in the solar interior.
(It may in the future be possible to also determine the Sun's deep
interior composition by exploiting neutrinos from the CN cycle and the
p-p chain to determine the primordial solar core abundances of C and N
at an interesting level of precision \cite{%
2008ApJ...687..678H
}.)
In the convection zone the gradual ionization of carbon, nitrogen, and
oxygen with depth influences the equation of state, and helioseismic
measurements of the effective ratio of specific heats of the gas can thus
provides constraints on the abundance of these elements also there \citep{%
2008PhR...457..217B,
2006ApJ...644.1292A
}.

The significant downward revision of solar abundances proposed in Ref.~\cite{%
2005ASPC..336...25A
} and even the somewhat more moderate revisions proposed more recently
by the same group \citep{%
2009ARA&A..47..481A} 
turned out to be difficult to reconcile with observational constraints from helioseismology,
despite many different attempts to do so; cf.\ Ref.~\cite{%
2008PhR...457..217B
} and references therein.
However, the downward revisions recommended by
\cite{2008A&A...488.1031C} 
are only about half as large and are in fact consistent with
helioseismic estimates of the heavy element abundance in the
solar convection zone, $Z=0.167$;
see Table~2 of Ref.~\cite{2006ApJ...644.1292A}. 

Due to gravitational settling the abundances of all elements
differ somewhat between the convection zone and the
radiative interior \citep{1996A&A...312.1000R}. 
Because of rapid mixing the abundance levels are constant in the
convection zone, but below the convection zone the chemical abundances
vary with radius in a manner that is influenced by how turbulence in
the convection zone generates weak overshooting motions in the
radiative zone, which result in a slow mixing over depth of chemical
elements \citep{2004ApJ...607.1046R}. 

There was always a small departure in sound speed between models and
helioseismic observations in a narrow region just below the convection zone.
With the revised abundance estimates by
\cite{2005ASPC..336...25A} 
this departure increased from about 0.3\% to about 1.2\% \cite{%
2008PhR...457..217B,
2005ApJ...618.1049B
}, while with the abundances recommended by
\cite{2008A&A...488.1031C} 
the discrepancy is of the order 0.6\%.
Even the smallest of these discrepancies is many times larger than
the helioseismic measurement uncertainties, and one should thus worry
less about the particular size of the discrepancy in any one case, and
more about the very existence of the discrepancy.
In general terms, the lack of a detailed quantitative understanding of the overshoot of
convection below the bottom of the convection zone and the associated slow
mixing seems to be a likely reason for the discrepancy
\citep{2002RvMP...74.1073C}. 

An important additional observational constraint on slow mixing below the
convection zone comes from the depletion of Lithium in
the Sun.  Lithium is destroyed at temperatures that are reached about one
pressure scale height (corresponding to about 1\% of the solar mass) below
the convection zone, and the observed depletion (a factor of about 160)
implies that mixing down to that temperature takes place on a time scale
considerably shorter than the age of the Sun, but still very large compared
to convection zone turnover times
\citep{2002RvMP...74.1073C,2010Ap&SS.328..193M}. 
Lithium depletion in other stars is now known to be essentially consistent
with the behavior expected from the differences in age and structure deduced
from standard stellar evolution theory \cite{%
2010Ap&SS.328..193M}.  

\subsection{Turbulence from the magneto-rotational instability}
\label{MagnetoRotational}

In the presence of shear and rotation, the slow magnetosonic waves
develop a long wavelength instability, where $\omega^2<0$ for
$v_{\rm A}^2k^2<2q\Omega^2$.
Here, $q=-\dd\ln\Omega/\dd\ln r$
quantifies the radial gradient of the angular velocity.
This is called the magneto-rotational instability (MRI).
It is particularly simple to analyze if the magnetic
field is vertical, in which case the instability is purely axisymmetric.
However, the same growth rates are obtained in the nonaxisymmetric
case, if $\BB$ points in the streamwise direction \cite{%
1998RvMP...70....1B
}.

\begin{figure}[t!]\begin{center}
\includegraphics[height=.5\textwidth]{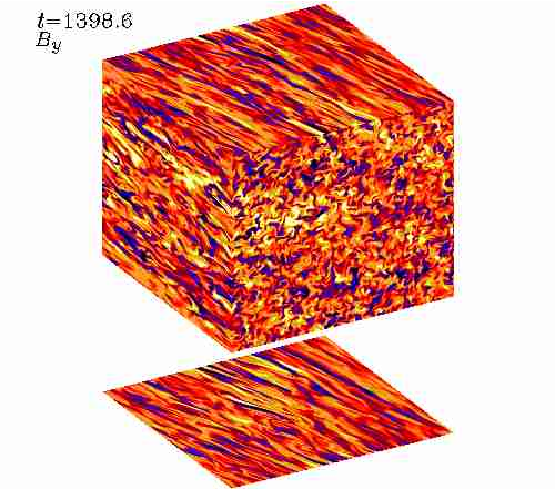}
\end{center}\caption[]{
Toroidal magnetic field component displayed on the
periphery of the computational domain (color coded).
The size of the box is $2\pi$ in all three directions
and the mesh size is $512^3$ meshpoints.
The gas is isothermal with a constant sound speed of $c_{\rm s}=5\Omega/k_1$.
Viscosity and magnetic diffusivity are
$\nu=\eta=2.5\times10^{-4}\Omega/k_1^2$.
Note that the magnetic field develops long thin structures aligned at
some angle relative to the toroidal direction.
Adapted from Ref.~\cite{%
2004AIPC..733..122B
}.
}\label{BH512By}\end{figure}

In the axisymmetric case the instability takes the form of so-called
channel flows.
In three dimensions the flow experiences strong shearing and hence
small length scales in the cross-stream direction.
This leads to the flow breaking up into what we
loosely call fully developed turbulence.
An example of such a flow is shown in \Fig{BH512By}, where periodic
boundary conditions have been used in the vertical and azimuthal
directions, and shearing-periodic boundary conditions in the cross-stream
direction.
No net magnetic flux has been applied \cite{%
2004AIPC..733..122B
}.
Numerical simulations show, however, that the MRI is no longer excited
when the magnetic Prandtl number is less than a critical value of the
order of unity \cite{%
2007A&A...476.1113F,
2007A&A...476.1123F
}.
A strong sensitivity on the magnetic Prandtl number has
also been found for magnetized Taylor-Couette flow \cite{%
2003PhRvE..67d6312R,
2005PhRvL..95l4501H,
2006PhRvL..97r4502S,
2007NJPh....9..295S,
2007AN....328.1158R
}.
At present it is still unclear whether there is a real problem with the MRI
in accretion discs when the magnetic Prandtl number is small.
This issue may well be connected with the difficulty to excite small-scale
dynamos at low magnetic Prandtl numbers \cite{%
2005ApJ...625L.115S,
1997PhRvE..56..417R,
2004PhRvL..92n4501B,
2007PhRvL..98t8501I,
2007NJPh....9..300S
}.
On the other hand, astrophysical dynamos are large-scale dynamos, and they
do not suffer from that particular difficulty \cite{%
2009ApJ...697.1206B,
2007PhRvE..76b6316M
}.
It would therefore be important to perform new MRI simulations in cases
where large-scale dynamos are possible, i.e.\ in the presence of vertical
density stratification, which can then lead to an $\alpha$ effect \cite{%
1995ApJ...446..741B,
2002GApFD..96..319B
}.
In another recent study it has been shown that even without stratification,
large-scale dynamo action is possible when pseudo-vacuum boundary conditions
are used at top and bottom of the rotating shearing box \cite{%
2010arXiv1004.2417K
}.
A similar generation of mean fields has also been found without rotation \cite{%
2005ApJ...625..539B,
2008PhRvL.100r4501Y,
2008AN....329..737Y,
2008ApJ...676..740B
}.
Possible candidates for explaining the origin of large-scale fields in this
case include the incoherent $\alpha$--shear dynamo \cite{%
1997ApJ...475..263V,
2007MNRAS.382L..39P
} and the shear--current effect \cite{%
2003PhRvE..68c6301R,
2004PhRvE..70d6310R
}.
For the latter effect to work, it is necessary that one of the off-diagonal
components of the magnetic diffusivity tensor has a suitable sign, which may
however not be the case \cite{%
2005AN....326..787B,
2008ApJ...676..740B,
2006PhRvE..73e6311R,
2006AN....327..298R
}.

\begin{figure}[t!]\begin{center}
\includegraphics[height=.5\textwidth]{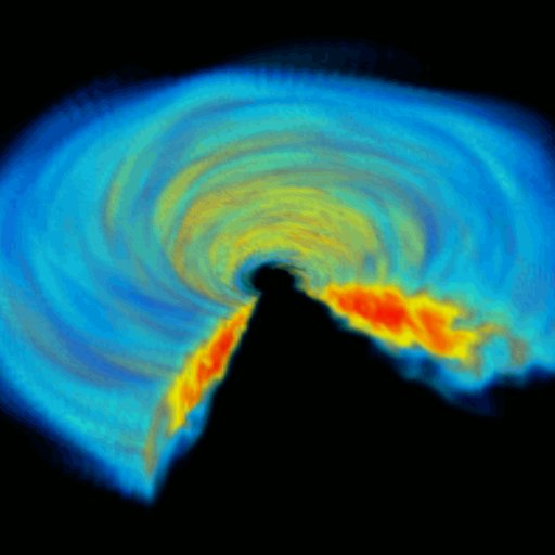}
\end{center}\caption[]{
Visualization of the logarithmic density of an accretion torus
around a black hole.
Courtesy of John F.\ Hawley \cite{%
2000ApJ...528..462H
}.
}\label{torus3d}\end{figure}

The MRI is generally thought to be responsible for driving turbulence in
accretion discs, where $q=3/2$.
A more accurate representation of accretion discs is obtained with the inclusion of
vertical and radial density stratification.
The former case can be treated within the shearing-box approximation \cite{%
1988AJ.....95..925W,
1995ApJ...440..742H
} while the latter requires a global treatment \cite{%
2000ApJ...528..462H,
2000ApJ...532L..67M,
2003ApJ...585..429M,
2003ApJ...599.1238D
}.
In \Fig{torus3d} we show a visualization of the logarithmic density
of an accretion torus around a black hole

An important diagnostic quantity of accretion disc simulations
is the dimensionless turbulent disc viscosity,
$\alpha_{\rm SS}=\nu_{\rm t}/c_{\rm s}H$.
Here, the subscripts SS refer to Shakura and Sunyaev \cite{%
1973A&A....24..337S
}, who employed this parameterization of turbulent viscosity $\nu_{\rm t}$
in terms of local sound speed $c_{\rm s}$ and pressure scale height $H$.
In the simulations, $\nu_{\rm t}$ is normally estimated by the mean
total horizontal stress,
$\overline{\Pi}_{R\phi}=\overline{b_Rb_\phi/\mu_0-\rho u_Ru_\phi}$,
divided by the mean
rate of strain resulting from the differential rotation,
$\rho R\partial\Omega/\partial R$, where cylindrical coordinates,
$(R,\phi,z)$, have been employed.

In comparison with local shearing box simulations, an important difference
is that global simulations are capable of producing about 10 times larger
values of $\alpha_{\rm SS}$.
This is an immediate consequence of the larger field strength
in global simulations rather than a difference in the intrinsic properties
of local versus global disc simulations \cite{%
2000ApJ...528..462H
}.
Another important outcome of global disc simulations is the fact that
$\overline{\Pi}_{R\phi}$ is finite at the innermost marginally stable
orbit.
This is a property that is not normally taken into account in analytic
models and continues to be debated in the literature \cite{%
2002ApJ...573..754K,
2008MNRAS.390...21B
}.

\begin{figure}[t!]\begin{center}
\includegraphics[width=.5\textwidth]{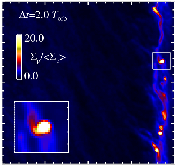}
\end{center}\caption[]{
Simulations showing the total column density
(gas plus all particle sizes) in the horizontal plane.
The insets show the column density in logarithmic scale centered around
the most massive cluster in the simulation.
Time is given in terms of the number of orbits after turning on self-gravity.
Courtesy of Anders Johansen.
}
\label{johansen1}
\end{figure}

A number of new simulations have emerged in recent years.
A major step was the combination of dust dynamics with self-gravity
in the shearing box approximation \cite{%
2007Natur.448.1022J,
2007arXiv0708.3893J
}.
One of the remarkable results is a rapid formation
of nearly Earth-sized bodies from boulders (\Fig{johansen1}).
Even though the mass of what one might call protoplanet
is growing, this body is also shedding mass during encounters
with ambient material as it flows by.
One might speculate that what is missing is the effect of
radiative cooling of the protoplanet.
This would allow the newly accreted material to lose entropy,
become denser, and hence fall deeper into its potential well.

The main reason the simulations presented in Ref.~\cite{
2007Natur.448.1022J
} produce rapid growth is connected
with the occurrence of sufficiently strong compressions caused by the
turbulence.
Once the compression is strong enough, self-gravity takes over
and leads to a fully developed nonlinear collapse.

\subsection{Effects of thermal and gravitational instabilities}
\label{ThermalInstability}

A thermal instability may arise if a cooling term, $\Lambda(T)$,
and a heating term, $\Gamma(T)$, are
included on the right hand side of the energy or entropy equation, i.e.\
\EQ
\rho T{\DD s\over\DD t}=...+\rho\Gamma(T)-\rho^2\Lambda(T).
\EN
It is convenient to abbreviate the combination of the two terms on the
right hand side by $\rho{\cal L}$, where ${\cal L}=\Gamma-\rho\Lambda$.
This allows us to state a sufficient condition for stability \cite{%
2007ApJ...654..945B,
1965ApJ...142..531F
}
\EQ
\left({\partial{\cal L}\over\partial T}\right)_p>0\quad\mbox{(stability)}.
\EN
This means that when the temperature is increased, the corresponding
cooling increases, bringing the temperature down again to the
original value.
In the presence of thermal diffusion, with $\FF_{\rm rad}=-K\nab T\neq0$,
the system can always be stabilized at small scales, i.e.\ for
large wavenumbers, where eventually
the thermal diffusion rate becomes faster than the cooling rate.
For $\Gamma=\const$ and $\Lambda\propto T^\beta$, the dispersion relation
$\omega(k)$, is on sufficiently large scales (small wavenumbers) of the form \cite{%
2007ApJ...654..945B,
1965ApJ...142..531F
}.
\EQ
\omega=c_{\rm iso}k\sqrt{1-\beta^{-1}},
\EN
where $c_{\rm iso}=c_{\rm s}/\sqrt{\gamma}$ is the isothermal sound speed.
Evidently, for $\beta<1$ sound waves become destabilized
($\omega$ becomes imaginary).

\begin{figure}[t!]\begin{center}
\includegraphics[width=\textwidth]{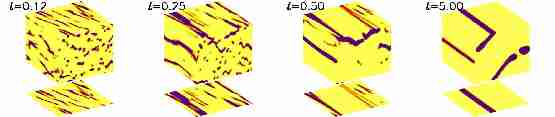}
\end{center}\caption[]{
Visualization of $\ln T$ on the periphery of the box at different times,
for $\nu=\chi=5\times10^{-4}\Gyr\,\km^2\s^{-2}$ and $256^3$ mesh points.
$\bra{\rho}\approx1.74\times10^{-24}\g\cm^{-3}$,
$\bra{p}\approx24.2\times10^{-14}\dyn$, and $\bra{T}\approx8200\K$.
Here $\Omega=100\Gyr^{-1}$ and $S=-\Omega$.
For this run $k_{\rm F}/k_1=32$ and $n_\rho/(c_{\rm s}k_{\rm F})=1.5$.
The growth rate is about $190\Gyr^{-1}$, which is somewhat larger
than for the corresponding non-shearing run.
Note that the initially produced structures are quickly sheared out.
Adapted from \cite{%
2007ApJ...654..945B
}
}\label{sh256b5}\end{figure}

Numerical simulations \cite{%
2002ApJ...569L.127K,
2007ApJ...654..945B
} have not been able to confirm alternative findings \cite{%
2007ASPC..365..162I
} that the thermal instability can lead to sustained turbulence.
This is demonstrated in \Fig{sh256b5}, which shows
(here in the presence of shear) that the thermal
instability leads to the development of patches with low temperature
(100\,K compared to 10,000\,K outside those patches), but over time
these patches merge until eventually a stable equilibrium is reached
where a few big patches continue to coexist.

Another instability where sound waves are destabilized is the
Jeans instability.
Here the dispersion relation can be written in the form \cite{%
1997ApJ...489L.179T,
1902RSPTA.199....1J,
1956MNRAS.116..351B,
2004ARA&A..42...79B
}
\EQ
\omega^2=c_{\rm s}^2k^2-4\pi G\rho,
\EN
where $\rho$ is the local density of the gas.
So, again, large scales become unstable.
In an asymptotically thin layer such as an accretion discs or galaxies,
the dispersion relation becomes \cite{%
1987gady.book.....B
}
\EQ
\omega^2=c_{\rm s}^2k^2-4\pi G\Sigma|\kk|,
\EN
where $\Sigma$ is the local surface density.
In the context of local accretion disc models, simulations suggest that
this process can indeed lead to sustained turbulence \cite{%
2008ApJ...673.1138C,
2001ApJ...553..174G,
2007prpl.conf..607D,%
2009arXiv0912.0288K
}.
In simulations of star formation \cite{%
2004RvMP...76..125M,
2003MNRAS.339..577B,
2007ApJ...661..972P,
2008MNRAS.389.1556B
},
the Jeans instability leads to a continuous production of gravitationally
bound structures corresponding to protostars.
The stars that form have a broad distribution of masses, determined mainly
by the statistics of mass fragmentation in supersonic MHD-turbulence \citep{%
2002ApJ...576..870P}.
The fraction of stars that are heavier than about 8 solar masses eventually
(after a delay of up to a few tens of Myr) explode as supernovae.
These supernovae contribute to sustaining the turbulence in the interstellar
medium that ultimately causes additional generations of stars to be born
\citep{2005A&A...436..585D,2007ApJ...665L..35D}.

\subsection{Supernova-driven turbulence}
\label{SNdriven}

Interstellar turbulence is an example of astrophysical turbulent flows
where the driving is usually modeled by a distributed body force.
As discussed in \Sec{InterstellarTurbulence}, the blast waves of supernova
explosions provide energy input to the surrounding gas.
These explosions drive gas flows with temperatures of around $10^8\K$,
but they also lead to strong compressions where the gas cools rapidly
to about $10^4\K$.
When the temperature is between $100\K$ and $10^4\K$ the gas may,
depending as details of the cooling curve $\Lambda(T)$,
be thermally unstable \cite{%
1965ApJ...142..531F
}.
This contributes to keeping the gas in the interstellar medium
preferentially in one of two distinct temperature regimes
(the so-called cold and warm phases; see \Sec{ThermalInstability}).
The hot phase at temperatures $>10^6\K$ is a direct result of
heating by supernova explosions combined with a low cooling efficiency
of the interstellar medium at that temperature.
This is also borne out by various simulations
\cite{%
1999ApJ...514L..99K,
2004A&A...425..899D,
2005ApJ...626..864M
}.
Simulations show that the filling factor of the hot gas ($T>10^6\K$)
grows with height from 0.2--0.3 at the midplane to about 0.5--0.6 at a
height of about 300\pc\ \cite{%
1999ApJ...514L..99K
}.
However, this result depends on the degree of correlation of supernovae
in space and can reach 0.6 at the midplane for completely uncorrelated
supernovae, as in an early analytic model \cite{%
1977ApJ...218..148M
}.
Simulations have also been able to demonstrate that significant amounts of
vorticity are being produced if the flow is sufficiently supersonic
and if the baroclinic term is important \cite{%
1999intu.conf..127K,
2010arXiv1008.5281D
}.
The presence of vorticity is advantageous for dynamo action; in fact, no
dynamos have yet been found when the turbulence is irrotational \cite{%
2006MNRAS.370..415M
}.

There is now mounting evidence that for large Mach numbers the energy
ratio of compressive to solenoidal velocities approaches 1/2 \cite{%
2002ApJ...576..870P,
1994PhFl....6.2133P,
1998PhFl...10..237P,
2004MNRAS.353..947H,
2009arXiv0912.0546K,
2010arXiv1008.0665P
}.
This can be explained if the mean square values of
longitudinal and transversal velocity derivatives were equal, i.e.\
$\bra{u_{x,x}^2}=\bra{u_{x,y}^2}$.
Assuming isotropy and that mixed terms cancel, this implies
$\bra{(\nab\cdot\uu)^2}\approx3\bra{u_{x,x}^2}$ and
$\bra{\oo^2}\approx6\bra{u_{x,y}^2}$, giving a ratio of $1/2$ \cite{%
2004MNRAS.353..947H
}.
Whether or not this behavior is really universal needs to be seen.
In the papers listed above the turbulence was forced with a substantial
solenoidal component, so the issue of vorticity production was not addressed.
In the following we discuss the opposite limit, where only compressive
modes are driven and where no vorticity is produced.

\subsection{Irrotational turbulence}
\label{Irro}

Turbulence is usually thought of as being an ensemble of interacting eddies.
If one associates eddies with vortices, then
``irrotational'' turbulence must be a contradiction in terms.
Nevertheless, irrotational turbulence can be regarded as an idealization
that can serve its purpose in illustrating the difference to regular
(vortical) turbulence.

\begin{figure}[t!]\begin{center}
\includegraphics[width=.244\textwidth]{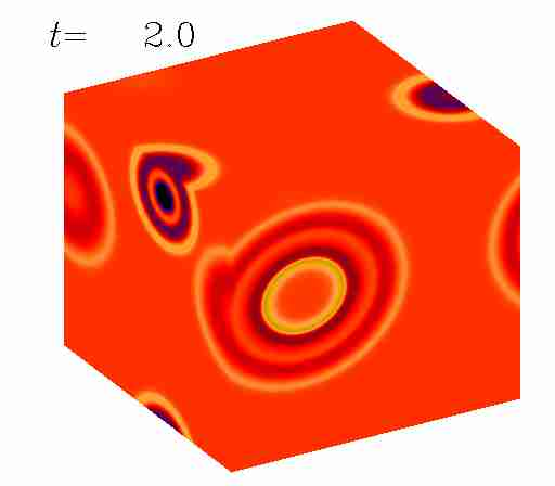}
\includegraphics[width=.244\textwidth]{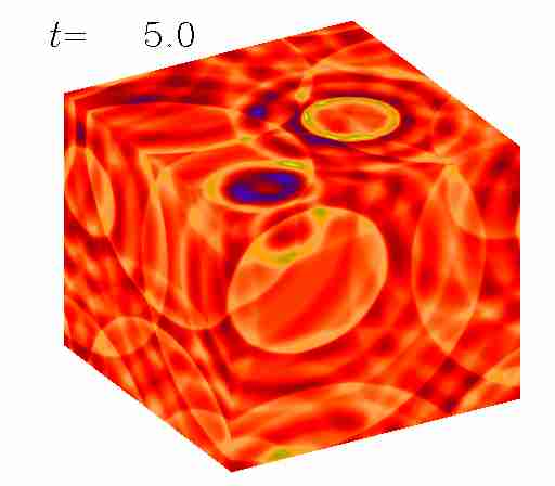}
\includegraphics[width=.244\textwidth]{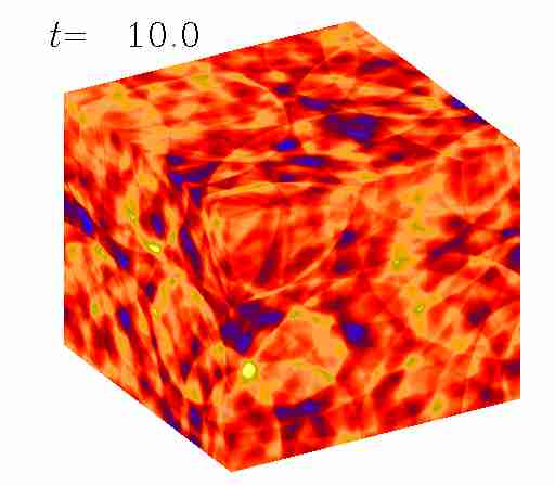}
\includegraphics[width=.244\textwidth]{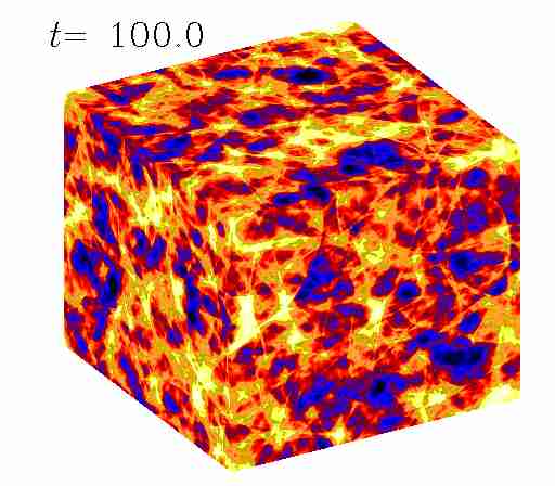}
\end{center}\caption[]{
Visualization of $\ln\rho$ on the periphery of the box at different times,
for $k_1 R=0.2$, $\mbox{Re}=50$, and $512^3$ mesh points.
Note that in the fully developed state individual expansion waves
can hardly be recognized.
Adapted from Ref.~\cite{%
2006MNRAS.370..415M
}.
}\label{img}\end{figure}

Irrotational turbulence means that $\oo=\nab\times\uu=0$.
As explained in \Sec{IncompressibleTurbulence}, the $\uu\times\oo$ nonlinearity
is absent and the only nonlinearity comes from the $\half\uu^2$ term.
This causes a significant modification of the turbulent cascade,
which is one of the reason why irrotational turbulence may be
a contradiction in terms.
Because of compressibility, however, vorticity can in principle be
generated via the viscous term.
Taking the curl of ${1\over\rho}\nab\cdot\ttau$ in \Eq{Fvisc},
and assuming $\nu=\const$, gives
\EQ
\nab\times\left({1\over\rho}\nab\cdot\ttau\right)
=\nu\nabla^2\oo+\nab\times\left[2\nu\SSSS\cdot\nab\ln(\rho\nu)\right].
\EN
Here, the first term vanishes if $\oo=\bm{0}$, but the second term
does not.
As mentioned in \Sec{SNdriven}, simulations show that this term remains
small in the limit $\nu\to0$ \cite{%
2006MNRAS.370..415M
}.
In \Fig{img} we show visualizations of the logarithmic density in a simulation,
which shows that the initially highly ordered expansion waves turn
rapidly into a complicated pattern.
The flow is here driven by a forcing function $\ff=-\nab\phi$, where
$\phi$ is a scalar function consisting of randomly placed Gaussians that
change in regular time intervals, $\Delta t$, such that
$\Delta t\urms\kf\approx0.25$.

Given that viscosity always perturbs the zero vorticity state slightly,
and because the vorticity equation is analogous to the induction equation,
one must ask whether a small initial vorticity could increase owing to
an instability.
However, at the Reynolds numbers achieved so far in simulations, neither
vorticity nor magnetic fields have been found to increase spontaneously \cite{%
2006MNRAS.370..415M
}.
The suggestion that purely irrotational turbulence cannot produce
dynamo action may be related to the finding that in vortical supersonic
turbulence the critical magnetic Reynolds number for small-scale
dynamo action shows a ``bimodal'' behavior with Mach number:
for Mach numbers below unity the critical magnetic Reynolds number
is about 35 to 40, and above unity it is about 70 to 80 \cite{%
2004MNRAS.353..947H
}.
Note, however, that the flow is here not purely irrotational,
and that the ratio of $\bra{(\nab\cdot\uu)^2}$ and $\bra{\oo^2}$
is about 1/2; see the discussion in \Sec{SNdriven}.

The results concerning vorticity production may be of relevance for
other flows that can be described by spherical expansion waves.
One example concerns phase transition bubbles that are believed to be
generated in connection with the electroweak phase transition in the
early universe \cite{%
1994PhRvD..49.3854I,
1986PhRvD..34.1719K
}.
Here the equation of state is that of a relativistic fluid,
$p=\rho c^2/3$, where $c$ is the speed of light.
Thus, again, there is no baroclinic term and no obvious source of vorticity.
However, the relativistic equation of state may be modified at small
length scales, but it is not clear that this can facilitate significant
vorticity production.

\section{Collective effects of turbulence}
\label{CollectiveEffects}

In this section we denote the velocity by a capital $\UU$.
Overbars indicate averages over one or two coordinate directions.
They are not therefore regarded as spatial filters that are often used
in the theory of large-eddy simulations \cite{%
1987QJRMS.113..413M,
1994BoLMe..71..247S
}.
The definition of averages in terms of coordinate averages
is convenient for interpreting simulation data.
Other definitions of averages are possible.
In analytic studies ensemble averages are commonly used.
Departures from these averages are denoted by lower case symbols,
i.e.\ $\uu=\UU-\meanUU$ and $\bb=\BB-\meanBB$ denote the fluctuating
components of the velocity and magnetic field vectors.
We discuss the properties of various correlators such as
$\overline{u_iu_j}$, $\overline{u_ib_j}$, and $\overline{b_ib_j}$.

In general turbulence is non-isotropic.
This can lead to the possibility of non-trivial components of the
correlations tensors $\overline{u_iu_j}$, $\overline{u_ib_j}$, and
$\overline{b_ib_j}$.
The effects of these correlations on the evolution of the mean flow,
$\meanUU$, and the mean magnetic field, $\meanBB$,
or the mean passive scalar concentration, $\meanC$,
are referred to as collective or mean-field effects of the turbulence.
Even in the special case where $\overline{u_iu_j}$ is a diagonal tensor there
is at least the phenomenon of turbulent diffusion, which will now be
illustrated in connection with the passive scalar field.

\subsection{Turbulent passive scalar diffusion}

The relevant dynamics comes from the nonlinearity.
In order to keep the discussion simple, we neglect the diffusion term.
The evolution equation of the passive scalar density
per unit volume, $C=\rho\theta$, is then
\EQ
{\partial C\over\partial t}=-{\partial\over\partial x_j}
\left(C U_j\right),
\label{dCdt}
\EN
cf.\ \Eq{drhoTheta}.
Again, we define the average concentration per unit volume as
$\meanC$ and write $C=\meanC+c$.
The evolution equation of $\meanC$ is obtained by averaging
\Eq{dCdt}, i.e.\
\EQ
{\partial\meanC\over\partial t}=-{\partial\over\partial x_j}
\left(\meanC\meanU_j+\overline{c u_j}\right).
\label{dmeanCdt}
\EN
The problematic term here is $\overline{c u_j}$, and the hope is that it
can be expressed in terms of mean fields such as $\meanC$ and $\meanUU$.

In order to derive an expression for $\overline{c u_j}$, we
consider its evolution equation,
\EQ
{\partial\over\partial t}\overline{c u_j}=
\overline{\dot{c}u_j}+\overline{c\dot{u}_j},
\EN
where dots denote partial time derivatives.
The evolution equation for $c$ is obtained by subtracting \Eq{dmeanCdt}
from \Eq{dCdt}, which yields
\EQ
{\partial c\over\partial t}=-{\partial\over\partial x_j}\left(
\meanC u_j+c \meanU_j+N_j^{(c)}\right),
\label{dcdt}
\EN
where $N_j^{(c)}=c u_j-\overline{c u_j}$ denotes nonlinear terms.
In the absence of rotation, shear, viscosity, or other linear effects,
the momentum equation takes the form
\EQ
{\partial u_j\over\partial t}=N_j^{(u)}.
\label{dujdt}
\EN
Assuming incompressibility and no mean flow, $\meanUU=0$, we have
\EQ
{\partial\over\partial t}\overline{cu_j}=
-\tilde\kappa_{ij}^{(c)}{\partial\meanC\over\partial x_j}+T_j^{(cu)},
\label{dcujdt}
\EN
where $\tilde\kappa_{ij}^{(c)}=\overline{u_iu_j}$
and $\TT^{(cu)}=-\overline{[\nab\NN^{(c)}]\uu}+\overline{c\NN^{(u)}}$
denotes a triple correlation term.

Clearly, in the statistically steady state the two terms on the right
hand side of \Eq{dcujdt} must balance to zero, suggesting that
$\TT^{(cu)}$ cannot be neglected, as is assumed in the commonly used
first order smoothing approximation, when it is applied to the case of
vanishing diffusivity ($\kappa_\theta=0$); see Ref.~\cite{%
2004PhFl...16.1020B
} for a more detailed discussion.
When $\kappa_\theta$ is large, the microscopic diffusion term involving
$\kappa_\theta\nabla^2\theta$ in \Eq{drhoTheta}
or $\kappa_C\nabla^2 C$ in \Eq{dCdt} needs to be restored.
Since it is applied to the small-scale field with typical wavenumber
$k_{\rm f}$, the inclusion of the $\kappa_\theta$ term corresponds
essentially to adding $-\kappa_\theta k_{\rm f}^2\,\overline{cu_j}$
on the right-hand side of \Eq{dcujdt}.
(This can be treated more accurately in Fourier space; see Ref.~\cite{%
2007MNRAS.376.1238S
} for a corresponding treatment in the magnetic case.)

The closure assumption used in the $\tau$ approximation consists of the
assumption that the triple correlations can be expressed in terms of
the quadratic correlation, i.e.\
\EQ
T_j^{(cu)}=-\overline{cu_j}/\tau\quad\mbox{(closure assumption)}.
\EN
Inserting this into \Eq{dcujdt} yields
\EQ
\left(1+\tau{\partial\over\partial t}\right)\overline{cu_j}=
-\kappa_{ij}^{(c)}{\partial\meanC\over\partial x_j},
\label{OnePlusddtc}
\EN
where $\kappa_{ij}^{(c)}=\tau\tilde\kappa_{ij}^{(c)}$
corresponds to the usual turbulent diffusivity.
This equation shows that, in the statistically steady state,
there is a flux of passive scalar concentration in the direction of
the negative gradient of $\meanC$.
Note that the effect described here works also when the turbulence is
isotropic, i.e.\ when $\overline{u_iu_j}=\onethird\delta_{ij}\overline{\uu^2}$.
In that case we have $\kappa_{ij}^{(c)}=\kappa_{\rm t}^{(c)}\delta_{ij}$,
where $\kappa_{\rm t}^{(c)}$ is the scalar turbulent diffusivity
of the mean passive scalar concentration.
By assuming $\tau=(\urms\kf)^{-1}$ we obtain
$\kappa_{\rm t}=\onethird\urms\kf^{-1}$.
On the other hand, if $\Pe\ll1$, $\kappa_{\rm t}^{(c)}$ is
small and increases linearly with $\Pe$ such that
$\kappa_{\rm t}=\onethird\Pe\,\urms\kf^{-1}$.

The effect discussed above is known as turbulent diffusion.
It is a very basic effect that characterizes an enhanced diffusion
experienced by the mean concentration.
It is present whenever the typical scale of the mean field is large
compared with the scale of the turbulence.
This is the requirement of scale separation that needs to be made
in order for a multiplicative relation in terms of the product of
$\kappa_{\rm t}$ and $\nab\meanC$ to be valid.
On the other hand, if the scale of the turbulence is comparable with the
system size, a local connection between flux and gradient becomes invalid,
and nonlocal expressions must be considered \cite{%
2008A&A...482..739B
}.

Let us now contrast the $\tau$ approximation
with the first order smoothing approximation,
where \Eq{dcdt} is still used, but the $N_j^{(c)}$ term is now neglected.
Again, assuming $\meanUU=0$ and integrating in time, we have
\EQ
c(\xx,t)=-\int_0^t
{\partial\meanC(\xx,t')\over\partial x_j}\,u_j(\xx,t')\,\dd t'.
\label{ct}
\EN
Thus,
\EQ
\overline{cu_i}=-\int_0^t\overline{u_i(t)u_j(t')}\,
{\partial\meanC(t')\over\partial x_j}\,\dd t',
\label{cut}
\EN
where we have dropped the common $\xx$ dependence of all variables
for clarity.
This expression would be identical to \Eq{OnePlusddtc} in the special case
where
\EQ
\overline{u_i(t)u_j(t')}=-\overline{u_i u_j}\exp[-(t-t')/\tau]
\quad\mbox{for $t>t'$}.
\EN
This type of agreement is restricted to the simplest case when there is
no contribution from the momentum equation.
Examples where such agreement is lost include cases with rotation or shear,
as well as analogous cases with magnetic field where there can be contributions
from the Lorentz force \cite{%
2005PhR...417....1B,
2009MNRAS.395.1599B
}.

The concept of turbulent diffusion carries over to vector fields such
as the velocity itself and the magnetic field.
In these cases one talks about turbulent viscosity, $\nu_{\rm t}$,
and turbulent magnetic diffusivity, $\eta_{\rm t}$.
The relevant correlations are then $\overline{u_iu_j}$
and $\overline{u_ib_j}$ that are being expressed in terms of negative
gradient terms, i.e.\
\EQ
\overline{u_iu_j}=-\nu_{\rm t}\left(
{\partial\meanU_i\over\partial x_j}+
{\partial\meanU_j\over\partial x_i}\right),
\label{turbvisc}
\EN
\EQ
\overline{u_ib_j}=-\eta_{\rm t}
{\partial\meanB_j\over\partial x_i}.
\label{turbeta}
\EN
This last formula is quite analogous to the passive scalar case discussed
in \Eq{OnePlusddtc},
\EQ
\overline{u_ic}=-\kappa_{\rm t}^{(c)}
{\partial\meanC\over\partial x_i},
\label{turbkappa}
\EN
where we have dropped the time derivative of $\overline{u_ic}$.
The term on the right-hand side of \Eq{turbvisc} is similar to the
expression for microscopic diffusion, see \Eq{StrainDef}.
The correlation that enters in the mean induction equation is
\EQ
\meanemf_i=(\overline{\uu\times\bb})_i
=\epsilon_{ijk}\overline{u_jb_k}
=-\eta_{\rm t}(\nab\times\meanBB)_i
=-\eta_{\rm t}\mu_0\meanJ_i,
\EN
which gives a contribution similar the microscopic diffusion term
in \Eq{Induction}.

\subsection{The $\alpha$ effect}
\label{Alpha}

Turbulence does not always act just diffusively.
There can be non-diffusive effects, especially if the turbulence lacks
local isotropy or at least parity invariance.
If the flow is statistically non-mirrorsymmetric
(for example helical) interesting effects can occur
in connection with the evolution of the mean magnetic field.
In particular, there are terms proportional to the mean magnetic
field itself, i.e.\
\EQ
\overline{\uu\times\bb}=\alpha\meanBB-\eta_{\rm t}\nab\times\meanBB.
\label{meanuxb}
\EN
This is the famous $\alpha$ effect \cite{%
1978mfge.book.....M,
1979cmft.book.....P,
1980mfmd.book.....K
} that is
often invoked in order to understand the generation of large-scale
magnetic fields in astrophysical bodies.
The possibility of magnetic field generation can be seen by inserting
\Eq{meanuxb} into the mean induction equation,
\EQ
{\partial\meanBB\over\partial t}=\nab\times\left(
\overline{\uu\times\bb}-\eta\nab\times\meanBB\right).
\label{dmeanB}
\EN
One can look for solutions proportional to $\exp(\ii\kk\cdot\xx+\lambda t)$
and finds that
\EQ
\lambda=\pm\alpha k-(\eta+\eta_{\rm t})k^2,
\EN
where $k=|\kk|$; see, e.g., Ref.~\cite{%
2005PhR...417....1B
} for details.
This shows that exponentially growing solutions exist on sufficiently
large scales, i.e.\ on sufficiently small wavenumbers, $k<\alpha/\eta_{\rm T}$.
Here we have introduced the
total magnetic diffusivity, $\eta_{\rm T}=\eta+\eta_{\rm t}$.

Although this topic has reached text book level already several decades ago
\cite{%
1978mfge.book.....M,
1979cmft.book.....P,
1980mfmd.book.....K
}, it continues to be a field of intense research -- especially with
regards to nonlinear feedback.
Basic aspects of the $\alpha$ effect can best be explained in the context of
isotropic turbulence.
In that case the following expression for $\alpha$ has been derived \cite{%
1976JFM....77..321P,
2002PhRvL..89z5007B,
2003GApFD..97..249R
}
\EQ
\alpha=-\onethird\tau\bra{\oo\cdot\uu}+\onethird\tau\bra{\jj\cdot\bb}/\rho,
\label{alphaM}
\EN
which shows that $\alpha$ is determined by the residual between kinetic
helicity of the small-scale velocity, $\bra{\oo\cdot\uu}$, and the normalized
small-scale current density, $\bra{\jj\cdot\bb}/\rho$.

The $\bra{\jj\cdot\bb}$ term contributes to the nonlinear saturation of
the dynamo.
This is because the $\alpha$ effect produces magnetic helicity both at
large and small scales such as to obey the magnetic helicity equation;
see \Sec{MagneticFields}.
While this can lead to a dramatic reduction of $\alpha$ in periodic or
closed domains \cite{%
2001ApJ...550..824B,
1996PhRvE..54.4532C
}, the quenching effect may be less extreme in the astrophysically relevant case
of open domains where magnetic helicity can be transported out of the
domain by magnetic helicity fluxes \cite{%
2000A&A...361L...5K,
2000MNRAS.318..724B
}.
The theory of these fluxes \cite{%
2006ApJ...648L..71S
} shows that there can be several contributions to the flux.
One such contribution is along the contours of constant shear \cite{%
2001ApJ...550..752V,
2004PhRvL..93t5001S
}, but recent work has cast some doubt on whether such shear-driven
magnetic helicity fluxes really exist \cite{%
2010arXiv1006.3549H
}.
Other contributions can come from advection \cite{%
2006A&A...448L..33S
} and diffusion \cite{%
2010AN....331..130M
}.
For further aspects regarding nonlinear dynamo theory we refer to a
review dedicated to recent developments; see Ref.~\cite{%
2005PhR...417....1B
}.

The presence of shear provides an additional induction effect that
usually contributes to the dynamo.
In order to estimate the relative importance of these effects, and
to estimate whether a large-scale dynamo is excited, one needs
to know the values of some relevant non-dimensional numbers that
characterize the magnitude of $\alpha$ effect and shear,
\EQ
C_\alpha=\alpha/\eta_{\rm T} k_{\rm m},\quad
C_\Omega=\Delta\Omega/\eta_{\rm T} k_{\rm m}^2,
\EN
where $k_{\rm m}$ is an estimate for the relevant wavenumber of
the dynamo that fits into the domain, and
$\Delta\Omega$ is the absolute differential rotation.
In the case of the Sun it is about 30\% of the average angular velocity.
Let us quantify the degree of helicity in the turbulence as
$\epsilon_{\rm f}=\bra{\oo\cdot\uu}/\kf\urms^2$, where $\kf=\orms/\urms$,
and assume $\eta_{\rm t}\gg\eta$, we find
\EQ
C_\alpha\approx\epsilon_{\rm f}\kf/k_{\rm m}.
\EN
Thus, the efficiency of the $\alpha$ effect depends on how helical
the turbulence is and on the amount of scale separation available.
A so-called $\alpha^2$ dynamo is possible when $C_\alpha$ exceeds a
critical value of the order of unity.

Often $C_\alpha$ is too small, and then the presence of shear can
help to produce large-scale dynamo action.
We assume that the shear is a significant fraction,
$q=\Delta\Omega/\Omega$, of the mean angular velocity $\Omega$
which, in turn, is often expressed as the Coriolis number,
$\mbox{Co}=2\Omega/\urms\kf$.
We may then estimate
\EQ
C_\Omega\approx\threehalf q\,\mbox{Co}\,(\kf/k_{\rm m})^2.
\EN
A large-scale dynamo of $\alpha\Omega$ type is excited when
the product $C_\alpha C_\Omega$
exceeds another critical value which is also of the order of unity.
For a homogeneous dynamo, the critical value of $C_\alpha C_\Omega$
for plane wave solutions is 2.

In conclusion, we see that the possibility of large-scale dynamo
action depends critically on the scale separation ratio, $\kf/k_{\rm m}$.
It is therefore important that the domain is big enough to contain
a significant number of turbulent eddies.
Simulations have now confirmed the possibility of large-scale
dynamo action in cases of forced turbulence, convective turbulence,
and for turbulence driven in turn by magnetic fields through the
magneto-rotational instability.

\subsection{Lambda effect}
\label{LambdaEffect}

An effect somewhat analogous to the $\alpha$ effect is the $\Lambda$ effect.
It parameterizes the dependence of the Reynolds stress on the mean angular
velocity \cite{%
1980GApFD..16..239R,
1989drsc.book.....R
} as
\EQ
\overline{u_iu_j}=\Lambda_{ijk}\meanO_k
+{\cal N}_{ijkl}{\partial\meanU_k\over\partial x_l},
\label{uiujLambda}
\EN
where $\meanO=\meanU_\phi/(r\sin\theta)$ is the {\it local} angular velocity
(not the $\Omega_0$ used earlier in connection with the transformation to a
rigidly rotating frame of reference).
The second term in \Eq{uiujLambda} is just the tensorial form of the
turbulent viscosity; see \Eq{turbvisc}.
The first one exists already in the presence of uniform rotation.
It is this term, balanced against the resulting turbulent viscosity term,
that drives and maintains non-uniformity in the mean angular velocity \cite{%
1995A&A...299..446K,
1993ApJ...407..367D,
1989ApJ...338..509D
}.
There are two important contributions to the $\Lambda$ effect,
a vertical and a horizontal one that quantify the
$r\phi$ and $\theta\phi$ components of the Reynolds stress, respectively.
In particular, we have, in spherical coordinates, $(r,\theta,\phi)$,
\EQ
\Lambda_{ijk}\Omega_k=\pmatrix{
0 & 0 & V\sin\theta\cr
0 & 0 & H\cos\theta\cr
V\sin\theta & H\cos\theta & 0}.
\EN
Here, $V$ and $H$ are functions of $r$ and $\theta$ that depend on the
anisotropy of the turbulence.
Using the first order smoothing approximation one finds \cite{%
1980GApFD..16..239R,
1989drsc.book.....R
}
\EQ
V=\tau\left(\overline{u_\phi^2}-\overline{u_r^2}\right),
\EN
\EQ
H=\tau\left(\overline{u_\phi^2}-\overline{u_\theta^2}\right).
\EN
For small turbulent Taylor numbers, $\Tat=\bra{2\Omega R^2/\nu_{\rm t}}^2$,
one finds for $H=0$ and $V\neq0$ that the $\Omega$ contours are purely
radial, while for $V=0$ and $H\neq0$ the $\Omega$ contours are purely
spoke-like.
For $V=\nu_{\rm t}(\sin^2\!\theta-1)$ and $H=\nu_{\rm t}\sin^2\!\theta$
one finds disk-shaped $\Omega$ contours.
For $V=\nu_{\rm t}({5\over4}\sin^2\!\theta-1)$ and
$H={5\over4}\nu_{\rm t}\sin^2\!\theta$
one finds approximately spoke-like contours.
However, those contours can change significantly with increasing values
of $\Tat$, which leads to the development of cylindrical $\Omega$ contours.
This is explained by the Taylor-Proudman theorem, as will be explained below.

The development of differential rotation is well established and is routinely
seen in direct simulations of convective turbulence in rotating shells
\cite{%
2000ApJ...532..593M,
1986ApJS...61..585G,
1994A&A...286..471R
}.
The existence of the $\Lambda$ effect has been verified in local
Cartesian simulations and the magnitude and spatial dependence has been
determined \cite{%
1993A&A...267..265P
}.
Solutions of the equations for $\meanUU$ have shown differential rotation
roughly similar to what is found for the Sun using helioseismology.
However, both DNS and solutions of the mean field equations show
a tendency toward $\Omega$ contours being constant along cylinders,
which is not the case in the Sun.
The cylindrical contours are the result of a feedback from the production
of meridional circulation modifying the angular velocity contours.
This leads to an approximate geostrophic balance, where
\EQ
\meanUU\cdot\nab\meanUU+{\overline{\rho}\,}^{-1}\nab\overline{p}=0.
\label{BaroMeanSphere}
\EN
In the barotropic case, when $\nab T$ and $\nab s$ are parallel to
each other, taking the curl of \Eq{BaroMeanSphere} yields
$\nab\times(\meanUU\cdot\nab\meanUU)=0$.
Assuming that the mean flow is purely toroidal, i.e.\
$\meanUU=(0,0,\meanO r\sin\theta)$, we have
\EQ
r\sin\theta{\partial\meanO^2\over\partial z}=0.
\label{OmegaTaylor}
\EN
So, if viscous and inertial terms are small, which is indeed the case
for rapid rotation, then $\partial\meanO^2/\partial z$ must be small,
so $\meanO$ would be constant along cylinders \cite{%
1995A&A...299..446K
}.
This is also what is seen in mean-field models with $\Lambda$ effect \cite{
1990SoPh..128..243B,
2005AN....326..379K
}.

\begin{figure}[t!]\begin{center}
\includegraphics[width=\textwidth]{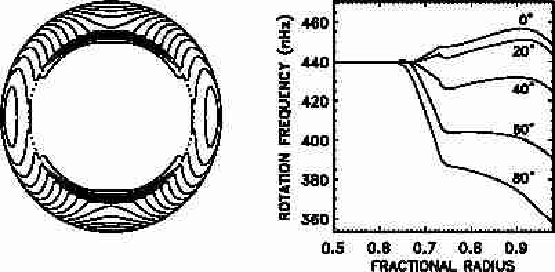}
\end{center}\caption[]{
Contours (left) and radial profiles (right) of
differential rotation in a model of the Sun.
Note the presence of radial deceleration near
the surface layers (fractional radius above 0.9).
Adapted from Ref.~\cite{%
2005AN....326..379K
}.
}\label{Kit+Rue05}\end{figure}

It is generally believed that the main reason for $\meanO$ not
having cylindrical contours in the Sun is connected with the
presence of the baroclinic term \cite{%
1992A&A...265..328B
1995A&A...299..446K
}.
The highest resolution simulations available today produce $\meanO$
contours that are still too close to being constant along cylinders \cite{%
2003ARA&A..41..599T,
2000ApJ...532..593M,
2004ApJ...614.1073B,
2002ApJ...570..865B,
2006ApJ...641..618M,
2007A&A...474..239J,
2007ApJ...669.1190B
}.
These simulations do not quite reach the solar surface,
so they cannot show the near-surface shear layer
where the rotation rate drops by more than $20\nHz$ over the last
$30\Mm$ just below the surface.
Nevertheless, these simulations reproduce some important features of
the Sun's differential rotation such as a more rapidly spinning equator.

Mean-field simulations using the $\Lambda$ effect
show surprisingly good agreement with the helioseismologically
inferred $\meanO$ pattern \cite{%
2005AN....326..379K,
1993A&A...279L...1K
}, and they are also beginning to reproduce
the near-surface shear layer; see \Fig{Kit+Rue05}.
In these models it is indeed the baroclinic term that is
responsible for causing the departure from cylindrical contours.
This, in turn, is caused by an anisotropy of the turbulent heat
conductivity which causes a slight enhancement in temperature and
specific entropy at the poles.
In the bulk of the convection zone the specific entropy is nearly constant
while the temperature varies significantly in the vertical direction.
It is therefore primarily the latitudinal specific entropy variation that
determines the baroclinic term.
This can be demonstrated by focusing on the contribution from the radial
temperature and the latitudinal specific entropy gradients; see
\Eqs{baroclinic}{curl_ugradu}, i.e.\
\EQ
r\sin\theta{\partial\meanO^2\over\partial z}\approx-\pp\cdot
\overline{\nab T\times\nab s}
\approx-{1\over r}
{\partial\overline{T}\over\partial r}
{\partial\overline{s}\over\partial\theta}<0.
\EN
The inequality shows that
negative values of $\partial\meanO^2/\partial z$
require that the pole is slightly warmer than the equator
($\partial\overline{s}/\partial\theta<0$).
However, this effect is so weak that it cannot at present be observed.
Allowing for these conditions in a simulation may require particular care
in the treatment of the outer boundary conditions,
or perhaps at the bottom of the convection zone in the tachocline.
Given that the turbulent convective flux is proportional
to $-\chi_{ij}\nabla_j\overline{s}$, a negative
$\partial\overline{s}/\partial\theta$ can be produced from a positive
enthalpy flux with a positive value of $\chi_{r\theta}$.
This is indeed compatible with theory that predicts a rotational influence
on the turbulence which makes $\chi_{ij}$ anisotropic \cite{%
2009MNRAS.395.1599B,
1994AN....315..157K
}.
One expects
\EQ
\chi_{ij}=\chi_{\rm t}\delta_{ij}
+\chi_{\Omega}\epsilon_{ijk}\Omega_k^{(0)}
+\chi_{\Omega\Omega}\Omega_i^{(0)}\Omega_j^{(0)},
\label{chiij}
\EN
where we have used superscripts (0) interchangeably with subscripts 0
to denote the rotation vector in a rotating frame of reference.
In spherical polar coordinates we have $\OO_0=(\cos\theta,-\sin\theta,0)$,
so $\chi_{r\theta}=-\chi_{\Omega\Omega}\sin\theta\cos\theta\,\Omega_0^2$.
Simulations confirm that $\chi_{\Omega\Omega}$ is negative, but only
when the scale of the mean field is comparable with that of the
fluctuating velocity field \cite{%
2009MNRAS.395.1599B
}, which is somewhat unexpected.
An alternative idea was advanced by Rempel \cite{%
2005ApJ...622.1320R
}, who was able to reproduce solar-like $\meanO$ contours by imposing a
suitable latitudinal $\overline{s}$ variation at the bottom of the
convection zone.

In the discussion above we ignored in the last step
the correlation between specific entropy and temperature fluctuations,
i.e.\ a contribution from the term $\overline{\nab T'\times\nab s'}$
where primes denote fluctuations.
Such correlations, if of suitable sign, might provide yet another
explanation for a non-zero value of $\partial\meanO^2/\partial z$.

It is in principle also possible that the differential rotation
could entirely be driven by the baroclinic term \cite{%
2009MNRAS.395.2056B,
2009MNRAS.400..176B
}.
However, quantitative calculations showed that this effect on its
own would be too small to \cite{%
1989drsc.book.....R,
1989A&A...217..217T
}.

\subsection{Turbulent transport coefficients from simulations}
\label{TurbulentTransportCoefficients}

In the past few years significant progress has been made in determining
tensor components such as $\kappa_{ij}$, $\alpha_{ij}$, and $\eta_{ijk}$
from local turbulence simulations.
The recommended approach is what is referred to as the test-field method \cite{%
2005AN....326..245S,
2007GApFD.101...81S
}.
This method is not to be confused with the test-field {\it model} that was
introduced by Kraichnan \cite{%
1972JFM....56..287K
} as a closure approach.

In the test-field method one solves numerically the evolution equation
\eq{dcdt} for the fluctuations of the passive scalar concentration $c$,
or a corresponding equation for fluctuations of the magnetic field $\bb$
to obtain the magnetic transport coefficients.
These equations are inhomogeneous in $c$ or $\bb$ and have terms of
the form $\nab\cdot(\uu\meanC)$ or $\nab\times(\uu\times\meanBB)$.
Here the mean fields $\meanC$ and $\meanBB$ are now replaced by test fields.
The best studied cases are for periodic boundary conditions and then
the test fields are taken to be $\meanC^{cx}=\cos kx$ or
$\meanC^{sx}=\sin kx$, and similarly for the $y$ and $z$ directions.
For each test field one evaluates the corresponding flux,
$\overline{u_jc^{pq}}$, and computes
\EQ
\kappa_{ij}=-\bra{\cos kx_j\meanFFF_i^{js}-\sin kx_j\meanFFF_i^{jc}}/k,
\label{Fxyz}
\EN
for $i,j=x,y,z$.
Here, angular brackets denote volume averages.
Using this method it has now been possible to compute the dependence
of the coefficients $\kappa_{\rm t}$, $\kappa_\Omega$, and
$\kappa_{\Omega\Omega}$ in an equation analogous to \Eq{chiij},
where $\chi$ has been replaced by $\kappa$.
A similar equation can also be written down for the case where
the anisotropy is caused by an applied magnetic field.

In the presence of a linear shear flow with $\meanU_{i,j}=\const$,
it has proved advantageous to express $\kappa_{ij}$ in terms of
the tensors $\meanSSS_{ij}=\half(\meanU_{i,j}+\meanU_{j,i})$
and $\meanAAA_{ij}=\half(\meanU_{i,j}-\meanU_{j,i})$.
The corresponding representation of $\kappa_{ij}$ has been found to
be of the form
\EQ
\kappa_{ij}
=\kappa_{\rm t}\delta_{ij}
+\kappa_{\sf S}\overline{\sf S}_{ij}
+\kappa_{\sf A}\overline{\sf A}_{ij}
+\kappa_{{\sf S}{\sf S}}(\meanSSSS\,\meanSSSS)_{ij}
+\kappa_{{\sf A}{\sf S}}(\meanAAAA\,\meanSSSS)_{ij}.
\label{kappaijAS}
\EN
There are indications that, in addition to $\kappa_{\rm t}$, only the
coefficients $\kappa_{\sf S}$ and $\kappa_{{\sf S}{\sf S}}$ are
important, while $\kappa_{\sf A}$ and $\kappa_{{\sf A}{\sf S}}$
are either small or become small at larger Peclet number \cite{%
2009arXiv0906.3314M
}.

The test-field method also allows one to compute turbulent transport
coefficients where the assumption of scale separation is not obeyed,
or where the mean quantities vary on time scales comparable to the
turnover time of the turbulence.
In those cases we have to replace the multiplications in
\Eqss{turbvisc}{turbkappa} by convolutions with integral kernels
of the corresponding transport coefficients, e.g.\
\EQ
\overline{u_ic}(\xx,t)=-\int\kappa_{\rm t}^{(c)}(\xx,\xx',t,t')
{\partial\meanC(\xx',t')\over\partial x_i'}\,\dd^3x'\,\dd t',
\EN
and likewise for the other equations.
If the system is homogeneous and statistically stationary, the
kernels depend only on the differences $\xx-\xx'$ and $t-t'$.
In such cases the convolution in real space becomes a multiplication
in Fourier space.
The test-field method yields directly the Fourier-transformed kernels
if the test fields consist of sine and cosine functions \cite{%
2008A&A...482..739B
}, and if they are made time-dependent \cite{%
2009ApJ...706..712H
}.
By changing the wavenumber and/or the frequency of the test fields
one can then obtain the full wavenumber and frequency dependence
of the Fourier-transformed kernel functions that enables one to
compute the kernels in real space via Fourier transformation.

It turns out that, for a range of quite different physical circumstances,
the $k$ dependence can well be fitted to the form of a Lorentzian
proportional to $[1+(ak/\kf)^2]^{-1}$, where $a$ is a fit parameter
of the order of unity.
The frequency dependence can be fitted to a function whose Fourier
transform corresponds to a multiplicative contribution to the kernel of
the form $e^{-t/\tau}\cos\omega_0 t$, where $\omega_0$ and $\tau^{-1}$
are fit parameters that are of the order of the inverse turnover
time, $\urms\kf$.

This type of analysis has been carried out both for turbulent
transport coefficients of both passive scalars and magnetic fields.
In the magnetic case there is, in addition to the turbulent magnetic
diffusivity tensor, also the $\alpha$ tensor \cite{%
2008ApJ...676..740B
}.
By using test fields proportional to sine and cosine functions,
one can compute $\alpha$ and $\eta_{\rm t}$ simultaneously.
It turns out that in the kinematic regime $\alpha$ and $\eta_{\rm t}$
reach asymptotic values for $\Rm>1$ where $\alpha\approx\alpha_0$
and $\eta_{\rm t}\approx\eta_{\rm t0}$ with \cite{%
2008MNRAS.385L..15S
}
\EQ
\alpha_0=\onethird\tau\bra{\oo\cdot\uu},\quad
\eta_{\rm t0}=\onethird\tau\bra{\uu^2},\quad
\mbox{where}\quad
\tau=(\urms\kf)^{-1}.
\EN
Both show similar behavior as far as their wavenumber and frequency
dependence is concerned \cite{%
2009A&A...495....1M
}.

\begin{figure}[t!]
\centering\includegraphics[width=.8\columnwidth]{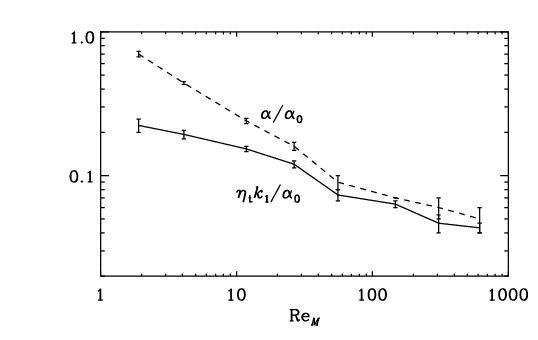}\caption{
$\Rm$-dependence of $\alpha$ and $\tilde{\eta}_{\rm t}$.
Both curves are normalized by $\alpha_0$.
Adapted from Ref.~\cite{%
2008ApJ...687L..49B
}
}\label{palpeta}\end{figure}

Anisotropies of the flow yield anisotropies in the $\alpha$ and $\eta$
tensors.
In addition, since the magnetic field is an active vector, it backreacts
on the flow and makes it anisotropic even if it was otherwise isotropic.
It turns out that the $\alpha$ and $\eta$ tensors are of the form
\EQ
\alpha_{ij}=\alpha_1\delta_{ij}+\alpha_2\hatB_i\hatB_j,\quad
\eta_{ij}=\eta_1\delta_{ij}+\eta_2\hatB_i\hatB_j,
\EN
where we have assumed that only the unit vector of the mean field,
$\hatBB=\meanBB/|\meanBB|$ is crucial.
Simulations with the test-field method have revealed that in the
quenched state $\alpha_1$ and $\alpha_2$ have opposite sign \cite{%
2008ApJ...687L..49B
}.
The test-field method has been used to compute the dependence of
$\etat$ and $\etat$ on $\Rm$ in the nonlinearly saturated state
for $\meanBB\sim B_{\rm eq}$, where $B_{\rm eq}$ is the field
strength where magnetic and kinetic energies are in equilibrium.
It turns out that at large $\Rm$ the effects of $\alpha$ and $\etat$
tend to balance each other.
Furthermore, $\etat$ does not show a sharp decline like $\Rm^{-1}$,
as would be the case in two dimensions, but, even though $\Rm$ is already
around 600, there remains a weak decrease of $\etat$, without any
obvious indications that this trend might level off (\Fig{palpeta}).

When the $\alpha$ and $\eta$ tensors are multiplied by $\meanBB$,
the result is \cite{%
2008ApJ...687L..49B
}
\EQ
\alpha_{ij}\meanB_j-\mu_0\eta_{ij}\meanJ_j
=(\alpha_1+\alpha_2-\eta_2 k_{\rm m})\meanB_i-\eta_1\mu_0\meanJ_i,
\EN
where $k_{\rm m}=\mu_0\meanJJ\cdot\meanBB/\meanBB^2$ is an effective
wavenumber of the mean field.
This shows that the tensorial nature of $\alpha$ is unimportant in this context.
However, this changes when considering passive vector equations that
are equivalent to the induction equation, with a passive vector field
$\tilde{\BB}$ that is similar to the actual magnetic field, but it has
no effect on the motions.
Such a passive vector field can display dynamo action and can continue
to grow even when the underlying velocity field corresponds to that
of a nonlinearly saturated dynamo.
This phenomenon was first observed for turbulent convective dynamos \cite{%
2009JFM...621..205C
} and then confirmed for laminar dynamos generating a mean field that is
an eigenvector of the matrix $\hatB_i\hatB_j$ with vanishing eigenvalue.
Thus, given that $\alpha_2/\alpha_1$ is negative, such fields remain
unquenched for a velocity field or an $\alpha$ tensor that correspond
to a saturated dynamo \cite{%
2008MNRAS.391.1477T
}.

\section{Concluding remarks}

Over the past few decades hydrodynamic and magnetohydrodynamic
simulations have become a frequently used tool in astrophysics research.
This trend is surely going to continue.
As an example of the importance of turbulence considerations
we mention here the well-established field of stellar structure,
which has recently been the subject of intense debate, because fits to
three-dimensional time-dependent turbulent model atmospheres have
led (mostly due to non-3D effects!) to a significantly
lower estimate of the solar abundance of heavier elements.
Although this issue is not yet settled, it is clear that the results from
three-dimensional turbulence simulations will continue to provide valuable
input to the debate.

Even the radially symmetric
(one-dimensional) models of stellar interiors are bound to be soon
superseded or amended by higher-dimensional models.
Clearly, the vast range of time and length scales between those of
turbulent convection of stars and those of stellar structure and evolution
necessitate a proper understanding of the collective or mean-field effects that
are controlled by various correlators discussed in \Sec{CollectiveEffects}.
Obviously, we were only able to expose a small part of the many recent
developments in this field.
Quite frequently astrophysical turbulence involves magnetic fields, and
often many more ingredients such as dust, chemicals, cosmic rays,
and coupling to radiation.
Instead of simply neglecting such additional features, one may attempt
to incorporate them into stellar evolution models using a mean-field approach.
The transport properties depend on rotation, shear, and magnetic field in ways
that are reasonably well understood now.
This is important for example in understanding the dependence of the Lithium abundance
of young stars on their rotation rate \cite{%
2001A&A...377...84T,
2001A&A...375..149R
}.

Astrophysical turbulence concerns usually extreme parameter regimes:
large Reynolds and/or Mach numbers, very large or very small Prandtl
numbers, as well as extreme density and temperature contrasts.
This motivates thorough studies of turbulence in regimes
that are not otherwise addressed.
This can either provide broader support for certain turbulence
theories, or it can more clearly highlight problems that would be otherwise
overlooked.
In this sense astrophysical turbulence research is not just the
application of regular turbulence theory, but it can also provide
complementary insights of broader relevance also for other research fields.

One of the aspects where astrophysical turbulence encounters an as yet
unsettled issue is the question how compressibility really enters the theory.
We have seen some ambiguity in the proper definition of the kinetic
energy where, empirically, the spectrum of $\rho^{1/3}\uu$ appears to
be closest to the case of incompressible turbulence.
There are several related issues in the context of mean-field theory.
For example, the equation for the magnetic $\alpha$ effect in \Eq{alphaM}
contains a $\rho$ factor, but since this equation was derived for the
compressible case, it is not clear whether $\rho$ should enter inside or
outside the average of $\jj\cdot\bb$ when $\rho$ is non-uniform
or strongly fluctuating.
Another occurrence of compressibility effects could be in the relation
between the enthalpy flux and the specific entropy gradient.
Finally, let us also mention here the issue of the baroclinic term,
which can be important in the production of vorticity and
shaping the form of the differential rotation contours in the Sun.
There could potentially be systematic corrections resulting from
the fluctuations of specific entropy and temperature.
This and other effects might be responsible for causing a departure from
cylindrical Taylor-Proudman contours of $\Omega(r,\theta)$ in the Sun.

There are several other quadratic correlation functions that need to be
modelled more accurately.
One is the current helicity flux involving terms of the form $\EE\times\JJ$,
for example.
Other examples include Reynolds and Maxwell stresses and their dependence
not only on the mean velocity, as discussed above, but also on the
magnetic field.
This quadratic nonlinearity means that the standard test-field method
cannot be used, but possibly some kind of modification of it might work.

\section*{Acknowledgements}

We thank Alexei Kritsuk for making useful suggestions to the manuscript.
We acknowledge the allocation of computing resources provided by the
Swedish National Allocations Committee at the Center for
Parallel Computers at the Royal Institute of Technology in
Stockholm and the National Supercomputer Centers in Link\"oping.
This work was supported in part by
the European Research Council under the AstroDyn Research Project 227952
and the Swedish Research Council grant 621-2007-4064.

\newcommand\newblock[1]{}
\def\apss   {Astrophys. Space Sci.}
\def\araa   {Ann.\ Rev.\ Astron.\ Astrophys.}
\def\pre    {Phys.\ Rev.\ E}
\def\apj    {Astrophys.\ J.}
\def\apjl   {Astrophys.\ J.\ Lett.}
\def\apjs   {Astrophys.\ J.\ Supp.}
\def\mnras  {Month.\ Not.\ Roy.\ Astron.\ Soc.}
\def\physrep{Phys.\ Rep.}
\def\aap    {Astron.\ Astrophys.}
\def\jgr    {J.\ Geophys.\ Res.}
\def\grl    {Geophys.\ Res.\ Lett.}
\bibliography{ref}

\begin{thebibliography}{100}

\bibitem{2003matu.book.....B}
D.~{Biskamp}.
\newblock {\em {Magnetohydrodynamic turbulence}}.
\newblock Cambridge University Press, Cambridge (UK). 2003.

\bibitem{1971ARA&A...9..323S}
E.~A. {Spiegel}.
\newblock {Convection in stars: I. Basic boussinesq convection}.
\newblock {\em \araa}, 9:323--352, 1971.

\bibitem{1972ARA&A..10..261S}
E.~A. {Spiegel}.
\newblock {Convection in stars. II. Special effects}.
\newblock {\em \araa}, 10:261--304, 1972.

\bibitem{1978RPPh...41.1929B}
F.~H. {Busse}.
\newblock {Non-linear properties of thermal convection}.
\newblock {\em Rep. Prog. Phys.}, 41:1929--1967, 1978.

\bibitem{1990ARA&A..28..263S}
H.~C. {Spruit}, A.~{Nordlund}, and A.~M. {Title}.
\newblock {Solar convection}.
\newblock {\em \araa}, 28:263--301, 1990.

\bibitem{1995tlan.book.....F}
U.~{Frisch}.
\newblock {\em {Turbulence. The legacy of A. N. Kolmogorov.}}
\newblock Cambridge University Press, Cambridge (UK). 1995.

\bibitem{1995ARA&A..33..283G}
M.~L. {Goldstein}, D.~A. {Roberts}, and W.~H. {Matthaeus}.
\newblock {Magnetohydrodynamic turbulence in the solar wind}.
\newblock {\em \araa}, 33:283--326, 1995.

\bibitem{2004tise.book.....D}
P.~A. {Davidson}.
\newblock {\em {Turbulence: an introduction for scientists and engineers}}.
\newblock Oxford University Press, Oxford (UK). 2004.

\bibitem{2004ARA&A..42..211E}
B.~G. {Elmegreen} and J.~{Scalo}.
\newblock {Interstellar turbulence I: observations and processes}.
\newblock {\em \araa}, 42:211--273, 2004.

\bibitem{2004ARA&A..42..275S}
J.~{Scalo} and B.~G. {Elmegreen}.
\newblock {Interstellar turbulence II: implications and effects}.
\newblock {\em \araa}, 42:275--316, 2004.

\bibitem{2005PhR...417....1B}
A.~{Brandenburg} and K.~{Subramanian}.
\newblock {Astrophysical magnetic fields and nonlinear dynamo theory}.
\newblock {\em \physrep}, 417:1--209, 2005.

\bibitem{1955IAUS....2..121L}
M.~J. {Lighthill}.
\newblock {The effect of compressibility on turbulence}.
\newblock volume~2 of {\em IAU Symp.}, pages 121--180, 1955.

\bibitem{2007ApJ...665..416K}
A.~G. {Kritsuk}, M.~L. {Norman}, P.~{Padoan}, and R.~{Wagner}.
\newblock {The statistics of supersonic isothermal turbulence}.
\newblock {\em \apj}, 665:416--431, 2007.

\bibitem{2002AN....323...99B}
A.~{Brandenburg}, W.~{Dobler}, and K.~{Subramanian}.
\newblock {Magnetic helicity in stellar dynamos: new numerical experiments}.
\newblock {\em Astron. Nachr.}, 323:99--122, 2002.

\bibitem{1978mfge.book.....M}
H.~K. {Moffatt}.
\newblock {\em {Magnetic field generation in electrically conducting fluids}}.
\newblock 1978.

\bibitem{1951JAP....22..469C}
S.~{Corrsin}.
\newblock {On the spectrum of isotropic temperature fluctuations in an
  isotropic turbulence}.
\newblock {\em J. Appl. Phys.}, 22:469--473, 1951.

\bibitem{2003PhRvE..68b6304D}
W.~{Dobler}, N.~E. {Haugen}, T.~A. {Yousef}, and A.~{Brandenburg}.
\newblock {Bottleneck effect in three-dimensional turbulence simulations}.
\newblock {\em \pre}, 68:026304, 2003.

\bibitem{2003PhFl...15L..21K}
Y.~{Kaneda}, T.~{Ishihara}, M.~{Yokokawa}, K.~{Itakura}, and A.~{Uno}.
\newblock {Energy dissipation rate and energy spectrum in high resolution
  direct numerical simulations of turbulence in a periodic box}.
\newblock {\em Phys. Fluids}, 15:L21--L24, 2003.

\bibitem{2006PhFl...18g5106H}
N.~E.~L. {Haugen} and A.~{Brandenburg}.
\newblock {Hydrodynamic and hydromagnetic energy spectra from large eddy
  simulations}.
\newblock {\em Phys. Fluids}, 18:075106, 2006.

\bibitem{1994PhFl....6.1411F}
G.~{Falkovich}.
\newblock {Bottleneck phenomenon in developed turbulence}.
\newblock {\em Phys. Fluids}, 6:1411--1414, 1994.

\bibitem{1968JPhE....1.1105W}
J.~C. {Wyngaard}.
\newblock {Measurement of small-scale turbulence structure with hot wires}.
\newblock {\em J. Phys. E Scientific Instruments}, 1:1105--1108, 1968.

\bibitem{1993PhRvE..48...29B}
R.~{Benzi}, S.~{Ciliberto}, R.~{Tripiccione}, C.~{Baudet}, F.~{Massaioli}, and
  S.~{Succi}.
\newblock {Extended self-similarity in turbulent flows}.
\newblock {\em \pre}, 48:R29--R32, 1993.

\bibitem{1994PhRvL..72..336S}
Z.-S. {She} and E.~{Leveque}.
\newblock {Universal scaling laws in fully developed turbulence}.
\newblock {\em Phys. Rev. Lett.}, 72:336--339, 1994.

\bibitem{2002ApJ...569..841B}
S.~{Boldyrev}.
\newblock {Kolmogorov-Burgers model for star-forming turbulence}.
\newblock {\em \apj}, 569:841--845, 2002.

\bibitem{2002PhRvL..89c1102B}
S.~{Boldyrev}, {\AA}.~{Nordlund}, and P.~{Padoan}.
\newblock {Supersonic turbulence and structure of interstellar molecular
  clouds}.
\newblock {\em Phys. Rev. Lett.}, 89:031102, 2002.

\bibitem{2002ApJ...573..678B}
S.~{Boldyrev}, {\AA}.~{Nordlund}, and P.~{Padoan}.
\newblock {Scaling relations of supersonic turbulence in star-forming molecular
  clouds}.
\newblock {\em \apj}, 573:678--684, 2002.

\bibitem{1995PhRvE..52..636P}
H.~{Politano} and A.~{Pouquet}.
\newblock {Model of intermittency in magnetohydrodynamic turbulence}.
\newblock {\em \pre}, 52:636--641, 1995.

\bibitem{1998PhRvE..57...21P}
H.~{Politano} and A.~{Pouquet}.
\newblock {von K{\'a}rm{\'a}n-Howarth equation for magnetohydrodynamics and its
  consequences on third-order longitudinal structure and correlation
  functions}.
\newblock {\em \pre}, 57:R21--R24, 1998.

\bibitem{2004PhRvE..70a5302S}
M.~{Siefert} and J.~{Peinke}.
\newblock {Different cascade speeds for longitudinal and transverse velocity
  increments of small-scale turbulence}.
\newblock {\em \pre}, 70:015302, 2004.

\bibitem{2004PhRvE..70e6301P}
B.~R. {Pearson}, T.~A. {Yousef}, N.~E.~L. {Haugen}, A.~{Brandenburg}, and
  P.-{\AA}. {Krogstad}.
\newblock {Delayed correlation between turbulent energy injection and
  dissipation}.
\newblock {\em \pre}, 70:056301, 2004.

\bibitem{2006Natur.441..953M}
J.~M. {Miller}, J.~{Raymond}, A.~{Fabian}, D.~{Steeghs}, J.~{Homan},
  C.~{Reynolds}, M.~{van der Klis}, and R.~{Wijnands}.
\newblock {The magnetic nature of disk accretion onto black holes}.
\newblock {\em \nat}, 441:953--955, 2006.

\bibitem{1958ApJ...128..664P}
E.~N. {Parker}.
\newblock {Dynamics of the interplanetary gas and magnetic fields.}
\newblock {\em \apj}, 128:664--676, 1958.

\bibitem{1992aita.book.....S}
S.~N. {Shore}.
\newblock {\em {An introduction to astrophysical hydrodynamics}}.
\newblock 1992.

\bibitem{1998pfp..book.....C}
A.~R. {Choudhuri}.
\newblock {\em {The physics of fluids and plasmas: an introduction for
  astrophysicists}}.
\newblock 1998.

\bibitem{1995AdSpR..16...85P}
J.~L. {Phillips}, S.~J. {Bame}, W.~C. {Feldman}, J.~T. {Gosling}, C.~M.
  {Hammond}, D.~J. {McComas}, B.~E. {Goldstein}, and M.~{Neugebauer}.
\newblock {ULYSSES solar wind plasma observations during the declining phase of
  solar cycle 22}.
\newblock {\em Adv. Space Res.}, 16:85--94, 1995.

\bibitem{1995SSRv...73....1T}
C.-Y. {Tu} and E.~{Marsch}.
\newblock {MHD structures, waves and turbulence in the solar wind: Observations
  and theories}.
\newblock {\em Space Science Reviews}, 73:1--210, 1995.

\bibitem{2006ApJ...645L..85S}
C.~W. {Smith}, K.~{Hamilton}, B.~J. {Vasquez}, and R.~J. {Leamon}.
\newblock {Dependence of the dissipation range spectrum of interplanetary
  magnetic fluctuations on the rate of energy cascade}.
\newblock {\em \apjl}, 645:L85--L88, 2006.

\bibitem{2008PhRvL.100f5004H}
G.~G. {Howes}, W.~{Dorland}, S.~C. {Cowley}, G.~W. {Hammett}, E.~{Quataert},
  A.~A. {Schekochihin}, and T.~{Tatsuno}.
\newblock {Kinetic simulations of magnetized turbulence in astrophysical
  plasmas}.
\newblock {\em Phys. Rev. Lett.}, 100:065004, 2008.

\bibitem{2009PhRvL.102w1102S}
F.~{Sahraoui}, M.~L. {Goldstein}, P.~{Robert}, and Y.~V. {Khotyaintsev}.
\newblock {Evidence of a cascade and dissipation of solar-wind turbulence at
  the electron gyroscale}.
\newblock {\em Phys. Rev. Lett.}, 102:231102, 2009.

\bibitem{1982PhRvL..48.1256M}
W.~H. {Matthaeus}, M.~L. {Goldstein}, and C.~{Smith}.
\newblock {Evaluation of magnetic helicity in homogeneous turbulence}.
\newblock {\em Phys. Rev. Lett.}, 48:1256--1259, 1982.

\bibitem{1982JGR....87.6011M}
W.~H. {Matthaeus} and M.~L. {Goldstein}.
\newblock {Measurement of the rugged invariants of magnetohydrodynamic
  turbulence in the solar wind}.
\newblock {\em \jgr}, 87:6011--6028, 1982.

\bibitem{1988GeoRL..15...88F}
J.~W. {Freeman}.
\newblock {Estimates of solar wind heating inside 0.3 AU}.
\newblock {\em \grl}, 15:88--91, 1988.

\bibitem{2001JGR...106.8253S}
C.~W. {Smith}, W.~H. {Matthaeus}, G.~P. {Zank}, N.~F. {Ness}, S.~{Oughton}, and
  J.~D. {Richardson}.
\newblock {Heating of the low-latitude solar wind by dissipation of turbulent
  magnetic fluctuations}.
\newblock {\em \jgr}, 106:8253--8272, 2001.

\bibitem{1996PhRvL..76.1264B}
D.~{Biskamp}, E.~{Schwarz}, and J.~F. {Drake}.
\newblock {Two-dimensional electron magnetohydrodynamic turbulence}.
\newblock {\em Phys. Rev. Lett.}, 76:1264--1267, 1996.

\bibitem{2009ApJS..182..310S}
A.~A. {Schekochihin}, S.~C. {Cowley}, W.~{Dorland}, G.~W. {Hammett}, G.~G.
  {Howes}, E.~{Quataert}, and T.~{Tatsuno}.
\newblock {Astrophysical gyrokinetics: kinetic and fluid turbulent cascades in
  magnetized weakly collisional plasmas}.
\newblock {\em \apjs}, 182:310--377, 2009.

\bibitem{2007ApJ...656..560G}
S.~{Galtier} and E.~{Buchlin}.
\newblock {Multiscale Hall-magnetohydrodynamic turbulence in the solar wind}.
\newblock {\em \apj}, 656:560--566, 2007.

\bibitem{2008ApJ...674.1153A}
O.~{Alexandrova}, V.~{Carbone}, P.~{Veltri}, and L.~{Sorriso-Valvo}.
\newblock {Small-scale energy cascade of the solar wind turbulence}.
\newblock {\em \apj}, 674:1153--1157, 2008.

\bibitem{2005SoPh..228..191V}
M.~{van Noort}, L.~{Rouppe van der Voort}, and M.~G. {L{\"o}fdahl}.
\newblock {Solar image restoration by use of multi-frame blind de-convolution
  with multiple objects and phase diversity}.
\newblock {\em \solphys}, 228:191--215, 2005.

\bibitem{2005ARA&A..43..481A}
M.~{Asplund}.
\newblock {New light on stellar abundance analyses: departures from LTE and
  Homogeneity}.
\newblock {\em \araa}, 43:481--530, 2005.

\bibitem{2008A&A...488.1031C}
E.~{Caffau}, {H.-G.} {Ludwig}, M.~{Steffen}, T.~R. {Ayres}, P.~{Bonifacio},
  R.~{Cayrel}, B.~{Freytag}, and B.~{Plez}.
\newblock {The photospheric solar oxygen project. I. Abundance analysis of
  atomic lines and influence of atmospheric models}.
\newblock {\em \aap}, 488:1031--1046, 2008.

\bibitem{2009ARA&A..47..481A}
M.~{Asplund}, N.~{Grevesse}, A.~J. {Sauval}, and P.~{Scott}.
\newblock {The chemical composition of the Sun}.
\newblock {\em \araa}, 47:481--522, 2009.

\bibitem{2005AN....326..681B}
A.~{Brandenburg}, K.~L. {Chan}, {\AA}.~{Nordlund}, and R.~F. {Stein}.
\newblock {Effect of the radiative background flux in convection}.
\newblock {\em Astron. Nachr.}, 326:681--692, 2005.

\bibitem{2009ASPC..416..421S}
R.~F. {Stein}, {\AA}.~{Nordlund}, D.~{Georgoviani}, D.~{Benson}, and
  W.~{Schaffenberger}.
\newblock {Supergranulation-scale convection simulations}.
\newblock volume 416 of {\em Astron. Soc. Pac. Conf. Ser.}, pages 421--426,
  2009.

\bibitem{1970JFM....44..441B}
F.~H. {Busse}.
\newblock {Thermal instabilities in rapidly rotating systems.}
\newblock {\em J. Fluid Mech.}, 44:441--460, 1970.

\bibitem{2000ApJ...532..593M}
M.~S. {Miesch}, J.~R. {Elliott}, J.~{Toomre}, T.~L. {Clune}, G.~A.
  {Glatzmaier}, and P.~A. {Gilman}.
\newblock {Three-dimensional spherical simulations of solar convection. I.
  Differential rotation and pattern evolution achieved with laminar and
  turbulent states}.
\newblock {\em \apj}, 532:593--615, 2000.

\bibitem{2010AN....331...73K}
P.~J. {K{\"a}pyl{\"a}}, M.~J. {Korpi}, A.~{Brandenburg}, D.~{Mitra}, and
  R.~{Tavakol}.
\newblock {Convective dynamos in spherical wedge geometry}.
\newblock {\em Astron. Nachr.}, 331:73--81, 2010.

\bibitem{2003ARA&A..41..599T}
M.~J. {Thompson}, J.~{Christensen-Dalsgaard}, M.~S. {Miesch}, and J.~{Toomre}.
\newblock {The Internal Rotation of the Sun}.
\newblock {\em \araa}, 41:599--643, 2003.

\bibitem{1995ApJ...440L.109P}
A.~A. {Pevtsov}, R.~C. {Canfield}, and T.~R. {Metcalf}.
\newblock {Latitudinal variation of helicity of photospheric magnetic fields}.
\newblock {\em \apjl}, 440:L109--L112, 1995.

\bibitem{1984ARA&A..22..593D}
F.-L. {Deubner} and D.~{Gough}.
\newblock {Helioseismology: oscillations as a daignostic of the solar
  interior}.
\newblock {\em \araa}, 22:593--619, 1984.

\bibitem{1985SoPh..100...65G}
D.~{Gough}.
\newblock {Inverting helioseismic data}.
\newblock {\em \solphys}, 100:65--99, 1985.

\bibitem{2002RvMP...74.1073C}
J.~{Christensen-Dalsgaard}.
\newblock {Helioseismology}.
\newblock {\em Rev. Mod. Phys.}, 74:1073--1129, 2002.

\bibitem{2008PhR...457..217B}
S.~{Basu} and H.~M. {Antia}.
\newblock {Helioseismology and solar abundances}.
\newblock {\em \physrep}, 457:217--283, 2008.

\bibitem{1988RvMP...60..297B}
J.~N. {Bahcall} and R.~K. {Ulrich}.
\newblock {Solar models, neutrino experiments, and helioseismology}.
\newblock {\em Rev. Mod. Phys.}, 60:297--372, 1988.

\bibitem{2001PhRvD..64a3009S}
H.~{Schlattl}.
\newblock {Three-flavor oscillation solutions for the solar neutrino problem}.
\newblock {\em \prd}, 64:013009, 2001.

\bibitem{2005MNRAS.356..587C}
J.~{Christensen-Dalsgaard}, M.~P. {Di Mauro}, H.~{Schlattl}, and A.~{Weiss}.
\newblock {On helioseismic tests of basic physics}.
\newblock {\em \mnras}, 356:587--595, 2005.

\bibitem{1986NCimC...9...17M}
S.~P. {Mikheev} and A.~I. {Smirnov}.
\newblock {Resonant amplification of neutrino oscillations in matter and
  solar-neutrino spectroscopy}.
\newblock {\em Nuovo Cimento C Geophys. Space Phys. C}, 9:17--26, 1986.

\bibitem{1978PhRvD..17.2369W}
L.~{Wolfenstein}.
\newblock {Neutrino oscillations in matter}.
\newblock {\em \prd}, 17:2369--2374, 1978.

\bibitem{2003Natur.421...43G}
L.~{Gizon}, T.~L. {Duvall}, and J.~{Schou}.
\newblock {Wave-like properties of solar supergranulation}.
\newblock {\em \nat}, 421:43--44, 2003.

\bibitem{1992A&A...265..106S}
E.~A. {Spiegel} and J.-P. {Zahn}.
\newblock {The solar tachocline}.
\newblock {\em \aap}, 265:106--114, 1992.

\bibitem{2007sota.conf.....H}
D.~W. {Hughes}, R.~{Rosner}, and N.~O. {Weiss}, editors.
\newblock {\em {The solar tachocline}}, 2007.

\bibitem{2002A&A...390..307O}
V.~{Ossenkopf} and {M.-M.} {Mac Low}.
\newblock {Turbulent velocity structure in molecular clouds}.
\newblock {\em \aap}, 390:307--326, 2002.

\bibitem{1979MNRAS.186..479L}
R.~B. {Larson}.
\newblock {Stellar kinematics and interstellar turbulence}.
\newblock {\em \mnras}, 186:479--490, 1979.

\bibitem{1981MNRAS.194..809L}
R.~B. {Larson}.
\newblock {Turbulence and star formation in molecular clouds}.
\newblock {\em \mnras}, 194:809--826, 1981.

\bibitem{2002ApJ...566..289B}
C.~M. {Brunt} and M.~H. {Heyer}.
\newblock {Interstellar turbulence. II. Energy spectra of molecular regions in
  the outer Galaxy}.
\newblock {\em \apj}, 566:289--301, 2002.

\bibitem{2002ApJ...566..276B}
C.~M. {Brunt} and M.~H. {Heyer}.
\newblock {Interstellar turbulence. I. Retrieval of velocity field statistics}.
\newblock {\em \apj}, 566:276--288, 2002.

\bibitem{2009A&A...504..883B}
C.~M. {Brunt}, M.~H. {Heyer}, and {M.-M.} {Mac Low}.
\newblock {Turbulent driving scales in molecular clouds}.
\newblock {\em \aap}, 504:883--890, 2009.

\bibitem{1990ARA&A..28..561R}
B.~J. {Rickett}.
\newblock {Radio propagation through the turbulent interstellar plasma}.
\newblock {\em \araa}, 28:561--605, 1990.

\bibitem{2003ApJ...584..791B}
S.~{Boldyrev} and C.~{Gwinn}.
\newblock {Scintillations and L{\'e}vy flights through the interstellar
  medium}.
\newblock {\em \apj}, 584:791--796, 2003.

\bibitem{2005A&A...436..585D}
M.~A. {de Avillez} and D.~{Breitschwerdt}.
\newblock {Global dynamical evolution of the ISM in star forming galaxies. I.
  High resolution 3D simulations: Effect of the magnetic field}.
\newblock {\em \aap}, 436:585--600, 2005.

\bibitem{2008ApJ...680..457T}
T.~H. {Troland} and R.~M. {Crutcher}.
\newblock {Magnetic fields in dark cloud cores: Arecibo OH Zeeman
  observations}.
\newblock {\em \apj}, 680:457--465, 2008.

\bibitem{2009ApJ...692..844C}
R.~M. {Crutcher}, N.~{Hakobian}, and T.~H. {Troland}.
\newblock {Testing magnetic star formation theory}.
\newblock {\em \apj}, 692:844--855, 2009.

\bibitem{2009RMxAC..36..107C}
R.~M. {Crutcher}.
\newblock {OH and CN Zeeman observations of magnetic fields in molecular
  clouds}.
\newblock volume~36 of {\em Revista Mexicana Astron. Astrofisica Conf. Ser.},
  pages 107--112, 2009.

\bibitem{1984Natur.310..557Y}
F.~{Yusef-Zadeh}, M.~{Morris}, and D.~{Chance}.
\newblock {Large, highly organized radio structures near the galactic centre}.
\newblock {\em \nat}, 310:557--561, 1984.

\bibitem{2004ApJS..155..421Y}
F.~{Yusef-Zadeh}, J.~W. {Hewitt}, and W.~{Cotton}.
\newblock {A 20 centimeter survey of the Galactic center region. I. Detection
  of numerous linear filaments}.
\newblock {\em \apjs}, 155:421--550, 2004.

\bibitem{1996ARA&A..34..155B}
R.~{Beck}, A.~{Brandenburg}, D.~{Moss}, A.~{Shukurov}, and D.~{Sokoloff}.
\newblock {Galactic magnetism: recent developments and perspectives}.
\newblock {\em \araa}, 34:155--206, 1996.

\bibitem{1998RvMP...70....1B}
S.~A. {Balbus} and J.~F. {Hawley}.
\newblock {Instability, turbulence, and enhanced transport in accretion disks}.
\newblock {\em Rev. Mod. Phys.}, 70:1--53, 1998.

\bibitem{2008ApJ...673.1138C}
K.~{Cai}, R.~H. {Durisen}, A.~C. {Boley}, M.~K. {Pickett}, and A.~C.
  {Mej{\'{\i}}a}.
\newblock {The thermal regulation of gravitational instabilities in
  protoplanetary disks. IV. Simulations with envelope irradiation}.
\newblock {\em \apj}, 673:1138--1153, 2008.

\bibitem{1992apa..book.....F}
J.~{Frank}, A.~{King}, and D.~{Raine}.
\newblock {\em {Accretion power in astrophysics.}}
\newblock 1992.

\bibitem{1995ApJ...445...80L}
C.~{Loken}, K.~{Roettiger}, J.~O. {Burns}, and M.~{Norman}.
\newblock {Radio jet propagation and wide-angle tailed radio sources in merging
  galaxy cluster environments}.
\newblock {\em \apj}, 445:80--97, 1995.

\bibitem{2008ApJ...680...17W}
D.~R. {Wik}, C.~L. {Sarazin}, P.~M. {Ricker}, and S.~W. {Randall}.
\newblock {The impact of galaxy cluster mergers on cosmological parameter
  estimation from surveys of the Sunyaev-Zel'dovich effect}.
\newblock {\em \apj}, 680:17--31, 2008.

\bibitem{1999PhR...310...97B}
M.~{Birkinshaw}.
\newblock {The Sunyaev-Zel'dovich effect}.
\newblock {\em \physrep}, 310:97--195, 1999.

\bibitem{1999ApJ...518..603R}
K.~{Roettiger}, J.~O. {Burns}, and J.~M. {Stone}.
\newblock {A cluster merger and the origin of the extended radio emission in
  Abell 3667}.
\newblock {\em \apj}, 518:603--612, 1999.

\bibitem{1999ApJ...518..594R}
K.~{Roettiger}, J.~M. {Stone}, and J.~O. {Burns}.
\newblock {Magnetic field evolution in merging clusters of galaxies}.
\newblock {\em \apj}, 518:594--602, 1999.

\bibitem{2006MNRAS.366.1437S}
K.~{Subramanian}, A.~{Shukurov}, and N.~E.~L. {Haugen}.
\newblock {Evolving turbulence and magnetic fields in galaxy clusters}.
\newblock {\em \mnras}, 366:1437--1454, 2006.

\bibitem{1980ApJ...241..925J}
W.~{Jaffe}.
\newblock {On the morphology of the magnetic field in galaxy clusters}.
\newblock {\em \apj}, 241:925--927, 1980.

\bibitem{1989MNRAS.241....1R}
A.~{Ruzmaikin}, D.~{Sokolov}, and A.~{Shukurov}.
\newblock {The dynamo origin of magnetic fields in galaxy clusters}.
\newblock {\em \mnras}, 241:1--14, 1989.

\bibitem{1991ApJ...380..344G}
I.~{Goldman} and Y.~{Rephaeli}.
\newblock {Turbulently generated magnetic fields in clusters of galaxies}.
\newblock {\em \apj}, 380:344--350, 1991.

\bibitem{2001PhR...348..163G}
D.~{Grasso} and H.~R. {Rubinstein}.
\newblock {Magnetic fields in the early Universe}.
\newblock {\em \physrep}, 348:163--266, 2001.

\bibitem{2001PhRvL..87y1302V}
T.~{Vachaspati}.
\newblock {Estimate of the primordial magnetic field helicity}.
\newblock {\em Phys. Rev. Lett.}, 87:251302, 2001.

\bibitem{1996PhRvD..54.1291B}
A.~{Brandenburg}, K.~{Enqvist}, and P.~{Olesen}.
\newblock {Large-scale magnetic fields from hydromagnetic turbulence in the
  very early universe}.
\newblock {\em \prd}, 54:1291--1300, 1996.

\bibitem{2001Sci...292.2440B}
A.~{Brandenburg}.
\newblock {Magnetic mysteries}.
\newblock {\em Science}, 292:2440--2441, 2001.

\bibitem{2001PhRvE..64e6405C}
M.~{Christensson}, M.~{Hindmarsh}, and A.~{Brandenburg}.
\newblock {Inverse cascade in decaying three-dimensional magnetohydrodynamic
  turbulence}.
\newblock {\em \pre}, 64:056405, 2001.

\bibitem{2003JFM...480..129K}
H.~S. {Kang}, S.~{Chester}, and C.~{Meneveau}.
\newblock {Decaying turbulence in an active-grid-generated flow and comparisons
  with large-eddy simulation}.
\newblock {\em J. Fluid Mech.}, 480:129--160, 2003.

\bibitem{2004PhRvE..70b6405H}
N.~E.~L. {Haugen} and A.~{Brandenburg}.
\newblock {Inertial range scaling in numerical turbulence with hyperviscosity}.
\newblock {\em \pre}, 70:026405, 2004.

\bibitem{1999PhRvL..83.2195B}
D.~{Biskamp} and W.-C. {M{\"u}ller}.
\newblock {Decay laws for three-dimensional magnetohydrodynamic turbulence}.
\newblock {\em Phys. Rev. Lett.}, 83:2195--2198, 1999.

\bibitem{2005AN....326..393C}
M.~{Christensson}, M.~{Hindmarsh}, and A.~{Brandenburg}.
\newblock {Scaling laws in decaying helical hydromagnetic turbulence}.
\newblock {\em Astron. Nachr.}, 326:393--399, 2005.

\bibitem{2004PhRvD..70l3003B}
R.~{Banerjee} and K.~{Jedamzik}.
\newblock {Evolution of cosmic magnetic fields: From the very early Universe,
  to recombination, to the present}.
\newblock {\em \prd}, 70:123003, 2004.

\bibitem{2002A&A...387..383D}
K.~{Dolag}, M.~{Bartelmann}, and H.~{Lesch}.
\newblock {Evolution and structure of magnetic fields in simulated galaxy
  clusters}.
\newblock {\em \aap}, 387:383--395, 2002.

\bibitem{2005JCAP...01..009D}
K.~{Dolag}, D.~{Grasso}, V.~{Springel}, and I.~{Tkachev}.
\newblock {Constrained simulations of the magnetic field in the local Universe
  and the propagation of ultrahigh energy cosmic rays}.
\newblock {\em J. Cosmology Astro-Particle Phys.}, 1:009, 2005.

\bibitem{2005PhRvD..72l3509T}
C.~G. {Tsagas}.
\newblock {Resonant amplification of magnetic seed fields by gravitational
  waves in the early universe}.
\newblock {\em \prd}, 72:123509, 2005.

\bibitem{2006PhRvD..74f3521C}
C.~{Caprini} and R.~{Durrer}.
\newblock {Gravitational waves from stochastic relativistic sources: Primordial
  turbulence and magnetic fields}.
\newblock {\em \prd}, 74:063521, 2006.

\bibitem{2008PhRvD..78l3006K}
T.~{Kahniashvili}, L.~{Campanelli}, G.~{Gogoberidze}, Y.~{Maravin}, and
  B.~{Ratra}.
\newblock {Gravitational radiation from primordial helical inverse cascade
  magnetohydrodynamic turbulence}.
\newblock {\em \prd}, 78:123006, 2008.

\bibitem{2008PhRvL.100w1301K}
T.~{Kahniashvili}, G.~{Gogoberidze}, and B.~{Ratra}.
\newblock {Gravitational radiation from primordial helical magnetohydrodynamic
  turbulence}.
\newblock {\em Phys. Rev. Lett.}, 100:231301, 2008.

\bibitem{2005ApJ...624..267T}
N.~J. {Turner}, O.~M. {Blaes}, A.~{Socrates}, M.~C. {Begelman}, and S.~W.
  {Davis}.
\newblock {The effects of photon bubble instability in radiation-dominated
  accretion disks}.
\newblock {\em \apj}, 624:267--288, 2005.

\bibitem{1994dmiv.book.....R}
R.~D. {Richtmyer} and K.~W. {Morton}.
\newblock {\em {Difference methods for initial-value problems}}.
\newblock 1994.

\bibitem{1962JAtS...19..173O}
Y.~{Ogura} and N.~A. {Phillips}.
\newblock {Scale analysis of deep and shallow convection in the atmosphere}.
\newblock {\em J. Atmos. Sci.}, 19:173--179, 1962.

\bibitem{1969JAtS...26..448G}
D.~O. {Gough}.
\newblock {The anelastic approximation for thermal convection.}
\newblock {\em J. Atmos. Sci.}, 26:448--456, 1969.

\bibitem{1981ApJS...45..335G}
P.~A. {Gilman} and G.~A. {Glatzmaier}.
\newblock {Compressible convection in a rotating spherical shell. I - Anelastic
  equations. II - A linear anelastic model. III - Analytic model for
  compressible vorticity waves}.
\newblock {\em \apjs}, 45:335--388, 1981.

\bibitem{1982A&A...107....1N}
A.~{Nordlund}.
\newblock {Numerical simulations of the solar granulation. I - Basic equations
  and methods}.
\newblock {\em \aap}, 107:1--10, 1982.

\bibitem{1984JCoPh..55..461G}
G.~A. {Glatzmaier}.
\newblock {Numerical simulations of stellar convective dynamos. I - The model
  and method}.
\newblock {\em J. Comp. Phys.}, 55:461--484, 1984.

\bibitem{2004ApJ...614.1073B}
A.~S. {Brun}, M.~S. {Miesch}, and J.~{Toomre}.
\newblock {Global-scale turbulent convection and magnetic dynamo action in the
  solar envelope}.
\newblock {\em \apj}, 614:1073--1098, 2004.

\bibitem{2005ApJ...620..432R}
T.~M. {Rogers} and G.~A. {Glatzmaier}.
\newblock {Penetrative convection within the anelastic approximation}.
\newblock {\em \apj}, 620:432--441, 2005.

\bibitem{1992A&A...265..328B}
A.~{Brandenburg}, D.~{Moss}, and I.~{Tuominen}.
\newblock {Stratification and thermodynamics in mean-field dynamos}.
\newblock {\em \aap}, 265:328--344, 1992.

\bibitem{1963MWRv...91...99S}
J.~{Smagorinsky}.
\newblock {General circulation experiments with the primitive equations}.
\newblock {\em Monthly Weather Review}, 91:99--164, 1963.

\bibitem{1993PhFl....5.2306S}
A.~{Scotti}, C.~{Meneveau}, and D.~K. {Lilly}.
\newblock {Generalized Smagorinsky model for anisotropic grids}.
\newblock {\em Phys. Fluids}, 5:2306--2308, 1993.

\bibitem{2004PhyA..338..379A}
S.~{Ansumali}, I.~V. {Karlin}, and S.~{Succi}.
\newblock {Kinetic theory of turbulence modeling: smallness parameter, scaling
  and microscopic derivation of Smagorinsky model}.
\newblock {\em Physica A Statistical Mechanics and its Applications},
  338:379--394, 2004.

\bibitem{2007JFM...570..491L}
E.~{L{\'e}v{\^e}que}, F.~{Toschi}, L.~{Shao}, and J.-P. {Bertoglio}.
\newblock {Shear-improved Smagorinsky model for large-eddy simulation of
  wall-bounded turbulent flows}.
\newblock {\em J. Fluid Mech.}, 570:491--502, 2007.

\bibitem{2008ApJS..178..137S}
J.~M. {Stone}, T.~A. {Gardiner}, P.~{Teuben}, J.~F. {Hawley}, and J.~B.
  {Simon}.
\newblock {Athena: A New Code for Astrophysical MHD}.
\newblock {\em \apjs}, 178:137--177, 2008.

\bibitem{2008A&A...488..429T}
O.~{Te{\c s}ileanu}, A.~{Mignone}, and S.~{Massaglia}.
\newblock {Simulating radiative astrophysical flows with the PLUTO code: a
  non-equilibrium, multi-species cooling function}.
\newblock {\em \aap}, 488:429--440, 2008.

\bibitem{2007CoPhC.176..652G}
O.~{Gressel} and U.~{Ziegler}.
\newblock {Shearingbox-implementation for the central-upwind,
  constraint-transport MHD-code NIRVANA}.
\newblock {\em Comp. Phys. Comm.}, 176:652--659, 2007.

\bibitem{2002A&A...385..337T}
R.~{Teyssier}.
\newblock {Cosmological hydrodynamics with adaptive mesh refinement. A new high
  resolution code called RAMSES}.
\newblock {\em \aap}, 385:337--364, 2002.

\bibitem{2000ApJS..131..273F}
B.~{Fryxell}, K.~{Olson}, P.~{Ricker}, F.~X. {Timmes}, M.~{Zingale}, D.~Q.
  {Lamb}, P.~{MacNeice}, R.~{Rosner}, J.~W. {Truran}, and H.~{Tufo}.
\newblock {FLASH: an adaptive mesh hydrodynamics code for modeling
  astrophysical thermonuclear flashes}.
\newblock {\em \apjs}, 131:273--334, 2000.

\bibitem{2005ApJS..160....1O}
B.~W. {O'Shea}, K.~{Nagamine}, V.~{Springel}, L.~{Hernquist}, and M.~L.
  {Norman}.
\newblock {Comparing AMR and SPH cosmological simulations. I. Dark matter and
  adiabatic simulations}.
\newblock {\em \apjs}, 160:1--27, 2005.

\bibitem{1992PhRvL..68.3156P}
D.~H. {Porter}, A.~{Pouquet}, and P.~R. {Woodward}.
\newblock {Three-dimensional supersonic homogeneous turbulence - A numerical
  study}.
\newblock {\em Phys. Rev. Lett.}, 68:3156--3159, 1992.

\bibitem{1992AnRFM..24..395F}
M.~{Farge}.
\newblock {Wavelet transforms and their applications to turbulence}.
\newblock {\em Ann. Rev. Fluid Mech.}, 24:395--457, 1992.

\bibitem{2007PhFl...19k5109O}
N.~{Okamoto}, K.~{Yoshimatsu}, K.~{Schneider}, M.~{Farge}, and Y.~{Kaneda}.
\newblock {Coherent vortices in high resolution direct numerical simulation of
  homogeneous isotropic turbulence: A wavelet viewpoint}.
\newblock {\em Phys. Fluids}, 19:115109, 2007.

\bibitem{1993AnRFM..25..539B}
G.~{Berkooz}, P.~{Holmes}, and J.~L. {Lumley}.
\newblock {The proper orthogonal decomposition in the analysis of turbulent
  flows}.
\newblock {\em Ann. Rev. Fluid Mech.}, 25:539--575, 1993.

\bibitem{2008PhPl...15g2305B}
W.~J.~T. {Bos}, S.~{Futatani}, S.~{Benkadda}, M.~{Farge}, and K.~{Schneider}.
\newblock {The role of coherent vorticity in turbulent transport in resistive
  drift-wave turbulence}.
\newblock {\em Phys. Plasmas}, 15:072305, 2008.

\bibitem{1995A&A...295L...1B}
P.~{Barge} and J.~{Sommeria}.
\newblock {Did planet formation begin inside persistent gaseous vortices?}
\newblock {\em \aap}, 295:L1--L4, 1995.

\bibitem{1996Icar..121..158T}
P.~{Tanga}, A.~{Babiano}, B.~{Dubrulle}, and A.~{Provenzale}.
\newblock {Forming planetesimals in vortices}.
\newblock {\em Icarus}, 121:158--170, 1996.

\bibitem{1998A&A...330.1169H}
L.~S. {Hodgson} and A.~{Brandenburg}.
\newblock {Turbulence effects in planetesimal formation}.
\newblock {\em \aap}, 330:1169--1174, 1998.

\bibitem{2004A&A...417..361J}
A.~{Johansen}, A.~C. {Andersen}, and A.~{Brandenburg}.
\newblock {Simulations of dust-trapping vortices in protoplanetary discs}.
\newblock {\em \aap}, 417:361--374, 2004.

\bibitem{2006ApJ...649..415I}
S.~{Inaba} and P.~{Barge}.
\newblock {Dusty vortices in protoplanetary disks}.
\newblock {\em \apj}, 649:415--427, 2006.

\bibitem{1990acr..book.....B}
V.~S. {Berezinskii}, S.~V. {Bulanov}, V.~A. {Dogiel}, and V.~S. {Ptuskin}.
\newblock {\em {Astrophysics of cosmic rays}}.
\newblock 1990.

\bibitem{2003A&A...412..331H}
M.~{Hanasz} and H.~{Lesch}.
\newblock {Incorporation of cosmic ray transport into the ZEUS MHD code.
  Application for studies of Parker instability in the ISM}.
\newblock {\em \aap}, 412:331--339, 2003.

\bibitem{1974ARA&A..12..407S}
R.~F. {Stein} and J.~{Leibacher}.
\newblock {Waves in the solar atmosphere}.
\newblock {\em \araa}, 12:407--435, 1974.

\bibitem{1988A&A...203..154B}
A.~{Brandenburg}.
\newblock {Hydrodynamic Green's functions for atmospheric oscillations}.
\newblock {\em \aap}, 203:154--161, 1988.

\bibitem{1989ApJ...342L..95S}
R.~F. {Stein} and A.~{Nordlund}.
\newblock {Topology of convection beneath the solar surface}.
\newblock {\em \apjl}, 342:L95--L98, 1989.

\bibitem{1996JFM...306..325B}
A.~{Brandenburg}, R.~L. {Jennings}, {\AA}.~{Nordlund}, M.~{Rieutord}, R.~F.
  {Stein}, and I.~{Tuominen}.
\newblock {Magnetic structures in a dynamo simulation}.
\newblock {\em J. Fluid Mech.}, 306:325--352, 1996.

\bibitem{1996ApJ...473..494B}
N.~H. {Brummell}, N.~E. {Hurlburt}, and J.~{Toomre}.
\newblock {Turbulent compressible convection with rotation. I. Flow structure
  and evolution}.
\newblock {\em \apj}, 473:494--513, 1996.

\bibitem{1984ApJ...282..557H}
N.~E. {Hurlburt}, J.~{Toomre}, and J.~M. {Massaguer}.
\newblock {Two-dimensional compressible convection extending over multiple
  scale heights}.
\newblock {\em \apj}, 282:557--573, 1984.

\bibitem{1968Sci...160..259M}
W.~V.~R. {Malkus}.
\newblock {Precession of the Earth as the cause of geomagnetism}.
\newblock {\em Science}, 160:259--264, 1968.

\bibitem{1975PEPI...11...43L}
D.~E. {Loper}.
\newblock {Torque balance and energy budget for the precessionally driven
  dynamo}.
\newblock {\em Phys. Earth Planet. Interiors}, 11:43--60, 1975.

\bibitem{1991GApFD..59..209V}
J.~P. {Vanyo}.
\newblock {A geodynamo powered by luni-solar precession}.
\newblock {\em Geophys. Astrophys. Fluid Dynam.}, 59:209--234, 1991.

\bibitem{2005PhFl...17c4104T}
A.~{Tilgner}.
\newblock {Precession driven dynamos}.
\newblock {\em Phys. Fluids}, 17:034104, 2005.

\bibitem{1970AmJPh..38..494S}
D.~P. {Stern}.
\newblock {Euler Potentials}.
\newblock {\em Am. J. Phys.}, 38:494--501, 1970.

\bibitem{2007MNRAS.379..915R}
S.~{Rosswog} and D.~{Price}.
\newblock {MAGMA: a three-dimensional, Lagrangian magnetohydrodynamics code for
  merger applications}.
\newblock {\em \mnras}, 379:915--931, 2007.

\bibitem{2009MNRAS.397..733K}
H.~{Kotarba}, H.~{Lesch}, K.~{Dolag}, T.~{Naab}, P.~H. {Johansson}, and F.~A.
  {Stasyszyn}.
\newblock {Magnetic field structure due to the global velocity field in spiral
  galaxies}.
\newblock {\em \mnras}, 397:733--747, 2009.

\bibitem{2010MNRAS.401..347B}
A.~{Brandenburg}.
\newblock {Magnetic field evolution in simulations with Euler potentials}.
\newblock {\em \mnras}, 401:347--354, 2010.

\bibitem{2006RPPh...69..563S}
S.~K. {Solanki}, B.~{Inhester}, and M.~{Sch{\"u}ssler}.
\newblock {The solar magnetic field}.
\newblock {\em Rep. Progr. Phys.}, 69:563--668, 2006.

\bibitem{1978stat.book.....M}
D.~{Mihalas}.
\newblock {\em {Stellar atmospheres}}.
\newblock 1978.

\bibitem{2002ApJ...569L.127K}
A.~G. {Kritsuk} and M.~L. {Norman}.
\newblock {Thermal instability-induced interstellar turbulence}.
\newblock {\em \apjl}, 569:L127--L131, 2002.

\bibitem{2007ApJ...654..945B}
A.~{Brandenburg}, M.~J. {Korpi}, and A.~J. {Mee}.
\newblock {Thermal instability in shearing and periodic turbulence}.
\newblock {\em \apj}, 654:945--954, 2007.

\bibitem{1998MNRAS.298...93B}
I.~A. {Bonnell}, M.~R. {Bate}, and H.~{Zinnecker}.
\newblock {On the formation of massive stars}.
\newblock {\em \mnras}, 298:93--102, 1998.

\bibitem{1997MNRAS.288.1060B}
M.~R. {Bate} and A.~{Burkert}.
\newblock {Resolution requirements for smoothed particle hydrodynamics
  calculations with self-gravity}.
\newblock {\em \mnras}, 288:1060--1072, 1997.

\bibitem{2002ApJ...576..870P}
P.~{Padoan} and {\AA}.~{Nordlund}.
\newblock {The stellar initial mass function from turbulent fragmentation}.
\newblock {\em \apj}, 576:870--879, 2002.

\bibitem{2003ApJ...585L.131V}
E.~{V{\'a}zquez-Semadeni}, J.~{Ballesteros-Paredes}, and R.~S. {Klessen}.
\newblock {A holistic scenario of turbulent molecular cloud evolution and
  control of the star formation efficiency: first tests}.
\newblock {\em \apjl}, 585:L131--L134, 2003.

\bibitem{2004ApJ...611..399K}
M.~R. {Krumholz}, C.~F. {McKee}, and R.~I. {Klein}.
\newblock {Embedding Lagrangian Sink Particles in Eulerian Grids}.
\newblock {\em \apj}, 611:399--412, 2004.

\bibitem{2007ApJ...656..959K}
M.~R. {Krumholz}, R.~I. {Klein}, and C.~F. {McKee}.
\newblock {Radiation-hydrodynamic simulations of collapse and fragmentation in
  massive protostellar cores}.
\newblock {\em \apj}, 656:959--979, 2007.

\bibitem{2009ApJ...694.1161J}
A.-K. {Jappsen}, M.-M.~M. {Low}, S.~C.~O. {Glover}, R.~S. {Klessen}, and
  S.~{Kitsionas}.
\newblock {Star formation at very low metallicity. V. The greater importance of
  initial conditions compared to metallicity thresholds}.
\newblock {\em \apj}, 694:1161--1170, 2009.

\bibitem{2009arXiv0907.0248P}
P.~{Padoan} and A.~{Nordlund}.
\newblock {The star formation rate of supersonic MHD turbulence}.
\newblock {\em arXiv:0907.0248}, 2009.

\bibitem{1992ApJS...80..753S}
J.~M. {Stone} and M.~L. {Norman}.
\newblock {ZEUS-2D: A radiation magnetohydrodynamics code for astrophysical
  flows in two space dimensions. I - The hydrodynamic algorithms and tests.}
\newblock {\em \apjs}, 80:753--790, 1992.

\bibitem{1992ARA&A..30..543M}
J.~J. {Monaghan}.
\newblock {Smoothed particle hydrodynamics}.
\newblock {\em \araa}, 30:543--574, 1992.

\bibitem{2002MNRAS.333..649S}
V.~{Springel} and L.~{Hernquist}.
\newblock {Cosmological smoothed particle hydrodynamics simulations: the
  entropy equation}.
\newblock {\em \mnras}, 333:649--664, 2002.

\bibitem{1996MNRAS.278.1005S}
M.~{Steinmetz}.
\newblock {GRAPESPH: cosmological smoothed particle hydrodynamics simulations
  with the special-purpose hardware GRAPE}.
\newblock {\em \mnras}, 278:1005--1017, 1996.

\bibitem{2010MNRAS.401..791S}
V.~{Springel}.
\newblock {E pur si muove: Galilean-invariant cosmological hydrodynamical
  simulations on a moving mesh}.
\newblock {\em \mnras}, 401:791--851, 2010.

\bibitem{1990Natur.344..226S}
{Z.-S.} {She}, E.~{Jackson}, and S.~A. {Orszag}.
\newblock {Intermittent vortex structures in homogeneous isotropic turbulence}.
\newblock {\em \nat}, 344:226--228, 1990.

\bibitem{1991JFM...225....1V}
A.~{Vincent} and M.~{Meneguzzi}.
\newblock {The spatial structure and statistical properties of homogeneous
  turbulence}.
\newblock {\em J. Fluid Mech.}, 225:1--20, 1991.

\bibitem{1994PhFl....6.2133P}
D.~H. {Porter}, A.~{Pouquet}, and P.~R. {Woodward}.
\newblock {Kolmogorov-like spectra in decaying three-dimensional supersonic
  flows}.
\newblock {\em Phys. Fluids}, 6:2133--2142, 1994.

\bibitem{2004PhRvE..70a6308H}
N.~E. {Haugen}, A.~{Brandenburg}, and W.~{Dobler}.
\newblock {Simulations of nonhelical hydromagnetic turbulence}.
\newblock {\em \pre}, 70:016308, 2004.

\bibitem{2004RvMP...76..125M}
M.-M. {Mac Low} and R.~S. {Klessen}.
\newblock {Control of star formation by supersonic turbulence}.
\newblock {\em Rev. Mod. Phys.}, 76:125--194, 2004.

\bibitem{1997MNRAS.288..145P}
P.~{Padoan}, A.~{Nordlund}, and B.~J.~T. {Jones}.
\newblock {The universality of the stellar initial mass function}.
\newblock {\em \mnras}, 288:145--152, 1997.

\bibitem{1999ApJ...526..279P}
P.~{Padoan} and {\AA}.~{Nordlund}.
\newblock {A super-Alfv{\'e}nic model of dark clouds}.
\newblock {\em \apj}, 526:279--294, 1999.

\bibitem{2006ApJ...653L.125P}
P.~{Padoan}, M.~{Juvela}, A.~{Kritsuk}, and M.~L. {Norman}.
\newblock {The power spectrum of supersonic turbulence in Perseus}.
\newblock {\em \apjl}, 653:L125--L128, 2006.

\bibitem{1973SPhD...18..115K}
B.~B. {Kadomtsev} and V.~I. {Petviashvili}.
\newblock {Acoustic turbulence}.
\newblock {\em Sov. Phys. Doklady}, 18:115--116, 1973.

\bibitem{2009ApJ...702.1190B}
A.~{Beresnyak} and A.~{Lazarian}.
\newblock {Comparison of spectral slopes of magnetohydrodynamic and
  hydrodynamic turbulence and measurements of alignment effects}.
\newblock {\em \apj}, 702:1190--1198, 2009.

\bibitem{2010PhRvE..81a6318L}
E.~{Lee}, M.~E. {Brachet}, A.~{Pouquet}, P.~D. {Mininni}, and D.~{Rosenberg}.
\newblock {Lack of universality in decaying magnetohydrodynamic turbulence}.
\newblock {\em \pre}, 81:016318, 2010.

\bibitem{1964SvA.....7..566I}
P.~S. {Iroshnikov}.
\newblock {Turbulence of a conducting fluid in a strong magnetic field}.
\newblock {\em Sov. Astron.}, 7:566--571, 1964.

\bibitem{1965PhFl....8.1385K}
R.~H. {Kraichnan}.
\newblock {Inertial-range spectrum of hydromagnetic turbulence}.
\newblock {\em Phys. Fluids}, 8:1385--1387, 1965.

\bibitem{2000JPlPh..63..447G}
S.~{Galtier}, S.~V. {Nazarenko}, A.~C. {Newell}, and A.~{Pouquet}.
\newblock {A weak turbulence theory for incompressible magnetohydrodynamics}.
\newblock {\em J. Plasma Phys.}, 63:447--488, 2000.

\bibitem{1995ApJ...438..763G}
P.~{Goldreich} and S.~{Sridhar}.
\newblock {Toward a theory of interstellar turbulence. 2. Strong Alfvenic
  turbulence}.
\newblock {\em \apj}, 438:763--775, 1995.

\bibitem{1997ApJ...485..680G}
P.~{Goldreich} and S.~{Sridhar}.
\newblock {Magnetohydrodynamic turbulence revisited}.
\newblock {\em \apj}, 485:680--688, 1997.

\bibitem{1994ApJ...432..612S}
S.~{Sridhar} and P.~{Goldreich}.
\newblock {Toward a theory of interstellar turbulence. 1. Weak Alfvenic
  turbulence}.
\newblock {\em \apj}, 432:612--621, 1994.

\bibitem{2007ApJ...655..269L}
Y.~{Lithwick}, P.~{Goldreich}, and S.~{Sridhar}.
\newblock {Imbalanced strong MHD turbulence}.
\newblock {\em \apj}, 655:269--274, 2007.

\bibitem{2003ApJ...582.1220L}
Y.~{Lithwick} and P.~{Goldreich}.
\newblock {Imbalanced weak magnetohydrodynamic turbulence}.
\newblock {\em \apj}, 582:1220--1240, 2003.

\bibitem{2003ApJ...597L.141H}
N.~E.~L. {Haugen}, A.~{Brandenburg}, and W.~{Dobler}.
\newblock {Is nonhelical hydromagnetic turbulence peaked at small scales?}
\newblock {\em \apjl}, 597:L141--L144, 2003.

\bibitem{2007ApJ...665L..35D}
M.~A. {de Avillez} and D.~{Breitschwerdt}.
\newblock {The generation and dissipation of interstellar turbulence: results
  from large-scale high-resolution simulations}.
\newblock {\em \apjl}, 665:L35--L38, 2007.

\bibitem{2009arXiv0904.4860P}
A.~{Pouquet}, J.~{Baerenzung}, J.~{Pietarila Graham}, P.~{Mininni},
  H.~{Politano}, and Y.~{Ponty}.
\newblock {Modeling of anisotropic turbulent flows with either magnetic fields
  or imposed rotation}.
\newblock {\em arXiv:0904.4860}, 2009.

\bibitem{2005ApJ...626L..37B}
S.~{Boldyrev}.
\newblock {On the spectrum of magnetohydrodynamic turbulence}.
\newblock {\em \apjl}, 626:L37--L40, 2005.

\bibitem{2006PhRvL..96k5002B}
S.~{Boldyrev}.
\newblock {Spectrum of magnetohydrodynamic turbulence}.
\newblock {\em Phys. Rev. Lett.}, 96:115002, 2006.

\bibitem{2006PhRvL..97y5002M}
J.~{Mason}, F.~{Cattaneo}, and S.~{Boldyrev}.
\newblock {Dynamic alignment in driven magnetohydrodynamic turbulence}.
\newblock {\em Phys. Rev. Lett.}, 97:255002, 2006.

\bibitem{2008ApJ...672L..61P}
J.~C. {Perez} and S.~{Boldyrev}.
\newblock {On weak and strong magnetohydrodynamic turbulence}.
\newblock {\em \apjl}, 672:L61--L64, 2008.

\bibitem{2009JPhCS.180a2020K}
A.~G. {Kritsuk}, S.~D. {Ustyugov}, M.~L. {Norman}, and P.~{Padoan}.
\newblock {Simulating supersonic turbulence in magnetized molecular clouds}.
\newblock {\em J. Phys. Conf. Ser.}, 180:012020, 2009.

\bibitem{2009ASPC..406...15K}
A.~G. {Kritsuk}, S.~D. {Ustyugov}, M.~L. {Norman}, and P.~{Padoan}.
\newblock {Simulations of supersonic turbulence in molecular clouds: evidence
  for a new universality}.
\newblock volume 406 of {\em Astron. Soc. Pac. Conf. Ser.}, pages 15--22, 2009.

\bibitem{2009ApJ...697.1206B}
A.~{Brandenburg}.
\newblock {Large-scale dynamos at low magnetic Prandtl numbers}.
\newblock {\em \apj}, 697:1206--1213, 2009.

\bibitem{2005PhRvL..94p4502P}
Y.~{Ponty}, P.~D. {Mininni}, D.~C. {Montgomery}, J.-F. {Pinton}, H.~{Politano},
  and A.~{Pouquet}.
\newblock {Numerical study of dynamo action at low magnetic Prandtl numbers}.
\newblock {\em Phys. Rev. Lett.}, 94:164502, 2005.

\bibitem{2007NJPh....9..296P}
Y.~{Ponty}, P.~D. {Mininni}, {J.-F.} {Pinton}, H.~{Politano}, and A.~{Pouquet}.
\newblock {Dynamo action at low magnetic Prandtl numbers: mean flow versus
  fully turbulent motions}.
\newblock {\em New J. Phys.}, 9:296, 2007.

\bibitem{2005ApJ...625L.115S}
A.~A. {Schekochihin}, N.~E.~L. {Haugen}, A.~{Brandenburg}, S.~C. {Cowley},
  J.~L. {Maron}, and J.~C. {McWilliams}.
\newblock {The onset of a small-scale turbulent dynamo at low magnetic Prandtl
  numbers}.
\newblock {\em \apjl}, 625:L115--L118, 2005.

\bibitem{1997PhRvE..56..417R}
I.~{Rogachevskii} and N.~{Kleeorin}.
\newblock {Intermittency and anomalous scaling for magnetic fluctuations}.
\newblock {\em \pre}, 56:417--426, 1997.

\bibitem{2004PhRvL..92n4501B}
S.~{Boldyrev} and F.~{Cattaneo}.
\newblock {Magnetic-field generation in Kolmogorov turbulence}.
\newblock {\em Phys. Rev. Lett.}, 92:144501, 2004.

\bibitem{2007PhRvL..98t8501I}
A.~B. {Iskakov}, A.~A. {Schekochihin}, S.~C. {Cowley}, J.~C. {McWilliams}, and
  M.~R.~E. {Proctor}.
\newblock {Numerical demonstration of fluctuation dynamo at low magnetic
  Prandtl numbers}.
\newblock {\em Phys. Rev. Lett.}, 98:208501, 2007.

\bibitem{2007NJPh....9..300S}
A.~A. {Schekochihin}, A.~B. {Iskakov}, S.~C. {Cowley}, J.~C. {McWilliams},
  M.~R.~E. {Proctor}, and T.~A. {Yousef}.
\newblock {Fluctuation dynamo and turbulent induction at low magnetic Prandtl
  numbers}.
\newblock {\em New J. Phys.}, 9:300, 2007.

\bibitem{2001ApJ...550..824B}
A.~{Brandenburg}.
\newblock {The inverse cascade and nonlinear alpha-effect in simulations of
  isotropic helical hydromagnetic turbulence}.
\newblock {\em \apj}, 550:824--840, 2001.

\bibitem{2007PhRvE..76e6313A}
A.~{Alexakis}, B.~{Bigot}, H.~{Politano}, and S.~{Galtier}.
\newblock {Anisotropic fluxes and nonlocal interactions in magnetohydrodynamic
  turbulence}.
\newblock {\em \pre}, 76:056313, 2007.

\bibitem{1996JFM...310..139K}
R.~M. {Kerr}.
\newblock {Rayleigh number scaling in numerical convection}.
\newblock {\em J. Fluid Mech.}, 310:139--179, 1996.

\bibitem{1996JFM...322..243J}
K.~{Julien}, S.~{Legg}, J.~{McWilliams}, and J.~{Werne}.
\newblock {Rapidly rotating turbulent Rayleigh-Benard convection}.
\newblock {\em J. Fluid Mech.}, 322:243--273, 1996.

\bibitem{2003PhRvL..91f4501H}
T.~{Hartlep}, A.~{Tilgner}, and F.~H. {Busse}.
\newblock {Large scale structures in Rayleigh-B{\'e}nard convection at high
  Rayleigh numbers}.
\newblock {\em Phys. Rev. Lett.}, 91:064501, 2003.

\bibitem{1991ApJ...370..282C}
F.~{Cattaneo}, N.~H. {Brummell}, J.~{Toomre}, A.~{Malagoli}, and N.~E.
  {Hurlburt}.
\newblock {Turbulent compressible convection}.
\newblock {\em \apj}, 370:282--294, 1991.

\bibitem{1998ApJ...499..914S}
R.~F. {Stein} and A.~{Nordlund}.
\newblock {Simulations of solar granulation. I. General properties}.
\newblock {\em \apj}, 499:914--933, 1998.

\bibitem{1986ApJ...311..563H}
N.~E. {Hurlburt}, J.~{Toomre}, and J.~M. {Massaguer}.
\newblock {Nonlinear compressible convection penetrating into stable layers and
  producing internal gravity waves}.
\newblock {\em \apj}, 311:563--577, 1986.

\bibitem{2004A&A...421..775D}
B.~{Dintrans} and A.~{Brandenburg}.
\newblock {Identification of gravity waves in hydrodynamical simulations}.
\newblock {\em \aap}, 421:775--782, 2004.

\bibitem{2005A&A...438..365D}
B.~{Dintrans}, A.~{Brandenburg}, {\AA}.~{Nordlund}, and R.~F. {Stein}.
\newblock {Spectrum and amplitudes of internal gravity waves excited by
  penetrative convection in solar-type stars}.
\newblock {\em \aap}, 438:365--376, 2005.

\bibitem{2005MNRAS.364.1135R}
T.~M. {Rogers} and G.~A. {Glatzmaier}.
\newblock {Gravity waves in the Sun}.
\newblock {\em \mnras}, 364:1135--1146, 2005.

\bibitem{2008MNRAS.387..616R}
T.~M. {Rogers}, K.~B. {MacGregor}, and G.~A. {Glatzmaier}.
\newblock {Non-linear dynamics of gravity wave driven flows in the solar
  radiative interior}.
\newblock {\em \mnras}, 387:616--630, 2008.

\bibitem{2009A&A...494..191B}
K.~{Belkacem}, R.~{Samadi}, M.~J. {Goupil}, M.~A. {Dupret}, A.~S. {Brun}, and
  F.~{Baudin}.
\newblock {Stochastic excitation of nonradial modes. II. Are solar asymptotic
  gravity modes detectable?}
\newblock {\em \aap}, 494:191--204, 2009.

\bibitem{2008A&A...491..353K}
P.~J. {K{\"a}pyl{\"a}}, M.~J. {Korpi}, and A.~{Brandenburg}.
\newblock {Large-scale dynamos in turbulent convection with shear}.
\newblock {\em \aap}, 491:353--362, 2008.

\bibitem{1983ApJS...53..243G}
P.~A. {Gilman}.
\newblock {Dynamically consistent nonlinear dynamos driven by convection in a
  rotating spherical shell. II - Dynamos with cycles and strong feedbacks}.
\newblock {\em \apjs}, 53:243--268, 1983.

\bibitem{1985ApJ...291..300G}
G.~A. {Glatzmaier}.
\newblock {Numerical simulations of stellar convective dynamos. II - Field
  propagation in the convection zone}.
\newblock {\em \apj}, 291:300--307, 1985.

\bibitem{2010ApJ...711..424B}
B.~P. {Brown}, M.~K. {Browning}, A.~S. {Brun}, M.~S. {Miesch}, and J.~{Toomre}.
\newblock {Persistent magnetic wreaths in a rapidly rotating Sun}.
\newblock {\em \apj}, 711:424--438, 2010.

\bibitem{1986Sci...234...61H}
J.~E. {Hart}, J.~{Toomre}, A.~E. {Deane}, N.~E. {Hurlburt}, G.~A. {Glatzmaier},
  G.~H. {Fichtl}, F.~{Leslie}, W.~W. {Fowlis}, and P.~A. {Gilman}.
\newblock {Laboratory experiments on planetary and stellar convection performed
  on Spacelab 3}.
\newblock {\em Science}, 234:61--64, 1986.

\bibitem{2009ApJ...697..923M}
D.~{Mitra}, R.~{Tavakol}, A.~{Brandenburg}, and D.~{Moss}.
\newblock {Turbulent dynamos in spherical shell segments of varying geometrical
  extent}.
\newblock {\em \apj}, 697:923--933, 2009.

\bibitem{2010ApJ...719L...1M}
D.~{Mitra}, R.~{Tavakol}, P.~J. {K{\"a}pyl{\"a}}, and A.~{Brandenburg}.
\newblock {Oscillatory migrating magnetic fields in helical turbulence in
  spherical domains}.
\newblock {\em \apjl}, 719:L1--L4, 2010.

\bibitem{2010AN....331..130M}
D.~{Mitra}, S.~{Candelaresi}, P.~{Chatterjee}, R.~{Tavakol}, and
  A.~{Brandenburg}.
\newblock {Equatorial magnetic helicity flux in simulations with different
  gauges}.
\newblock {\em Astron. Nachr.}, 331:130--135, 2010.

\bibitem{2006ApJ...641L..77G}
C.~A. {Green} and A.~G. {Kosovichev}.
\newblock {Traveling convective modes in the Sun's subsurface shear layer}.
\newblock {\em \apjl}, 641:L77--L80, 2006.

\bibitem{2007ApJ...665L..75G}
C.~A. {Green} and A.~G. {Kosovichev}.
\newblock {Magnetic effect on wave-like properties of solar supergranulation}.
\newblock {\em \apjl}, 665:L75--L78, 2007.

\bibitem{2007SoPh..245...27B}
F.~H. {Busse}.
\newblock {Convection in the presence of an inclined axis of rotation with
  applications to the Sun}.
\newblock {\em \solphys}, 245:27--36, 2007.

\bibitem{2007IAUS..239..457B}
A.~{Brandenburg}.
\newblock {Near-surface shear layer dynamics}.
\newblock In {T.~Kuroda, H.~Sugama, R.~Kanno, \& M.~Okamoto}, editor, {\em IAU
  Symp.}, volume 239 of {\em IAU Symp.}, pages 457--466, 2007.

\bibitem{1999ApJ...517L.163B}
E.~E. {Benevolenskaya}, J.~T. {Hoeksema}, A.~G. {Kosovichev}, and P.~H.
  {Scherrer}.
\newblock {The interaction of new and old magnetic fluxes at the beginning of
  Solar Cycle 23}.
\newblock {\em \apjl}, 517:L163--L166, 1999.

\bibitem{2005ApJ...625..539B}
A.~{Brandenburg}.
\newblock {The case for a distributed solar dynamo shaped by near-surface
  shear}.
\newblock {\em \apj}, 625:539--547, 2005.

\bibitem{2001ApJ...556L..63A}
C.~{Allende Prieto}, D.~L. {Lambert}, and M.~{Asplund}.
\newblock {The forbidden abundance of oxygen in the Sun}.
\newblock {\em \apjl}, 556:L63--L66, 2001.

\bibitem{2004A&A...417..751A}
M.~{Asplund}, N.~{Grevesse}, A.~J. {Sauval}, C.~{Allende Prieto}, and
  D.~{Kiselman}.
\newblock {Line formation in solar granulation. IV. [O I], O I and OH lines and
  the photospheric O abundance}.
\newblock {\em \aap}, 417:751--768, 2004.

\bibitem{2005ASPC..336...25A}
M.~{Asplund}, N.~{Grevesse}, and A.~J. {Sauval}.
\newblock {The solar chemical composition}.
\newblock In {T.~G.~Barnes III \& F.~N.~Bash}, editor, {\em Cosmic abundances
  as records of stellar evolution and nucleosynthesis}, volume 336 of {\em
  Astron. Soc. Pac. Conf. Ser.}, pages 25--38, 2005.

\bibitem{2009A&A...498..877C}
E.~{Caffau}, E.~{Maiorca}, P.~{Bonifacio}, R.~{Faraggiana}, M.~{Steffen},
  {H.-G.} {Ludwig}, I.~{Kamp}, and M.~{Busso}.
\newblock {The solar photospheric nitrogen abundance. Analysis of atomic
  transitions with 3D and 1D model atmospheres}.
\newblock {\em \aap}, 498:877--884, 2009.

\bibitem{1989GeCoA..53..197A}
E.~{Anders} and N.~{Grevesse}.
\newblock {Abundances of the elements - meteoritic and solar}.
\newblock {\em \gca}, 53:197--214, 1989.

\bibitem{2001ApJ...555..990B}
J.~N. {Bahcall}, M.~H. {Pinsonneault}, and S.~{Basu}.
\newblock {Solar models: current epoch and time dependences, neutrinos, and
  helioseismological properties}.
\newblock {\em \apj}, 555:990--1012, 2001.

\bibitem{2006ApJS..165..400B}
J.~N. {Bahcall}, A.~M. {Serenelli}, and S.~{Basu}.
\newblock {10,000 standard solar models: a Monte Carlo simulation}.
\newblock {\em \apjs}, 165:400--431, 2006.

\bibitem{2008ApJ...687..678H}
W.~C. {Haxton} and A.~M. {Serenelli}.
\newblock {CN cycle solar neutrinos and the Sun's primordial core metallicity}.
\newblock {\em \apj}, 687:678--691, 2008.

\bibitem{2006ApJ...644.1292A}
H.~M. {Antia} and S.~{Basu}.
\newblock {Determining solar abundances using helioseismology}.
\newblock {\em \apj}, 644:1292--1298, 2006.

\bibitem{1996A&A...312.1000R}
O.~{Richard}, S.~{Vauclair}, C.~{Charbonnel}, and W.~A. {Dziembowski}.
\newblock {New solar models including helioseismological constraints and
  light-element depletion.}
\newblock {\em \aap}, 312:1000--1011, 1996.

\bibitem{2004ApJ...607.1046R}
M.~{Rempel}.
\newblock {Overshoot at the base of the solar convection zone: a semianalytical
  approach}.
\newblock {\em \apj}, 607:1046--1064, 2004.

\bibitem{2005ApJ...618.1049B}
J.~N. {Bahcall}, S.~{Basu}, M.~{Pinsonneault}, and A.~M. {Serenelli}.
\newblock {Helioseismological implications of recent solar abundance
  determinations}.
\newblock {\em \apj}, 618:1049--1056, 2005.

\bibitem{2010Ap&SS.328..193M}
J.~{Mel{\'e}ndez}, I.~{Ram{\'{\i}}rez}, L.~{Casagrande}, M.~{Asplund},
  B.~{Gustafsson}, D.~{Yong}, J.~D. {Do Nascimento}, M.~{Castro}, and
  M.~{Bazot}.
\newblock {The solar, exoplanet and cosmological lithium problems}.
\newblock {\em \apss}, 328:193--200, 2010.

\bibitem{2004AIPC..733..122B}
A.~{Brandenburg}, B.~{Dintrans}, and N.~E.~L. {Haugen}.
\newblock {Shearing and embedding box simulations of the magnetorotational
  instability}.
\newblock In {R.~Rosner, G.~R{\"u}diger, \& A.~Bonanno}, editor, {\em MHD
  Couette Flows: Experiments and Models}, volume 733 of {\em Am. Inst. Phys.
  Conf. Ser.}, pages 122--136, 2004.

\bibitem{2007A&A...476.1113F}
S.~{Fromang} and J.~{Papaloizou}.
\newblock {MHD simulations of the magnetorotational instability in a shearing
  box with zero net flux. I. The issue of convergence}.
\newblock {\em \aap}, 476:1113--1122, 2007.

\bibitem{2007A&A...476.1123F}
S.~{Fromang}, J.~{Papaloizou}, G.~{Lesur}, and T.~{Heinemann}.
\newblock {MHD simulations of the magnetorotational instability in a shearing
  box with zero net flux. II. The effect of transport coefficients}.
\newblock {\em \aap}, 476:1123--1132, 2007.

\bibitem{2003PhRvE..67d6312R}
G.~{R{\"u}diger}, M.~{Schultz}, and D.~{Shalybkov}.
\newblock {Linear magnetohydrodynamic Taylor-Couette instability for liquid
  sodium}.
\newblock {\em \pre}, 67:046312, 2003.

\bibitem{2005PhRvL..95l4501H}
R.~{Hollerbach} and G.~{R{\"u}diger}.
\newblock {New Type of Magnetorotational Instability in Cylindrical
  Taylor-Couette Flow}.
\newblock {\em Phys. Rev. Lett.}, 95:124501, 2005.

\bibitem{2006PhRvL..97r4502S}
F.~{Stefani}, T.~{Gundrum}, G.~{Gerbeth}, G.~{R{\"u}diger}, M.~{Schultz},
  J.~{Szklarski}, and R.~{Hollerbach}.
\newblock {Experimental evidence for magnetorotational instability in a
  Taylor-Couette flow under the influence of a helical magnetic field}.
\newblock {\em Phys. Rev. Lett.}, 97:184502, 2006.

\bibitem{2007NJPh....9..295S}
F.~{Stefani}, T.~{Gundrum}, G.~{Gerbeth}, G.~{R{\"u}diger}, J.~{Szklarski}, and
  R.~{Hollerbach}.
\newblock {Experiments on the magnetorotational instability in helical magnetic
  fields}.
\newblock {\em New J. Phys.}, 9:295, 2007.

\bibitem{2007AN....328.1158R}
G.~{R{\"u}diger}, R.~{Hollerbach}, M.~{Gellert}, and M.~{Schultz}.
\newblock {The azimuthal magnetorotational instability (AMRI)}.
\newblock {\em Astron. Nachr.}, 328:1158--1161, 2007.

\bibitem{2007PhRvE..76b6316M}
P.~D. {Mininni}.
\newblock {Inverse cascades and {$\alpha$} effect at a low magnetic Prandtl
  number}.
\newblock {\em \pre}, 76:026316, 2007.

\bibitem{1995ApJ...446..741B}
A.~{Brandenburg}, A.~{Nordlund}, R.~F. {Stein}, and U.~{Torkelsson}.
\newblock {Dynamo-generated turbulence and large-scale magnetic fields in a
  Keplerian shear flow}.
\newblock {\em \apj}, 446:741--754, 1995.

\bibitem{2002GApFD..96..319B}
A.~{Brandenburg} and D.~{Sokoloff}.
\newblock {Local and nonlocal magnetic diffusion and alpha-effect tensors in
  shear flow turbulence}.
\newblock {\em Geophys. Astrophys. Fluid Dynam.}, 96:319--344, 2002.

\bibitem{2010arXiv1004.2417K}
P.~J. {K{\"a}pyl{\"a}} and M.~J. {Korpi}.
\newblock {Magnetorotational instability driven dynamos at low magnetic Prandtl
  numbers}.
\newblock {\em arXiv:1004.2417}, 2010.

\bibitem{2008PhRvL.100r4501Y}
T.~A. {Yousef}, T.~{Heinemann}, A.~A. {Schekochihin}, N.~{Kleeorin},
  I.~{Rogachevskii}, A.~B. {Iskakov}, S.~C. {Cowley}, and J.~C. {McWilliams}.
\newblock {Generation of magnetic field by combined action of turbulence and
  shear}.
\newblock {\em Phys. Rev. Lett.}, 100:184501, 2008.

\bibitem{2008AN....329..737Y}
T.~A. {Yousef}, T.~{Heinemann}, F.~{Rincon}, A.~A. {Schekochihin},
  N.~{Kleeorin}, I.~{Rogachevskii}, S.~C. {Cowley}, and J.~C. {McWilliams}.
\newblock {Numerical experiments on dynamo action in sheared and rotating
  turbulence}.
\newblock {\em Astron. Nachr.}, 329:737--749, 2008.

\bibitem{2008ApJ...676..740B}
A.~{Brandenburg}, K.-H. {R{\"a}dler}, M.~{Rheinhardt}, and P.~J.
  {K{\"a}pyl{\"a}}.
\newblock {Magnetic diffusivity tensor and dynamo effects in rotating and
  shearing turbulence}.
\newblock {\em \apj}, 676:740--751, 2008.

\bibitem{1997ApJ...475..263V}
E.~T. {Vishniac} and A.~{Brandenburg}.
\newblock {An incoherent alpha--Omega dynamo in accretion disks}.
\newblock {\em \apj}, 475:263--274, 1997.

\bibitem{2007MNRAS.382L..39P}
M.~R.~E. {Proctor}.
\newblock {Effects of fluctuation on {$\alpha$}{$\Omega$} dynamo models}.
\newblock {\em \mnras}, 382:L39--L42, 2007.

\bibitem{2003PhRvE..68c6301R}
I.~{Rogachevskii} and N.~{Kleeorin}.
\newblock {Electromotive force and large-scale magnetic dynamo in a turbulent
  flow with a mean shear}.
\newblock {\em \pre}, 68:036301, 2003.

\bibitem{2004PhRvE..70d6310R}
I.~{Rogachevskii} and N.~{Kleeorin}.
\newblock {Nonlinear theory of a ``shear-current'' effect and mean-field
  magnetic dynamos}.
\newblock {\em \pre}, 70:046310, 2004.

\bibitem{2005AN....326..787B}
A.~{Brandenburg}.
\newblock {Turbulence and its parameterization in accretion discs}.
\newblock {\em Astron. Nachr.}, 326:787--797, 2005.

\bibitem{2006PhRvE..73e6311R}
{K.-H.} {R{\"a}dler} and R.~{Stepanov}.
\newblock {Mean electromotive force due to turbulence of a conducting fluid in
  the presence of mean flow}.
\newblock {\em \pre}, 73:056311, 2006.

\bibitem{2006AN....327..298R}
G.~{R{\"u}diger} and L.~L. {Kitchatinov}.
\newblock {Do mean-field dynamos in nonrotating turbulent shear-flows exist?}
\newblock {\em Astron. Nachr.}, 327:298--303, 2006.

\bibitem{1988AJ.....95..925W}
J.~{Wisdom} and S.~{Tremaine}.
\newblock {Local simulations of planetary rings}.
\newblock {\em \aj}, 95:925--940, 1988.

\bibitem{1995ApJ...440..742H}
J.~F. {Hawley}, C.~F. {Gammie}, and S.~A. {Balbus}.
\newblock {Local three-dimensional magnetohydrodynamic simulations of accretion
  disks}.
\newblock {\em \apj}, 440:742--763, 1995.

\bibitem{2000ApJ...528..462H}
J.~F. {Hawley}.
\newblock {Global magnetohydrodynamical simulations of accretion tori}.
\newblock {\em \apj}, 528:462--479, 2000.

\bibitem{2000ApJ...532L..67M}
M.~{Machida}, M.~R. {Hayashi}, and R.~{Matsumoto}.
\newblock {Global simulations of Differentially Rotating Magnetized Disks:
  Formation of Low-{$\beta$} Filaments and Structured Coronae}.
\newblock {\em \apjl}, 532:L67--L70, 2000.

\bibitem{2003ApJ...585..429M}
M.~{Machida} and R.~{Matsumoto}.
\newblock {Global three-dimensional magnetohydrodynamic simulations of black
  hole accretion disks: X-ray flares in the plunging region}.
\newblock {\em \apj}, 585:429--442, 2003.

\bibitem{2003ApJ...599.1238D}
J.-P. {De Villiers}, J.~F. {Hawley}, and J.~H. {Krolik}.
\newblock {Magnetically driven accretion flows in the Kerr metric. I. Models
  and overall structure}.
\newblock {\em \apj}, 599:1238--1253, 2003.

\bibitem{1973A&A....24..337S}
N.~I. {Shakura} and R.~A. {Sunyaev}.
\newblock {Black holes in binary systems. Observational appearance.}
\newblock {\em \aap}, 24:337--355, 1973.

\bibitem{2002ApJ...573..754K}
J.~H. {Krolik} and J.~F. {Hawley}.
\newblock {Where is the inner edge of an accretion disk around a black hole?}
\newblock {\em \apj}, 573:754--763, 2002.

\bibitem{2008MNRAS.390...21B}
K.~{Beckwith}, J.~F. {Hawley}, and J.~H. {Krolik}.
\newblock {Where is the radiation edge in magnetized black hole accretion
  discs?}
\newblock {\em \mnras}, 390:21--38, 2008.

\bibitem{2007Natur.448.1022J}
A.~{Johansen}, J.~S. {Oishi}, M.-M.~M. {Low}, H.~{Klahr}, T.~{Henning}, and
  A.~{Youdin}.
\newblock {Rapid planetesimal formation in turbulent circumstellar disks}.
\newblock {\em \nat}, 448:1022--1025, 2007.

\bibitem{2007arXiv0708.3893J}
A.~{Johansen}, J.~S. {Oishi}, M.-M. {Mac Low}, H.~{Klahr}, T.~{Henning}, and
  A.~{Youdin}.
\newblock {Supplementary Information for ``Rapid planetesimal formation in
  turbulent circumstellar discs''}.
\newblock {\em arXiv:0708.3893}, 2007.

\bibitem{1965ApJ...142..531F}
G.~B. {Field}.
\newblock {Thermal instability}.
\newblock {\em \apj}, 142:531--567, 1965.

\bibitem{2007ASPC..365..162I}
S.~{Inutsuka} and H.~{Koyama}.
\newblock {Dynamics of a Multi-Phase Interstellar Medium}.
\newblock In {M.~Haverkorn \& W.~M.~Goss}, editor, {\em SINS - Small Ionized
  and Neutral Structures in the Diffuse Interstellar Medium}, volume 365 of
  {\em Astron. Soc. Pac. Conf. Ser.}, pages 162--165, 2007.

\bibitem{1997ApJ...489L.179T}
J.~K. {Truelove}, R.~I. {Klein}, C.~F. {McKee}, J.~H. {Holliman}, II, L.~H.
  {Howell}, and J.~A. {Greenough}.
\newblock {The Jeans condition: a new constraint on spatial resolution in
  simulations of isothermal self-gravitational hydrodynamics}.
\newblock {\em \apjl}, 489:L179+, 1997.

\bibitem{1902RSPTA.199....1J}
J.~H. {Jeans}.
\newblock {The Stability of a Spherical Nebula}.
\newblock {\em Roy. Soc. Lond. Phil. Trans. Ser. A}, 199:1--53, 1902.

\bibitem{1956MNRAS.116..351B}
W.~B. {Bonnor}.
\newblock {Boyle's Law and gravitational instability}.
\newblock {\em \mnras}, 116:351--359, 1956.

\bibitem{2004ARA&A..42...79B}
V.~{Bromm} and R.~B. {Larson}.
\newblock {The first stars}.
\newblock {\em \araa}, 42:79--118, 2004.

\bibitem{1987gady.book.....B}
J.~{Binney} and S.~{Tremaine}.
\newblock {\em {Galactic dynamics}}.
\newblock 1987.

\bibitem{2001ApJ...553..174G}
C.~F. {Gammie}.
\newblock {Nonlinear outcome of gravitational instability in cooling, gaseous
  disks}.
\newblock {\em \apj}, 553:174--183, 2001.

\bibitem{2007prpl.conf..607D}
R.~H. {Durisen}, A.~P. {Boss}, L.~{Mayer}, A.~F. {Nelson}, T.~{Quinn}, and
  W.~K.~M. {Rice}.
\newblock {Gravitational instabilities in gaseous protoplanetary disks and
  implications for giant planet formation}.
\newblock {\em Protostars and Planets V}, pages 607--622, 2007.

\bibitem{2009arXiv0912.0288K}
R.~S. {Klessen} and P.~{Hennebelle}.
\newblock {Accretion-driven turbulence as universal process: galaxies,
  molecular clouds, and protostellar disks}.
\newblock {\em \aap}, 520:A17+, September 2010.

\bibitem{2003MNRAS.339..577B}
M.~R. {Bate}, I.~A. {Bonnell}, and V.~{Bromm}.
\newblock {The formation of a star cluster: predicting the properties of stars
  and brown dwarfs}.
\newblock {\em \mnras}, 339:577--599, 2003.

\bibitem{2007ApJ...661..972P}
P.~{Padoan}, {\AA}.~{Nordlund}, A.~G. {Kritsuk}, M.~L. {Norman}, and P.~S.
  {Li}.
\newblock {Two regimes of turbulent fragmentation and the stellar initial mass
  function from primordial to present-day star formation}.
\newblock {\em \apj}, 661:972--981, 2007.

\bibitem{2008MNRAS.389.1556B}
I.~A. {Bonnell}, P.~{Clark}, and M.~R. {Bate}.
\newblock {Gravitational fragmentation and the formation of brown dwarfs in
  stellar clusters}.
\newblock {\em \mnras}, 389:1556--1562, 2008.

\bibitem{1999ApJ...514L..99K}
M.~J. {Korpi}, A.~{Brandenburg}, A.~{Shukurov}, I.~{Tuominen}, and
  {\AA}.~{Nordlund}.
\newblock {A supernova-regulated interstellar medium: simulations of the
  turbulent multiphase medium}.
\newblock {\em \apjl}, 514:L99--L102, 1999.

\bibitem{2004A&A...425..899D}
M.~A. {de Avillez} and D.~{Breitschwerdt}.
\newblock {Volume filling factors of the ISM phases in star forming galaxies.
  I. The role of the disk-halo interaction}.
\newblock {\em \aap}, 425:899--911, 2004.

\bibitem{2005ApJ...626..864M}
M.-M. {Mac Low}, D.~S. {Balsara}, J.~{Kim}, and M.~A. {de Avillez}.
\newblock {The Distribution of Pressures in a Supernova-driven Interstellar
  Medium. I. Magnetized Medium}.
\newblock {\em \apj}, 626:864--876, 2005.

\bibitem{1977ApJ...218..148M}
C.~F. {McKee} and J.~P. {Ostriker}.
\newblock {A theory of the interstellar medium - three components regulated by
  supernova explosions in an inhomogeneous substrate}.
\newblock {\em \apj}, 218:148--169, 1977.

\bibitem{1999intu.conf..127K}
M.~{Korpi}, A.~{Brandenburg}, A.~{Shukurov}, and I.~{Tuominen}.
\newblock {Vortical motions driven by supernova explosions}.
\newblock pages 127--131, 1999.

\bibitem{2010arXiv1008.5281D}
F.~{Del Sordo} and A.~{Brandenburg}.
\newblock {Vorticity production through rotation, shear and baroclinicity}.
\newblock {\em arXiv:1008.5281}, 2010.

\bibitem{2006MNRAS.370..415M}
A.~J. {Mee} and A.~{Brandenburg}.
\newblock {Turbulence from localized random expansion waves}.
\newblock {\em \mnras}, 370:415--419, 2006.

\bibitem{1998PhFl...10..237P}
D.~H. {Porter}, P.~R. {Woodward}, and A.~{Pouquet}.
\newblock {Inertial range structures in decaying compressible turbulent flows}.
\newblock {\em Phys. Fluids}, 10:237--245, 1998.

\bibitem{2004MNRAS.353..947H}
N.~E.~L. {Haugen}, A.~{Brandenburg}, and A.~J. {Mee}.
\newblock {Mach number dependence of the onset of dynamo action}.
\newblock {\em \mnras}, 353:947--952, 2004.

\bibitem{2009arXiv0912.0546K}
A.~G. {Kritsuk}, S.~D. {Ustyugov}, M.~L. {Norman}, and P.~{Padoan}.
\newblock {Self-organization in Turbulent Molecular Clouds: Compressional
  Versus Solenoidal Modes}.
\newblock 429:15, September 2010.

\bibitem{2010arXiv1008.0665P}
L.~{Pan} and E.~{Scannapieco}.
\newblock {Mixing in Supersonic Turbulence}.
\newblock {\em \apj}, 721:1765--1782, October 2010.

\bibitem{1994PhRvD..49.3854I}
J.~{Ignatius}, K.~{Kajantie}, H.~{Kurki-Suonio}, and M.~{Laine}.
\newblock {Growth of bubbles in cosmological phase transitions}.
\newblock {\em \prd}, 49:3854--3868, 1994.

\bibitem{1986PhRvD..34.1719K}
K.~{Kajantie} and H.~{Kurki-Suonio}.
\newblock {Bubble growth and droplet decay in the quark-hadron phase transition
  in the early Universe}.
\newblock {\em \prd}, 34:1719--1738, 1986.

\bibitem{1987QJRMS.113..413M}
P.~J. {Mason} and D.~J. {Thomson}.
\newblock {Large-eddy simulations of the neutral-static-stability planetary
  boundary layer}.
\newblock {\em Quart. J. Roy. Met. Soc.}, 113:413--443, 1987.

\bibitem{1994BoLMe..71..247S}
P.~P. {Sullivan}, J.~C. {McWilliams}, and C.-H. {Moeng}.
\newblock {A subgrid-scale model for large-eddy simulation of planetary
  boundary-layer flows}.
\newblock {\em Boundary-Layer Meteorology}, 71:247--276, 1994.

\bibitem{2004PhFl...16.1020B}
A.~{Brandenburg}, P.~J. {K{\"a}pyl{\"a}}, and A.~{Mohammed}.
\newblock {Non-Fickian diffusion and tau approximation from numerical
  turbulence}.
\newblock {\em Phys. Fluids}, 16:1020--1027, 2004.

\bibitem{2007MNRAS.376.1238S}
S.~{Sur}, K.~{Subramanian}, and A.~{Brandenburg}.
\newblock {Kinetic and magnetic {$\alpha$}-effects in non-linear dynamo
  theory}.
\newblock {\em \mnras}, 376:1238--1250, 2007.

\bibitem{2008A&A...482..739B}
A.~{Brandenburg}, K.-H. {R{\"a}dler}, and M.~{Schrinner}.
\newblock {Scale dependence of alpha effect and turbulent diffusivity}.
\newblock {\em \aap}, 482:739--746, 2008.

\bibitem{2009MNRAS.395.1599B}
A.~{Brandenburg}, A.~{Svedin}, and G.~M. {Vasil}.
\newblock {Turbulent diffusion with rotation or magnetic fields}.
\newblock {\em \mnras}, 395:1599--1606, 2009.

\bibitem{1979cmft.book.....P}
E.~N. {Parker}.
\newblock {\em {Cosmical magnetic fields: their origin and their activity}}.
\newblock 1979.

\bibitem{1980mfmd.book.....K}
F.~{Krause} and K.~H. {Raedler}.
\newblock {\em {Mean-field magnetohydrodynamics and dynamo theory}}.
\newblock 1980.

\bibitem{1976JFM....77..321P}
A.~{Pouquet}, U.~{Frisch}, and J.~{Leorat}.
\newblock {Strong MHD helical turbulence and the nonlinear dynamo effect}.
\newblock {\em J. Fluid Mech.}, 77:321--354, 1976.

\bibitem{2002PhRvL..89z5007B}
E.~G. {Blackman} and G.~B. {Field}.
\newblock {New dynamical mean-field dynamo theory and closure approach}.
\newblock {\em Phys. Rev. Lett.}, 89:265007, 2002.

\bibitem{2003GApFD..97..249R}
K.-H. {R{\"a}dler}, N.~{Kleeorin}, and I.~{Rogachevskii}.
\newblock {The mean electromotive force for MHD turbulence: the case of a weak
  mean magnetic field and slow rotation}.
\newblock {\em Geophys. Astrophys. Fluid Dynam.}, 97:249--274, 2003.

\bibitem{1996PhRvE..54.4532C}
F.~{Cattaneo} and D.~W. {Hughes}.
\newblock {Nonlinear saturation of the turbulent {$\alpha$} effect}.
\newblock {\em \pre}, 54:R4532--R4535, 1996.

\bibitem{2000A&A...361L...5K}
N.~{Kleeorin}, D.~{Moss}, I.~{Rogachevskii}, and D.~{Sokoloff}.
\newblock {Helicity balance and steady-state strength of the dynamo generated
  galactic magnetic field}.
\newblock {\em \aap}, 361:L5--L8, 2000.

\bibitem{2000MNRAS.318..724B}
E.~G. {Blackman} and G.~B. {Field}.
\newblock {Coronal activity from dynamos in astrophysical rotators}.
\newblock {\em \mnras}, 318:724--732, 2000.

\bibitem{2006ApJ...648L..71S}
K.~{Subramanian} and A.~{Brandenburg}.
\newblock {Magnetic helicity density and its flux in weakly inhomogeneous
  turbulence}.
\newblock {\em \apjl}, 648:L71--L74, 2006.

\bibitem{2001ApJ...550..752V}
E.~T. {Vishniac} and J.~{Cho}.
\newblock {Magnetic helicity conservation and astrophysical dynamos}.
\newblock {\em \apj}, 550:752--760, 2001.

\bibitem{2004PhRvL..93t5001S}
K.~{Subramanian} and A.~{Brandenburg}.
\newblock {Nonlinear current helicity fluxes in turbulent dynamos and alpha
  quenching}.
\newblock {\em Phys. Rev. Lett.}, 93:205001, 2004.

\bibitem{2010arXiv1006.3549H}
A.~{Hubbard} and A.~{Brandenburg}.
\newblock {Magnetic helicity flux in the presence of shear}.
\newblock {\em arXiv:1006.3549}, 2010.

\bibitem{2006A&A...448L..33S}
A.~{Shukurov}, D.~{Sokoloff}, K.~{Subramanian}, and A.~{Brandenburg}.
\newblock {Galactic dynamo and helicity losses through fountain flow}.
\newblock {\em \aap}, 448:L33--L36, 2006.

\bibitem{1980GApFD..16..239R}
G.~{R\"udiger}.
\newblock {Reynolds stresses and differential rotation. I - On recent
  calculations of zonal fluxes in slowly rotating stars}.
\newblock {\em Geophys. Astrophys. Fluid Dynam.}, 16:239--261, 1980.

\bibitem{1989drsc.book.....R}
G.~{R\"udiger}.
\newblock {\em {Differential rotation and stellar convection. Sun and the solar
  stars}}.
\newblock 1989.

\bibitem{1995A&A...299..446K}
L.~L. {Kitchatinov} and G.~{R\"udiger}.
\newblock {Differential rotation in solar-type stars: revisiting the
  Taylor-number puzzle.}
\newblock {\em \aap}, 299:446--452, 1995.

\bibitem{1993ApJ...407..367D}
B.~R. {Durney}.
\newblock {On the solar differential rotation - Meridional motions associated
  with a slowly varying angular velocity}.
\newblock {\em \apj}, 407:367--379, 1993.

\bibitem{1989ApJ...338..509D}
B.~R. {Durney}.
\newblock {On the behavior of the angular velocity in the lower part of the
  solar convection zone}.
\newblock {\em \apj}, 338:509--527, 1989.

\bibitem{1986ApJS...61..585G}
P.~A. {Gilman} and J.~{Miller}.
\newblock {Nonlinear convection of a compressible fluid in a rotating spherical
  shell}.
\newblock {\em \apjs}, 61:585--608, 1986.

\bibitem{1994A&A...286..471R}
M.~{Rieutord}, A.~{Brandenburg}, A.~{Mangeney}, and P.~{Drossart}.
\newblock {Reynolds stresses and differential rotation in Boussinesq convection
  in a rotating spherical shell}.
\newblock {\em \aap}, 286:471--480, 1994.

\bibitem{1993A&A...267..265P}
P.~{Pulkkinen}, I.~{Tuominen}, A.~{Brandenburg}, A.~{Nordlund}, and R.~F.
  {Stein}.
\newblock {Rotational effects on convection simulated at different latitudes}.
\newblock {\em \aap}, 267:265--274, 1993.

\bibitem{1990SoPh..128..243B}
A.~{Brandenburg}, I.~{Tuominen}, D.~{Moss}, and G.~{R\"udiger}.
\newblock {The nonlinear solar dynamo and differential rotation - A Taylor
  number puzzle?}
\newblock {\em \solphys}, 128:243--251, 1990.

\bibitem{2005AN....326..379K}
L.~L. {Kitchatinov} and G.~{R{\"u}diger}.
\newblock {Differential rotation and meridional flow in the solar convection
  zone and beneath}.
\newblock {\em Astron. Nachr.}, 326:379--385, 2005.

\bibitem{2002ApJ...570..865B}
A.~S. {Brun} and J.~{Toomre}.
\newblock {Turbulent convection under the influence of rotation: sustaining a
  strong differential rotation}.
\newblock {\em \apj}, 570:865--885, 2002.

\bibitem{2006ApJ...641..618M}
M.~S. {Miesch}, A.~S. {Brun}, and J.~{Toomre}.
\newblock {Solar differential rotation influenced by latitudinal entropy
  variations in the tachocline}.
\newblock {\em \apj}, 641:618--625, 2006.

\bibitem{2007A&A...474..239J}
L.~{Jouve} and A.~S. {Brun}.
\newblock {On the role of meridional flows in flux transport dynamo models}.
\newblock {\em \aap}, 474:239--250, 2007.

\bibitem{2007ApJ...669.1190B}
J.~{Ballot}, A.~S. {Brun}, and S.~{Turck-Chi{\`e}ze}.
\newblock {Simulations of turbulent convection in rotating young solar-like
  stars: differential rotation and meridional circulation}.
\newblock {\em \apj}, 669:1190--1208, 2007.

\bibitem{1993A&A...279L...1K}
M.~{K\"uker}, G.~{R\"udiger}, and L.~L. {Kitchatinov}.
\newblock {An alpha Omega-model of the solar differential rotation}.
\newblock {\em \aap}, 279:L1--L4, 1993.

\bibitem{1994AN....315..157K}
L.~L. {Kitchatinov}, V.~V. {Pipin}, and G.~{R\"udiger}.
\newblock {Turbulent viscosity, magnetic diffusivity, and heat conductivity
  under the influence of rotation and magnetic field}.
\newblock {\em Astron. Nachr.}, 315:157--170, 1994.

\bibitem{2005ApJ...622.1320R}
M.~{Rempel}.
\newblock {Solar differential rotation and meridional flow: the role of a
  subadiabatic tachocline for the Taylor-Proudman balance}.
\newblock {\em \apj}, 622:1320--1332, 2005.

\bibitem{2009MNRAS.395.2056B}
S.~A. {Balbus}.
\newblock {A simple model for solar isorotational contours}.
\newblock {\em \mnras}, 395:2056--2064, 2009.

\bibitem{2009MNRAS.400..176B}
S.~A. {Balbus}, J.~{Bonart}, H.~N. {Latter}, and N.~O. {Weiss}.
\newblock {Differential rotation and convection in the Sun}.
\newblock {\em \mnras}, 400:176--182, 2009.

\bibitem{1989A&A...217..217T}
I.~{Tuominen} and G.~{R\"udiger}.
\newblock {Solar differential rotation as a multiparameter turbulence problem}.
\newblock {\em \aap}, 217:217--228, 1989.

\bibitem{2005AN....326..245S}
M.~{Schrinner}, K.-H. {R{\"a}dler}, D.~{Schmitt}, M.~{Rheinhardt}, and
  U.~{Christensen}.
\newblock {Mean-field view on rotating magnetoconvection and a geodynamo
  model}.
\newblock {\em Astron. Nachr.}, 326:245--249, 2005.

\bibitem{2007GApFD.101...81S}
M.~{Schrinner}, K.-H. {R{\"a}dler}, D.~{Schmitt}, M.~{Rheinhardt}, and U.~R.
  {Christensen}.
\newblock {Mean-field concept and direct numerical simulations of rotating
  magnetoconvection and the geodynamo}.
\newblock {\em Geophys. Astrophys. Fluid Dynam.}, 101:81--116, 2007.

\bibitem{1972JFM....56..287K}
R.~H. {Kraichnan}.
\newblock {Test-field model for inhomogeneous turbulence}.
\newblock {\em J. Fluid Mech.}, 56:287--304, 1972.

\bibitem{2009arXiv0906.3314M}
E.~J.~M. {Madarassy} and A.~{Brandenburg}.
\newblock {Calibrating passive scalar transport in shear-flow turbulence}.
\newblock {\em \pre}, 82:016304, 2010.

\bibitem{2009ApJ...706..712H}
A.~{Hubbard} and A.~{Brandenburg}.
\newblock {Memory effects in turbulent transport}.
\newblock {\em \apj}, 706:712--726, 2009.

\bibitem{2008MNRAS.385L..15S}
S.~{Sur}, A.~{Brandenburg}, and K.~{Subramanian}.
\newblock {Kinematic {$\alpha$}-effect in isotropic turbulence simulations}.
\newblock {\em \mnras}, 385:L15--L19, 2008.

\bibitem{2009A&A...495....1M}
D.~{Mitra}, P.~J. {K{\"a}pyl{\"a}}, R.~{Tavakol}, and A.~{Brandenburg}.
\newblock {Alpha effect and diffusivity in helical turbulence with shear}.
\newblock {\em \aap}, 495:1--8, 2009.

\bibitem{2008ApJ...687L..49B}
A.~{Brandenburg}, K.-H. {R{\"a}dler}, M.~{Rheinhardt}, and K.~{Subramanian}.
\newblock {Magnetic quenching of {$\alpha$} and diffusivity tensors in helical
  turbulence}.
\newblock {\em \apjl}, 687:L49--L52, 2008.

\bibitem{2009JFM...621..205C}
F.~{Cattaneo} and S.~M. {Tobias}.
\newblock {Dynamo properties of the turbulent velocity field of a saturated
  dynamo}.
\newblock {\em J. Fluid Mech.}, 621:205--214, 2009.

\bibitem{2008MNRAS.391.1477T}
A.~{Tilgner} and A.~{Brandenburg}.
\newblock {A growing dynamo from a saturated Roberts flow dynamo}.
\newblock {\em \mnras}, 391:1477--1481, 2008.

\bibitem{2001A&A...377...84T}
R.~{Tsch{\"a}pe} and G.~{R{\"u}diger}.
\newblock {Rotation-induced lithium depletion of solar-type stars in open
  stellar clusters}.
\newblock {\em \aap}, 377:84--89, 2001.

\bibitem{2001A&A...375..149R}
G.~{R{\"u}diger} and V.~V. {Pipin}.
\newblock {Lithium as a passive tracer probing the rotating solar tachocline
  turbulence}.
\newblock {\em \aap}, 375:149--154, 2001.

\end{thebibliography}

\vfill\bigskip\noindent\tiny\begin{verbatim}
$Header: /var/cvs/brandenb/tex/hydro/rop/paper.tex,v 1.125 2010-12-01 14:54:41 brandenb Exp $
\end{verbatim}

\end{document}